%% file: spinpump_q.tex
\documentclass[aps,prb,preprint]{revtex4-1}
\usepackage{amsmath}
\usepackage{graphicx}
\usepackage{amssymb}

\usepackage{color} 
\usepackage{graphicx}
\usepackage{bm}
\usepackage{amsmath}

\usepackage{gt15}

\DeclareMathAlphabet\mathbfcal{OMS}{cmsy}{b}{n}

\begin{document}
\title{
Consistent microscopic analysis of spin pumping effects
}
\author{Gen Tatara}
\affiliation{RIKEN Center for Emergent Matter Science (CEMS), 
2-1 Hirosawa, Wako, Saitama, 351-0198 Japan}
\author{Shigemi Mizukami}
\affiliation{WPI - Advanced Insitute for Materials Research, Tohoku University
Katahira 2-1-1, Sendai,  Japan}

\date{\today}

\begin{abstract}
We present a consistent microscopic study of spin pumping effects for both metallic and insulating ferromagnets. 
As for metallic case, we present a simple quantum mechanical picture of the effect as due to the electron spin flip as a result of a nonadiabatic (off-diagonal) spin gauge field. The effect of interface spin-orbit interaction is briefly discussed. We also carry out field-theoretic calculation to discuss on the equal footing  the spin current generation and  
 torque effects such as enhanced Gilbert damping constant and shift of precession frequency both in metallic and insulating cases.  
 For thick ferromagnetic metal, our study reproduces results of previous theories such as the correspondence between the dc component of the spin current and enhancement of the damping.
For thin metal and insulator, the relation turns out to be modified. 
For the insulating case, driven locally by interface $sd$ exchange interaction due to magnetic proximity effect,  physical mechanism is distinct from the metallic case. 
Further study of proximity effect and interface spin-orbit interaction would be crucial to interpret experimental results in particular for insulators. 
\end{abstract}  

\maketitle

\def\Ascal#1#2{{\cal A}_{{\rm s},#1}^{#2}}
\def\Ascalv#1{ \boldsymbol{\mathcal A}_{{\rm s},#1}} 
\newcommand{\imag}{\eta_\tau}
\newcommand{\boson}{b}
\newcommand{\Jint}{J_{\rm I}}
\newcommand{\tso}{t}
\newcommand{\tsotil}{\widetilde{t}}

\section{Introduction}

\begin{figure}[tbh]
  \begin{center}
  \includegraphics[width=0.4\hsize]{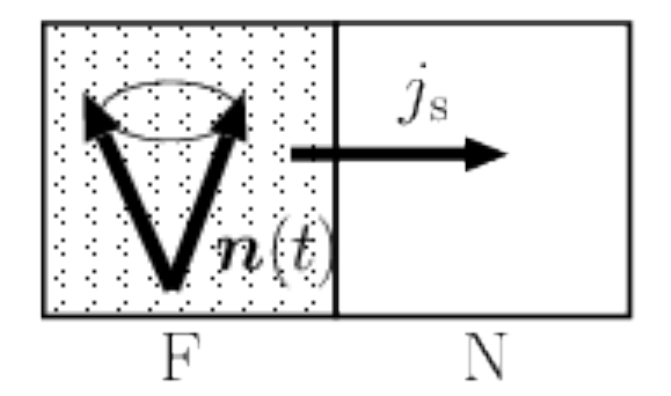}
  \end{center}
\caption{Spin pumping effect in a junction of ferromagnet (F) and normal metal (N). Dynamic magnetization $\nv(t)$ generates a spin current $j_{\rm s}$ through the interface. 
\label{FIGFN0}}
\end{figure}

Spin current generation is of a fundamental importance in spintronics. 
A dynamic method using magnetization precession induced by an applied magnetic field, called the spin pumping effect, turns out to be particularly useful \cite{Saitoh06} and is widely used in a junction of a ferromagnet (F) and a normal metal (N)(Fig. \ref{FIGFN0}).
The generated spin current density (in unit of A/m$^2$) has two independent components, proportional to $\dot{\nv}$ and $\nv\times\dot{\nv}$, where $\nv$ is a unit vector describing the direction of localized spin, and thus is represented phenomenologically as 
\begin{equation}
{\jv}_s = \frac{e}{4\pi} \left( A_{\rm r} \nv \times \dot{\nv}+ A_{\rm i} \dot{\nv} \right),\label{Jsphenom}
\end{equation}
where $e$ is the elementally electric charge and $A_{\rm r} $ and $A_{\rm i}$ are phenomenological constants having unit of $1/$m$^2$.
Spin pumping effect was theoretically formulated by Tserkovnyak et al. \cite{Tserkovnyak02}
 by use of scattering matrix approach \cite{Moskalets12}.
This approach, widely applied in mesoscopic physics, describes transport phenomena in terms of transmission and reflection amplitudes (scattering matrix), and provides quantum mechanical pictures of the phenomena without calculating explicitly the amplitudes. 
Tserkovnyak et al. applied the scattering matrix formulation of general adiabatic pumping \cite{Buttiker94,Brouwer98} to the spin-polarized case. 
The spin pumping effect was described in Ref. \cite{Tserkovnyak02} in terms of spin-dependent transmission and reflection coefficients at the FN interface,
and  it was demonstrated that the two parameters, $A_{\rm r} $ and $A_{\rm i}$, are the real and the imaginary part of a complex parameter called the spin mixing conductance.
The spin mixing conductance, which is represented by transmission and reflection coefficients, turned out to be a convenient parameter for discussing spin current generation and other effects like the inverse spin-Hall effect. 
Nevertheless, scattering approach hides microscopic physical pictures of what is going on, as the scattering coefficients are not fundamental material parameters but are composite quantities of Fermi wave vector,  electron effective mass and the interface properties.

Effects of slowly-varying potential is described in a physically straightforward and clear manner by use of a unitary transformation that represents the time-dependence. (See Sec. \ref{SECQMF} for details.)
The laboratory frame wave function under time-dependent potential,  $\ket{\psi(t)}$, is written in terms of a static ground state ('rotated frame' wave function) $\ket{\phi}$ and a unitary matrix $U(t)$ as 
$\ket{\psi(t)}=U(t)\ket{\phi}$. 
The time-derivative $\partial_t$ is then replaced by a covariant derivative, $\partial_t+(U^{-1}\partial_t U)$, and 
the effects of time-dependence are represented by (the time-component of) an effective gauge field, ${\cal A}\equiv -i(U^{-1}\partial_t U)$ 
(See Eq. (\ref{Seqrot})). 
In the same manner as the electromagnetic gauge field, the effective gauge field generates a current if spatial homogeneity is present  (like in junctions) and this is a physical origin of adiabatic pumping effect in metals. 

In the perturbative regime or in insulators, a simple picture instead of effective gauge field can be presented.
Let us focus on the case driven by an $sd$ exchange interaction, $\Jsd\nv(t)\cdot\sigmav$, where $\Jsd$ is a coupling constant and $\sigmav$ is the electron spin.
Considering the second-order effect of the $sd$ exchange interaction,  the electron wave function has a contribution of a time-dependent amplitude 
\begin{align}
 {\cal U}(t_1,t_2)&= (\Jsd)^2(\nv(t_1)\cdot\sigmav)(\nv(t_2)\cdot\sigmav)
 = (\Jsd)^2[(\nv(t_1)\cdot\nv(t_2))+i[\nv(t_1)\times\nv(t_2)]\cdot\sigmav] 
,
\end{align}
where  $t_1$ and $t_2$ are the time of the interactions.
The first term on the right-hand side, representing the amplitude for charge degrees of freedom, is neglected.
The spin contribution vanishes for static spin configuration, as is natural, while for slowly varying case, it reads 
\begin{align}
{\cal U}(t_1,t_2)&\simeq 
  -i(t_1-t_2) (\Jsd)^2 (\nv\times\dot{\nv})(t_1)\cdot\sigmav.
\end{align}
As a result of this amplitude, spin accumulation and spin current is induced proportional to $\nv\times\dot{\nv}$.
The fact indicates that  $\nv\times\dot{\nv}$ plays a role of an effective scalar potential or voltage in electromagnetism, as we shall demonstrate in Sec. \ref{SEC:interfacespincurrent} for insulators.
(The factor of time difference is written in terms of derivative with respect to energy or angular frequency in a rigorous derivation. See for example,  Eqs. (\ref{tildecalDs})(\ref{calDs}).)
The essence of spin pumping effect is therefore the non-commutativity of spin operators.
The above picture in the perturbative regime naturally leads to an effective gauge field in the strong coupling limit \cite{TK03}. 

The same scenario applies for cases of spatial variation of spin, and an equilibrium spin current proportional to $\nv\times\nabla_i \nv$ emerges, where $i$ denotes the direction of spatial variation \cite{TKS_PR08}. 
The spin pumping effect is therefore the time analog of the equilibrium spin current induced by vector spin chirality. 
Moreover, charge current emerges from the third-order process from the identity \cite{TK03}
\begin{align}
 \tr[(\nv_1\cdot\sigmav)(\nv_2\cdot\sigmav)(\nv_3\cdot\sigmav)]
 &= 2i \nv_1\cdot(\nv_2\times\nv_3),
\end{align}
and this factor, a scalar spin chirality, is the analog of the spin Berry phase in the perturbative regime. 
The spin pumping effect and spin Berry's phase and spin motive force have the same physical root, namely the non-commutative spin algebra.

From the scattering matrix theory view point the cases of metallic and insulating ferromagnet make no difference as what conduction electrons in the normal metal see is the interface.  
From physical viewpoints, such treatment appears too crude. 
Unlike the metallic case discussed above, in the case of insulator ferromagnet, the coupling between the magnetization and the conduction electron in normal metal occurs due to a magnetic proximity effect at the interface. 
Thus the spin pumping by an insulator ferromagnet seems to be a locally-induced perturbative effect rather than a transport induced by a driving force due to a generalized gauge field.
We therefore need to apply different approaches for the two cases as briefly argued above. 
In the insulating case, one may think that magnon spin current is generated inside the ferromagnet because magnon itself  couples to an effective gauge field \cite{Dugaev05} similarly to the electrons in metallic case. 
This is not, however, true, because the gauge field for magnon is abelian (U(1)).
Although scattering matrix approach apparently seems to apply to both metallic and insulating cases, it would be instructive to present in this paper a consistent microscopic description of the effects to see different physics governing the two cases.

\subsection{Brief overview of theories and scope of the paper}

Before carrying out calculation, let us overview history of theoretical studies of spin pumping effect. 
Spin current generation in a metallic junction was originally discussed by Silsbee \cite{Silsbee79}  before Tserkovnyak et al. 
It was shown there that  dynamic magnetization induces spin accumulation at the interface, resulting in a diffusive flow of spin in the normal metal.  
Although  of experimental curiosity at that time was the interface spin accumulation, which enhances the signal of conduction electron spin resonance, 
it would be fair to say that Silsbee  pointed out the \textquoteleft spin pumping effect\textquoteright. 

In Ref. \cite{Tserkovnyak02}, spin pumping effect was originally argued in the context of enhancement of Gilbert damping in FN junction, which had been a hot issue after the study by Berger \cite{Berger96}, who studied the case of FNF junction based on a quantum mechanical argument.
Berger discussed that when a normal metal is attached to a ferromagnet, the damping of ferromagnet is enhanced as a result of spin polarization formed in the normal metal, and the effect was experimentally confirmed by Mizukami \cite{Mizukami01}.
Tserkovnyak et al. pointed out that the effect has a different interpretation of the counter action of spin current generation, because the spin current injected into the normal metal indicates  a change of spin angular momentum or a  torque on ferromagnet. 
In fact, the equation of motion for the magnetization of ferromagnet reads 
\begin{align}
 \dot{\nv}=-\gamma \Bv\times\nv-\alpha \nv\times\dot{\nv}-\frac{a^3}{eSd}\jv_{\rm s},
\end{align}
where $\gamma$ is the gyromagnetic ratio, $\alpha$ is the Gilbert damping coefficient, $d$ is the thickness of the ferromagnet, $S$ is the magnetude of localized spin, and $a$ is the lattice constant.
Spin current of Eq. (\ref{Jsphenom}) thus indicates that the  gyromagnetic ratio and the the Gilbert damping coefficient are modified by the spin pumping effect to be  \cite{Tserkovnyak02}
\begin{align}
 \tilde{\alpha} &= \alpha+\frac{a^3}{4\pi Sd}A_{\rm r} \nnr 
 \tilde{\gamma} &= \gamma \lt[1+\frac{a^3}{4\pi Sd}A_{\rm i}\rt]^{-1} . \label{alphag}
\end{align}
The spin pumping effect is therefore detected by measuring the effective  damping constant and  gyromagnetic ratio.
The formula (\ref{alphag}) is, however, based on a naive picture neglecting the position-dependence of the damping torque  and the relation between the pumped spin current amplitude and damping or $\gamma$ would not be so simple in reality. (See Sec. \ref{SEC:damping}.) 

The issue of damping in FN junction was formulated based on linear-response theory by Simanek and Heinirch \cite{SimanekHeinrich03,Simanek03}. They showed that the damping coefficient is given by the first-order derivative with respect to the angular frequency $\omega$ of the imaginary part of the spin correlation function and argued that the damping effect is consistent with the Tserkovnyak's spin pumping effect. 
Recently, a microscopic formulation of spin pumping effect in metallic junction was provided by Chen and Zhang \cite{Chen15} and one of the author \cite{TataraSP16} by use of the Green's functions, and  a transparent microscopic  picture of pumping effect was provided. 
Scattering representation and Green's function one are related \cite{Chen15} because the asymptotic behaviors of the Green's functions  at long distance are governed by transmission coefficient \cite{Fisher81}.
In the study of Ref. \cite{TataraSP16}, the uniform ferromagnet was treated as a dot having only two degrees of freedom of spin. Such simplification neglects the dependence on electron wave vectors in ferromagnets and thus cannot discuss the  the case of inhomogeneous magnetization or position-dependence of spin damping. 

The aim of this paper is to provide a microscopic and consistent theoretical formulation of spin pumping effect for metallic and insulating  ferromagnets. 
We do not rely on the scattering approach.
Instead we provide elementary quantum mechanical argument to demonstrated that spin current generation is a natural consequence of magnetization dynamics (Sec. \ref{SECQM}).
Based on the formulation, the effect of interface spin-orbit interaction is discussed in Sec. \ref{SECQMSO}. 
We also provide a rigorous formulation based on field-theoretic approach emploied in Ref. \cite{TataraSP16} in Sec. \ref{SEC:GF}.
We also reproduce within the same framework Berger's result \cite{Berger96} that the spin pumping effect is equivalent to the enhancement of the spin damping (Sec. \ref{SEC:damping}). 
Effect of inhomogeneous magnetization is briefly discussed in Sec. \ref{SEC:structure}. 

Case of insulating ferromagnet is studied in Sec. \ref{SEC:insulator} assuming that the pumping is induced by an interface  exchange interaction between the magnetization and conduction electron in normal metal, namely, by magnetic proximity effect. 
The interaction is treated perturbatively similarly to Refs. \cite{Takeuchi08,Hosono_LT09}.
The dominant contribution to the spin current, the one linear in the interface exchange interaction, turns out to be proportional to $\dot{\nv}$, while the one proportional to $\nv\times\dot{\nv}$ is weaker if the proximity effect is weak.

The contribution from the magnon, magnetization fluctuation, is also studied. 
As has been argued \cite{Dugaev05}, a gauge field for magnon emerges from magnetization dynamics. It is, however, an  adiabatic one diagonal in spin, which acts as chemical potential for magnon giving rise only to adiabatic spin polarization proportional to $\nv$.
This is in sharp contrast to the metallic case, where electrons are directly driven by spin-flip component of spin gauge field, resulting in perpendicular  spin accumulation, i.e., along $\dot{\nv}$ and $\nv\times\dot{\nv}$. 
The excitation in ferromagnet when magnetization is time-dependent is therefore different for metallic and insulating cases.
We show that magnon excitation nevertheless generates perpendicular spin current, $\nv\times \dot{\nv}$, in the normal metal as a result of annihilation and creation at the interface, which in turn flips electron spin.
The result of magnon-driven contribution agrees with the one in previous study \cite{Adachi11} carried out in the context of thermally-driven spin pumping ('spin Seebeck' effect).
It is demonstrated that the magnon-induced spin current depends linearly on the temperature at high temperature compared to magnon energy.
The amplitude of magnon-driven spin current provides the magnitude of magnetic proximity effect. 
  
In our analysis, we calculate consistently the pumped spin current and change of the Gilbert damping and resonant frequency and obtain the relations among them. 
It is shown that the spin mixing conductance scenario saying that the magnitude of spin current proportional to $\nv\times\dot{\nv}$ is given by the enhancement factor of the Gilbert damping constant \cite{Tserkovnyak02}, applies only the case of thick ferromagnetic metal.
For thin metallic case and insulator case, different relations hold (See Sec. \ref{SECdiscussion}.).

\section{Quantum mechanical description of metallic case \label{SECQM}}

In this section, we derive the spin current generated by the magnetization dynamics of metallic ferromagnet by a quantum mechanical argument. It is sometimes useful for intuitive understanding, although the description may lack clearness as it cannot handle many-particle nature like particle distributions.
In Sec. \ref{SEC:GF} we formulate the problem in the field-theoretic language.

\subsection{Electrons in ferromagnet with dynamic magnetization \label{SECQMF}}
The model we consider is a junction of metallic ferromagnet (F) and a normal metal (N). 
The magnetization (or localized spins) in the ferromagnet is treated as spatially uniform but changing with time slowly.
As a result of strong $sd$ exchange interaction, the conduction electron's spin follows instantaneous directions of localized spins, i.e., the system is in the adiabatic limit.
The quantum mechanical Hamiltonian for the ferromagnet is
\begin{align}
 H_{\rm F} 
  &=  -\frac{\nabla^2}{2m} -\ef-\spol \nv(t)\cdot\sigmav ,
\end{align}
where $m$ is the electron's mass, $\sigmav$ is a vector of Pauli matrices, $\spol$ represents the energy splitting due to the $sd$ exchange interaction and $\nv(t)$ is a time-dependent unit vector  denoting the localized spin direction. 
The energy is measured from the Fermi energy $\ef$.

\begin{figure}[tbh]
  \begin{center}
  \includegraphics[width=0.4\hsize]{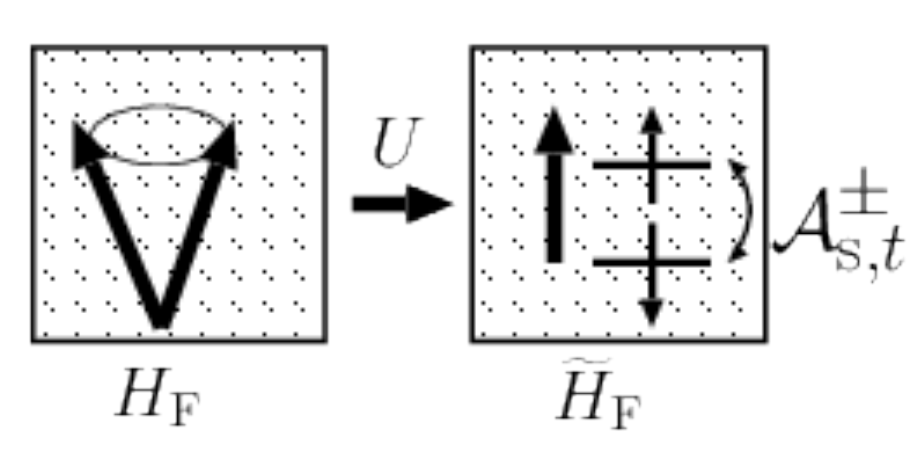}
  \end{center}
\caption{ Unitary transformation $U$ for conduction electron in ferromagnet converts the original Hamiltonian ${H}_{\rm F}$ into a diagonalized uniformly spin-polarized Hamiltonian $\widetilde{H}_{\rm F}$ and an interaction with spin gauge field, $\Ascalv{t}\cdot\sigmav$.
\label{FIGFNQM}}
\end{figure}

\begin{figure}[tbh]
  \begin{center}
    \includegraphics[height=6\baselineskip]{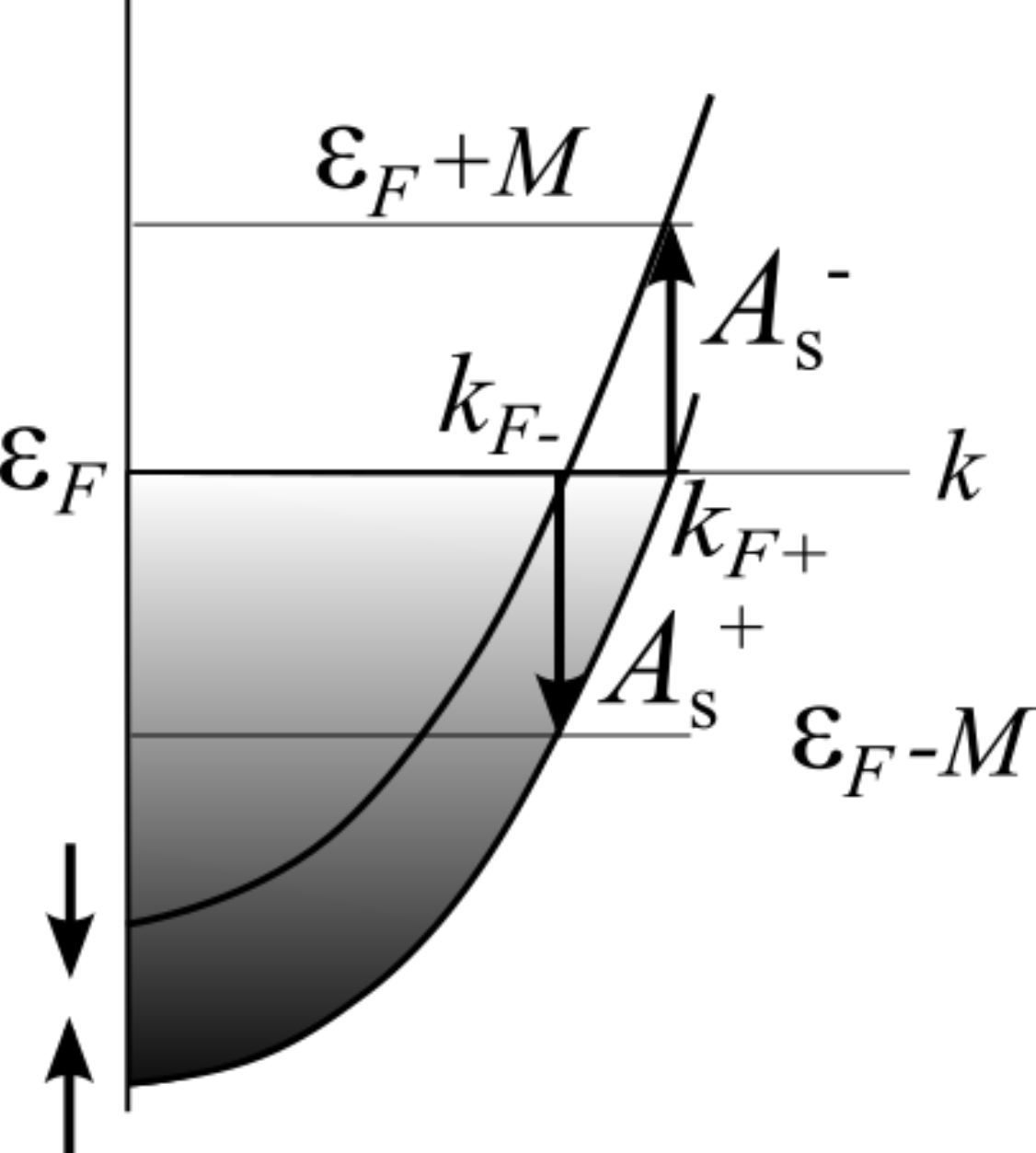}
  \end{center}
\caption{For uniform magnetization, the non-adiabatic components of the gauge field, $\Ascal{t}{\pm}$, induces a spin flip conserving the momentum. 
\label{FIGelectronband_sp}}
\end{figure}

As a result of the $sd$ exchange interaction, the electron's spin wave function is given by \cite{Sakurai94}
\begin{align}
 \ket{\nv}\equiv \cos\frac{\theta}{2}|\!\uparrow\rangle+\sin\frac{\theta}{2}e^{i\phi}|\!\downarrow\rangle                                                                                                    
\end{align}
where $\ket{\uparrow}$ and $\ket{\downarrow}$ represent the spin up and down states, respectively, and $(\theta,\phi)$ are polar coordinates for $\nv$.
To treat slowly varying localized spin, we switch to a rotating frame where the spin direction is defined with respect to instantaneous direction $\nv$ \cite{TKS_PR08}. 
This corresponds to diagonalizing the Hamiltonian at each time by  introducing a unitary matrix $U(t)$ as 
\begin{align}
 \ket{\nv(t)}\equiv U(t)|\!\uparrow\rangle,                                                                                             
\end{align}
where 
\begin{align}
   U(\rv)
= \lt(\begin{array}{cc} \cos\frac{\theta}{2} & \sin\frac{\theta}{2}e^{-i\phi} \\
        \sin\frac{\theta}{2}e^{i\phi} & -\cos\frac{\theta}{2} \end{array} \rt) ,
\label{Udef}
\end{align}
where states are in vector representation, i.e., 
$|\!\uparrow\rangle=\lt(\begin{array}{c} 1 \\ 0 \end{array} \rt)$ and 
$|\!\downarrow\rangle=\lt(\begin{array}{c} 0 \\ 1 \end{array} \rt)$.
The rotated Hamiltonian  is diagonalized as (in the momentum representation)   
\begin{align}
 \widetilde{H}_{\rm F}\equiv U^{-1}H_{\rm F} U= \ekv-\spol \sigma_z ,
\end{align}
where $\ekv\equiv \frac{k^2}{2m}-\ef$ is the kinetic energy in the momentum representation (Fig. \ref{FIGFNQM}).
%
In general, when a state $\ket{\psi}$  for a time-dependent Hamiltonian $H(t)$, satisfying  the Schr\"odinger equation  $i\delpo{t}\ket{\psi}=H(t)\ket{\psi}$, is written in terms of a state $\ket{\psi}$ connected by a unitary transformation  $\ket{\phi}\equiv U^{-1}\ket{\psi}$, the new state satisfies a modified Schr\"odinger equation \begin{align}
\lt(i\delpo{t}+i U^{-1}\delpo{t}U\rt)\ket{\phi}=\tilde{H}\ket{\phi},\label{Seqrot}
\end{align}
where $\tilde{H}\equiv U^{-1}H U$.
Namely, there arises a gauge field  $-i U^{-1}\delpo{t}U$ in the new frame $\ket{\phi}$. 
In the present case of dynamic localized spin, the gauge field has three  components (suffix $t$ denotes the time-component); 
\begin{align}
 \Ascal{t}{} & \equiv -i U^{-1}\delpo{t}U  
 \equiv \Ascalv{t} \cdot \sigmav,
 \label{gaugepotential}
\end{align}
explicitly given as \cite{TKS_PR08}
\begin{align}
\Ascalv{t}&= \frac{1}{2}
\lt( \begin{array}{c}
-\partial_t \theta \sin \phi -\sin\theta \cos\phi \partial_t \phi \\
\partial_t \theta \cos \phi -\sin\theta \sin\phi \partial_t \phi \\
  (1-\cos\theta)\partial_t \phi        
     \end{array}
\rt).
\label{Aexpression}
\end{align}
Including the gauge field in the Hamiltonian, the effective Hamiltonian in the rotated frame reads 
\begin{align}
 \widetilde{H}_{\rm F}^{\rm eff}\equiv\widetilde{H}_{\rm F} +  {\bf {\cal A}}_{{\rm s},t} \cdot \sigmav
 = \lt( \begin{array}{cc} 
         \epsilon_{k}-\spol -\Ascal{t}{z}  & \Ascal{t}{-} \\
                    \Ascal{t}{+}  & \epsilon_{k}+\spol +\Ascal{t}{z} 
        \end{array} \rt)
        \label{HF}
\end{align}
where 
$\Ascal{t}{\pm}\equiv \Ascal{t}{x}\pm i \Ascal{t}{y}$.
We see that the adiabatic ($z$) component of the gauge field, $\Ascal{t}{z}$, acts as a spin-dependent chemical potential (spin chemical potential) generated by dynamic magnetization, while non-adiabatic ($x$ and $y$) components causes spin mixing.
In the case of uniform magnetization we consider, the mixing is between the electrons with different spin $\uparrow$ and $\downarrow$ but having the same wave vector $\kv$, because the gauge field  $\Ascal{t}{\pm} $ carries no momentum.
This leads to a mixing of states having an excitation energy of $\spol$ as shown in Fig. \ref{FIGelectronband_sp}.
In low energy transport effects, what concern are the electrons at the Fermi energy; The wave vector $\kv$ should be chosen as $\kfu$ and $\kfd$, the Fermi wave vectors for $\uparrow$ and $\downarrow$ electrons, respectively. (Effects of finite momentum transfer is discussed in Sec. \ref{SEC:structure}. ) 

The Hamiltonian Eq. (\ref{HF}) is diagonalized to obtain energy eigenvalues of  
$\tilde{\epsilon}_{k\spinindex}=\ekv-\spinindex\sqrt{(\spol+\Ascal{t}{z})^2+|\Ascal{t}{\perp}|^2}$, where 
$|\Ascal{t}{\perp}|^2\equiv \Ascal{t}{+}\Ascal{t}{-}$ and $\sigma=\pm$ represents spin ($\uparrow$ and $\downarrow$ correspond to $+$ and $-$, respectively).
We are interested in the adiabatic limit, and so the contribution lowest-order, namely, the first order, in the perpendicular component, 
$\Ascal{t}{\perp}$, is sufficient. 
In the present rotating-frame approach, the gauge field is treated as a static potential, since it already include time-derivative to the linear order (Eq. (\ref{Aexpression})). 
Moreover, the adiabatic component of the gauge field, $\Ascal{t}{z}$,  is neglected, as it modifies the spin pumping only at the second-order of time-derivative. 
The energy eigenvalues, $\ekvs\simeq \ekv-\spinindex\spol$,  are thus unaffected by the gauge field, while the eigenstates to the linear order read 
\begin{align}
\ket{{{k\uparrow}}}_{\rm F} & \equiv \ket{k\uparrow} -\frac{\Ascal{t}{+}}{\spol}\ket{k\downarrow} \nnr
\ket{{{k\downarrow}}}_{\rm F} & \equiv \ket{k\downarrow} +\frac{\Ascal{t}{-}}{\spol}\ket{k\uparrow} ,
\label{Fstates}
\end{align}
corresponding to energy of $\epsilon_{\kv+}$ and $\epsilon_{\kv-}$, respectively.
For low energy transport, states we need to consider are  the following two having spin-dependent Fermi wave vectors, $\kfs$ for $\sigma=\uparrow,\downarrow$, namely 
\begin{align}
\ket{{k_{{\rm F}\uparrow}\uparrow}}_{\rm F} &= \ket{k_{{\rm F}\uparrow}\uparrow} -\frac{\Ascal{t}{+}}{\spol}\ket{k_{{\rm F}\uparrow}\downarrow} \nnr
\ket{{k_{{\rm F}\downarrow}\downarrow}}_{\rm F} &= \ket{k_{{\rm F}\downarrow}\downarrow} +\frac{\Ascal{t}{-}}{\spol}\ket{k_{{\rm F}\downarrow}\uparrow} .
\label{FSstates}
\end{align}

\subsection{Spin current induced in the normal metal \label{SEC:QMN}}
Spin pumping effect is now studied by taking account of the interface hopping effects on states in Eq. (\ref{FSstates}). 
The interface hopping amplitude of electron in F to N with spin $\spinindex$  is denoted by $\ttil_\spinindex$ and the amplitude from N to F is  $\ttil_\spinindex^*$. 
We assume that the spin-dependence of electron state in F is governed by the relative angle to the magnetization vector, and hence the spin $\spinindex$ is the one in the rotated frame.
Assuming moreover that there is no spin flip scattering at the interface, the amplitude 
$\ttil_\spinindex$ is diagonal in spin. (Interface spin-orbit interaction is considered in Sec. \ref{SECQMSO}.) 
The spin wave function formed in the N region at the interface as a result of the state in F (Eq. (\ref{FSstates})) is then 
\begin{align}
\ket{{\kf\uparrow}}_{\rm N} &\equiv {\ttil} \ket{\kf\uparrow} = {\ttil}_\uparrow \ket{\kf\uparrow} -{\ttil}_\downarrow\frac{\Ascal{t}{+}}{\spol}\ket{\kf\downarrow} \nnr
\ket{{\kf\downarrow}}_{\rm N} &\equiv {\ttil} \ket{\kf\downarrow} = {\ttil}_\downarrow\ket{\kf\downarrow} +{\ttil}_\uparrow\frac{\Ascal{t}{-}}{\spol}\ket{\kf\uparrow} ,
\label{Nstates}
\end{align}
where $\kf$ is the Fermi wave vector of N electron.
The spin density induced in N region at the interface is therefore 
\begin{align}
 \widetilde{\sv}^{\rm (N)}=\frac{1}{2}\lt(
  {}_{\rm N}\!\bra{{\kf\uparrow}}\sigmav \ket{{\kf\uparrow}}_{\rm N}  \dos_\uparrow
 + {}_{\rm N}\!\bra{{\kf\downarrow}}\sigmav \ket{{\kf\downarrow}}_{\rm N} \dos_\downarrow\rt)
\end{align}
where $\dos_\spinindex$ is the spin-dependent density of states of F electron at the Fermi energy. 
It reads 
\begin{align}
 \widetilde{\sv}^{\rm (N)}=\frac{1}{2}\sum_\spinindex \dos_\spinindex T_{\spinindex\spinindex} \zvhat
 -\frac{\dos_\uparrow-\dos_\downarrow}{\spol} 
 \lt(\Re[T_{\uparrow\downarrow}]\Ascal{t}{\perp}+\Im[T_{\uparrow\downarrow}](\zvhat\times \Ascal{t}{\perp}) \rt) \label{sevrot}
 \end{align}
where $\Ascalv{t}^{\perp}=(\Ascal{t}{x},\Ascal{t}{y},0)=\Ascalv{t}-\zvhat \Ascal{t}{z}$
 is the transverse (non-adiabatic) components of spin gauge field and 
 \begin{align}
T_{\spinindex\spinindex'}\equiv {\ttil}^*_\spinindex {\ttil}_{\spinindex'}.\label{Tdef}
\end{align}

Spin density of Eq. (\ref{sevrot}) is in the rotated frame.
The spin polarization in the laboratory frame is obtained by a rotation matrix
${\cal R}_{ij}$, defined by 
\begin{align}
U^{-1} \sigma_i U \equiv {\cal R}_{ij}\sigma_j,                  \label{Rdef}                                            
\end{align}
 as 
 \begin{align}
{s}_i^{\rm (N)}={\cal R}_{ij} \widetilde{s}_j^{\rm (N)}.    
\end{align}

Explicitly, ${\cal R}_{ij}=2m_im_j-\delta_{ij}$, where 
$\mv\equiv \lt(\sin\frac{\theta}{2}\cos\phi, \sin\frac{\theta}{2}\sin\phi,\cos\frac{\theta}{2}\rt)$ \cite{TKS_PR08}.
Using 
\begin{align}
 {\cal R}_{ij}(\Ascalv{t}^{\perp})_j &= -\frac{1}{2}(\nv\times \dot{\nv})_i \nnr
 {\cal R}_{ij}(\zvhat\times \Ascalv{t}^{\perp})_j &= -\frac{1}{2} \dot{\nv}_i ,
\end{align}
and ${\cal R}_{iz}=n_i$, the induced interface spin density is finally obtained as
\begin{align}
 \sv^{\rm (N)}=\zeta_0^{\rm s}\nv+\Re[\zeta^{\rm s}](\nv\times\dot{\nv})+\Im[\zeta^{\rm s}] \dot{\nv}
 \end{align}
where 
\begin{align}
     \zeta_0^{\rm s}&\equiv    \frac{1}{2}\sum_\spinindex \dos_\spinindex T_{\spinindex\spinindex} \nnr
    \zeta^{\rm s}&\equiv  \frac{\dos_\uparrow-\dos_\downarrow}{2\spol} T_{\uparrow\downarrow}.
    \label{zetadef}
\end{align}

Since the N electrons contributing to induced spin density is those at the Fermi energy, the spin current is simply proportional to the induced spin density as 
${\jsv}^{\rm N}=\frac{\kf}{m}{\sv}^{\rm (N)}$, resulting in  
\begin{align}
 \jsv^{\rm (N)}= \frac{\kf}{m}  \zeta^{\rm s}_0\nv+ \frac{\kf}{m} \Re[\zeta^{\rm s}](\nv\times\dot{\nv})+ \frac{\kf}{m} \Im[\zeta^{\rm s}]\dot{\nv}.
 \label{jsresultQM}
 \end{align}

This is the result of spin current at the interface. The pumping efficiency is determined by the product of hopping amplitudes $t_\uparrow$ and  $t_\downarrow^*$.
The spin mixing conductance defined in Ref. \cite{Tserkovnyak02} corresponds to $iT_{\uparrow\downarrow}$. 
If spin mixing effects due to spin-orbit interaction is neglected at the interface, the hopping amplitudes $t_\spinindex$ are chosen as real, and $\Im[\zeta^{\rm s}]=0$. 
If spin current proportional to $\dot{\nv}$ is measured, it would be useful tool to estimate the strength of interface spin-orbit interaction, as discussed in Sec. \ref{SECQMSO}.

It should be noted that the spin pumping effect at the linear order in time-derivative is mapped to a static problem of spin polarization formed by a static spin-mixing potential in the rotated frame as was mentioned in Ref. \cite{TataraSP16}.
The rotate frame approach employed here provides clear physical picture, as it grasps the low energy dynamics in a mathematically proper manner. 
In this approach, as we have seen, it is clearly seen that pumping of spin current arises as a result of off-diagonal components of the spin gauge field that cause electron spin flip.
Important role  of nonadiabaticity is also indicated in a recent analysis based on the full counting statistics \cite{Hashimoto17}.
In the strict sense, spin pumping effect is a result of a non-adiabatic process including state change.
The same goes for general adiabatic pumping; 
Some sort of state change is necessary for current generation, 
although the nonadiabaticity is obscured in the conventional ``adiabatic`` argument focusing on the wave function in the laboratory frame. 
In the case of slowly-varying external potential with frequency $\Omega$ acting on electrons, the state change is represented by the Fermi distribution difference, $f(\omega+\Omega)-f(\omega)\simeq \Omega f'(\omega)$, where $\omega$ is the electron frequency \cite{Buttiker94,Moskalets12}. The existence of a factor of $f'$ clearly indicates that a state change or nonadiabaticity is necessary for current pumping.

\section{Effects of interface spin-orbit interaction \label{SECQMSO}}
In this section, we discuss the effect of spin-orbit interaction at the interface, which modifies  hopping amplitude ${\ttil}_\spinindex$. 
We particularly focus on that linear in the wave vector, namely the interaction represented in the continuum representation  by a Hamiltonian 
\begin{align}
 H_{\rm so} &= a^2 \delta(x) \sum_{ij}\gamma_{ij} k_i \sigma_j ,
\end{align}
where $\gamma_{ij}$ is a coefficient having the unit of energy representing the spin-orbit interaction, $a$ is the lattice constant, and  the interface is chosen as at $x=0$.
Assuming that spin-orbit interaction is weaker than the $sd$ exchange interaction in F,   we carry out a unitary transformation to which diagonalize the $sd$ interaction to obtain  
\begin{align}
 H_{\rm so} &= a^2 \delta(x) \sum_{ij}\widetilde{\gamma}_{ij} k_i \sigma_j ,
\end{align}
where $\widetilde{\gamma}_{ij} \equiv \sum_{l}\gamma_{il} {\cal R}_{lj}$, with ${\cal R}_{ij}$ being a rotation matrix defined by Eq. (\ref{Rdef}).
This spin-orbit interaction modifies diagonal hopping amplitude $\ttil_{i}$ in the direction $i$ at the interface to become a complex as 
\begin{align}
 \tsotil_{i} &= \ttil^{0}_{i} -i\sum_{j}{\widetilde{\gamma}_{ij}} \sigma_j .
\end{align}
(In this section, we denote the total hopping amplitude including the interface spin-orbit interaction by $\tsotil$ and the one without by $\tsotil^0$.) 
We consider the hopping amplitude perpendicular to the interface, i.e., along the $x$ direction, and suppress the suffix $i$ representing the direction.
In the  matrix representation for spin the hopping amplitude is 
\begin{align}
 \tsotil &(\equiv \tsotil_x)= \lt( \begin{array}{cc}
                   \tsotil_{\uparrow} & \tsotil_{\uparrow\downarrow} \\
                   \tsotil_{\downarrow\uparrow} &  \tsotil_{\downarrow} \end{array} \rt),
                   \label{tsotilmatrix}
\end{align}
where 
\begin{align}
  \tsotil_{\uparrow} &=  \ttil^0_{\uparrow} -i{\widetilde{\gamma}_{xz}} &
  \tsotil_{\downarrow} =  \ttil^0_{\downarrow} +i{\widetilde{\gamma}_{xz}} \nnr 
    \tsotil_{\uparrow\downarrow} &= i({\widetilde{\gamma}_{xx}+i\widetilde{\gamma}_{xy}}) 
    &
  \tsotil_{\downarrow\uparrow} = i({\widetilde{\gamma}_{xx}-i\widetilde{\gamma}_{xy}}).
\end{align}

Let us discuss how the spin pumping effect discussed in Sec. \ref{SEC:QMN} is modified when the hopping amplitude is a matrix of Eq. (\ref{tsotilmatrix}).
The spin pumping efficiency is written as in Eqs. (\ref{Tdef})(\ref{zetadef}).  
In the absence of spin-orbit interaction hopping amplitude $\ttil$ is chosen as real, and thus the contribution proportional to $\nv\times \dot{\nv}$ in Eq. (\ref{jsresultQM}) is dominant. 
Spin-orbit interaction enhances the other contribution proportional to $\dot{\nv}$ because it gives rise to an imaginary part.
Moreover, it leads to spin mixing at the interface, modifying the spin accumulation formed in the N region at the interface. 

The electron states in the N region at the interface are now given instead of Eq. (\ref{Nstates}) by the following two states (choosing basis as $\lt( \begin{array}{c} \ket{k_{F}\uparrow} \\ \ket{{k_F}\downarrow} \end{array} \rt)$)
\begin{align}
\ket{{k_{{\rm F}}\uparrow}}_{\rm N} &\equiv  \tsotil \ket{{k_{{\rm F}\uparrow}\uparrow}}_{\rm F} 
= \lt( \begin{array}{c} 
{\tsotil}_\uparrow - {\tsotil}_{\uparrow\downarrow} \frac{\Ascal{t}{+}}{\spol} \\
{\tsotil}_{\downarrow\uparrow}-{\tsotil}_\downarrow\frac{\Ascal{t}{+}}{\spol} \end{array} \rt)
\nnr
\ket{{k_{{\rm F}}\downarrow}}_{\rm N}  &\equiv  \tsotil \ket{{k_{{\rm F}\downarrow}\downarrow}}_{\rm F}
=
\lt( \begin{array}{c} 
{\tsotil}_{\uparrow\downarrow}+{\tsotil}_\uparrow\frac{\Ascal{t}{-}}{\spol} \\ 
{\tsotil}_\downarrow + {\tsotil}_{\downarrow\uparrow} \frac{\Ascal{t}{-}}{\spol}  \end{array} \rt)
.
\end{align}
The pumped (i.e.,  linear in the gauge field) spin density for these two states are
\begin{align}
  \ {}_{\rm N}\!\bra{{\kf\uparrow}}\sigmav \ket{{\kf\uparrow}}_{\rm N} 
    &= - \frac{2}{\spol} ( \Ascalv{t}^{\perp} \Re[T^{\rm tot}_{\uparrow\downarrow}]
   +(\zvhat\times \Ascalv{t}^{\perp}) \Im[T^{\rm tot}_{\uparrow\downarrow}]
  \nnr 
& \lt. 
  +  \Re[({\tsotil}_{\uparrow\downarrow})^*{\tsotil}_{\downarrow\uparrow}] \lt(\begin{array}{c}\Ascal{t}{x} \\ -\Ascal{t}{y}\\0 \end{array} \rt)  
    +  \Im[({\tsotil}_{\uparrow\downarrow})^*{\tsotil}_{\downarrow\uparrow}] \lt(\begin{array}{c}\Ascal{t}{y} \\ \Ascal{t}{x}\\0 \end{array} \rt)  
\rt) \nnr
&
 - \zvhat (\Ascal{t}{x} 
       \Re[ ({\tsotil}_\uparrow) ^* {\tsotil}_{\uparrow\downarrow}- {\tsotil}_{\downarrow}({\tsotil}_{\downarrow\uparrow})^*] - \Ascal{t}{y} 
       \Im[  ({\tsotil}_\uparrow)^* {\tsotil}_{\uparrow\downarrow}- {\tsotil}_{\downarrow}({\tsotil}_{\downarrow\uparrow})^*] )
\label{spindensitymatrix}
\end{align}
\begin{align}
  \ {}_{\rm N}\!\bra{{\kf\downarrow}}\sigmav \ket{{\kf\downarrow}}_{\rm N} 
    &=  \frac{2}{\spol} ( \Ascalv{t}^{\perp} \Re[T^{\rm tot}_{\uparrow\downarrow}]
   +(\zvhat\times \Ascalv{t}^{\perp}) \Im[T^{\rm tot}_{\uparrow\downarrow}]
  \nnr 
& \lt. 
  +  \Re[({\tsotil}_{\uparrow\downarrow})^*{\tsotil}_{\downarrow\uparrow}] \lt(\begin{array}{c}\Ascal{t}{x} \\ -\Ascal{t}{y}\\0 \end{array} \rt)  
    +  \Im[({\tsotil}_{\uparrow\downarrow})^*{\tsotil}_{\downarrow\uparrow}] \lt(\begin{array}{c}\Ascal{t}{y} \\ \Ascal{t}{x}\\0 \end{array} \rt)  
\rt) \nnr
&
 + \zvhat (\Ascal{t}{x} 
       \Re[ ({\tsotil}_\uparrow) ^* {\tsotil}_{\uparrow\downarrow}- {\tsotil}_{\downarrow}({\tsotil}_{\downarrow\uparrow})^*] - \Ascal{t}{y} 
       \Im[  ({\tsotil}_\uparrow)^* {\tsotil}_{\uparrow\downarrow}- {\tsotil}_{\downarrow}({\tsotil}_{\downarrow\uparrow})^*] )
\label{spindensitymatrix2}
\end{align}

We here focus on the linear effect of interface spin-orbit interaction and neglect the spin polarization along the magnetization direction, $\nv$. 
The expression for the pumped spin current then agrees with Eq. (\ref{jsresultQM}) with the amplitude $\zeta^{\rm s}$ written in terms of hopping including the interface spin-orbit,  
\begin{align}
 T_{\uparrow\downarrow} &= ( (\ttil^0_{\uparrow})^* +i({\widetilde{\gamma}_{xz}})^*)  (\ttil^0_{\downarrow} +i{\widetilde{\gamma}_{xz}}).
\end{align}
If bulk spin-orbit interaction is neglected, bare hopping amplitude $\ttil^0_{\spinindex}$ is real and we may reasonably assume that $\widetilde{\gamma}_{ij}$ is real.
The interface spin-orbit then leads to an imaginary part as (using $\widetilde{\gamma}_{xz}=n_i \gamma_{xi}$) 
\begin{align}
\Im[\zeta^{\rm s}] &= \frac{\dos_\uparrow-\dos_\downarrow}{2\spol}(\ttil^0_{\uparrow}+\ttil^0_{\downarrow}){\gamma}_{xi}n_i.
\end{align}
The amplitude of spin current proportional to $\dot{\nv}$ thus works as a probe for interface spin-orbit interaction strength, $\gamma_{xi}$.

Let us discuss some examples.
Of recent particular interest is the interface Rashba interaction, represented by antisymmetric coefficient 
\begin{align}
\gamma_{ij}^{\rm (R)} = \epsilon_{ijk}\alpha^{\rm R}_k,
\end{align}
where $\alphav^{\rm R}$ is a vector representing the Rashba field.
In the case of interface, $\alphav^{\rm R}$ is perpendicular to the interface, i.e., $\alphav^{\rm R}\parallel \hat{\xv}$.
Therefore the interface Rashba interaction leads to 
$\gamma_{xj}^{\rm (R)}=0$ and  does not modify spin pumping effect at the linear order.
(It contributes at the second order as discussed in Ref. \cite{Chen15}.)
In other words, vector coupling between the wave vector and spin in the form of $\kv\times\sigmav$ exists only along the $x$-direction, and does not affect the interface hopping (i.e., does not include $k_x$). 

In contrast, a scalar coupling $\eta^{\rm (D)} (\kv\cdot\sigmav)$ ($\eta^{\rm (D)}$ is a coefficient), called the Dirac type spin-orbit interaction, leads to $\gamma_{ij}^{\rm (D)}=\eta^{\rm (D)}\delta_{ij}$.
The spin current along $\dot{\nv}$ then reads 
\begin{align}
 \jsv^{{\dot{\nv}}} = \eta^{\rm (D)}\frac{\kf(\dos_\uparrow-\dos_\downarrow)}{2m\spol}
  (\ttil^0_{\uparrow}+\ttil^0_{\downarrow})n_x \dot{\nv}.
\end{align}
For the case of in-plane easy axis along the $z$ direction and magnetization precession given by
$\nv(t)=(\sin\theta \cos\omega t,\sin\theta\sin \omega t,\cos\theta)$, where $\theta$ is the precession angle and $\omega$ is the angular frequency, we expect to  have a dc spin current along the $y$ direction, as $\overline{n_x \dot{\nv}}=-\frac{\omega}{2}\sin^2\theta \hat{\yv}$ ($\overline{ n_x \dot{\nv} }$ denotes time average).

\section{Field theoretic description of metallic case \label{SEC:GF}}
Here we present a field-theoretic description of spin pumping effect of metallic ferromagnet. 
The many-body approach has an advantage of taking account of particle distributions automatically.
Moreover, it describes propagation of particle density in terms of the Green's functions, and thus is suitable for studying spatial propagation as well as for intuitive understanding of transport phenomena. 
All the transport coefficients are  determined by material constants.

The formalism presented here is essentially the same as in Ref. \cite{TataraSP16}, but  treating the ferromagnet of a finite size and taking account of electron states with different wave vectors.
Interface spin-orbit interaction  is not considered here.

Conduction electron in ferromagnetic and normal metals are denoted by field operators $d$, $d^\dagger$ and $c$, $c^\dagger$, respectively. 
These operators are vectors with two spin components, i.e., $d\equiv (d_\uparrow,d_\downarrow)$.
The Hamiltonian describing the F and N electrons is $ H_{\rm F}+H_{\rm N} $, where 
\begin{align}
 H_{\rm F} &\equiv \int_{\rm F}d^3r d^\dagger
   \lt( -\frac{\nabla^2}{2m}-\ef -\spol \nv(t)\cdot \sigmav \rt) d \nnr
 H_{\rm N} &\equiv    \int_{\rm N}d^3r c^\dagger
   \lt( -\frac{\nabla^2}{2m}-\ef\rt) c .
\end{align}
We set the Fermi energies for ferromagnet and normal metal equal.
The hopping through the interface is described by the Hamiltonian
\begin{align}
 H_{\rm I} & \equiv \int_{{\rm I}_{\rm F}} d^3r \int_{{\rm I}_{\rm N}} d^3\rv'
 \lt( 
c^\dagger(\rv')  t(\rv',\rv,t)  d(\rv) + 
  d^\dagger(\rv) t^*(\rv',\rv,t) c(\rv') \rt),
 \label{FNhoppingreal}
\end{align}
where $t(\rv',\rv,t)$ represents the hopping amplitude of electron  from $\rv$ in ferromagnetic regime to a site $\rv'$ in the normal region and the integrals are over the interface (denoted by ${\rm I}_{\rm F}$ and ${\rm I}_{\rm N}$ for F and N regions, respectively).
The hopping amplitude is generally a matrix depending on magnetization direction $\nv(t)$, and thus depends on time $t$.
Hopping is treated as energy-conserving. Assuming sharp interface at $x=0$, the momentum perpendicular to the interface is not conserved on hopping.

\begin{figure}[tbh]
  \begin{center}
  \includegraphics[width=0.3\hsize]{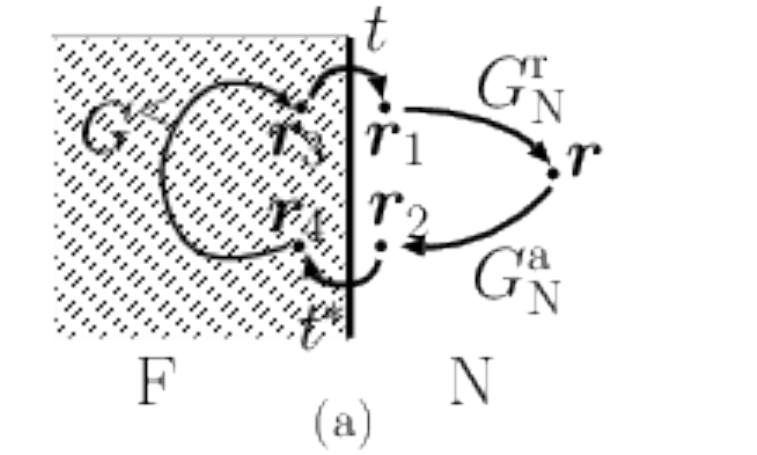}
  \includegraphics[width=0.3\hsize]{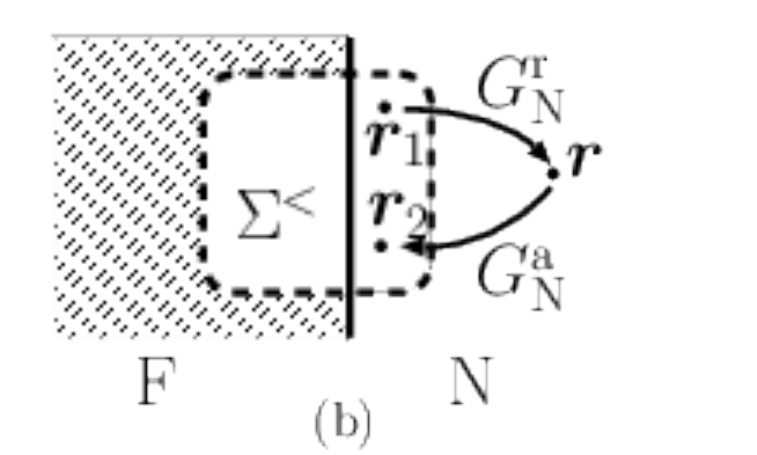}
  \includegraphics[width=0.3\hsize]{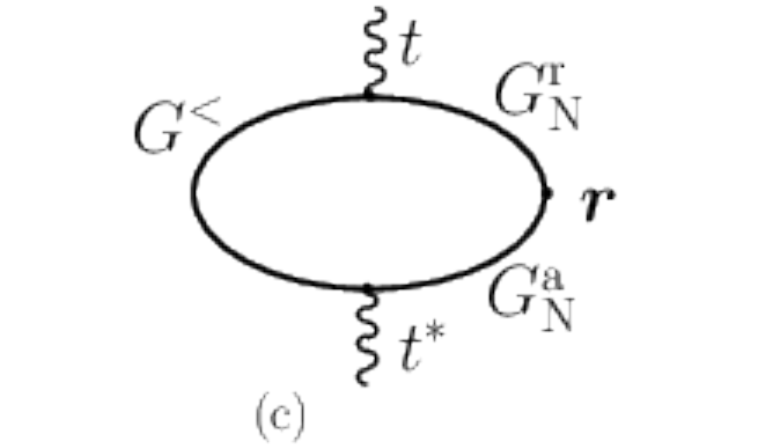}
  \end{center}
\caption{ (a) Schematic diagramatic representations of the lessor Green's function for N electron connecting the same position $\rv$, $ G_{\rm N}^<(\rv,\rv)
 \simeq G^\ret_{\rm N}\Sigma_{\rm N}^<G^\adv_{\rm N}$ representing propagation of electron density.
It is decomposed into a propagation of N electron from $\rv$ to the interface at $\rv_2$, then hopping to $\rv_4$ in the F side, a propagation inside F, followed by a hopping to N side (to $\rv_1$) and propagation back to $\rv$. (Position labels are as in Eqs. (\ref{DysonGN})(\ref{SigmaNdef}).) 
(b): The self energy $\Sigma_{\rm N}^<$ represents all the effects of the ferromagnet.
(c) Standard Feynman diagram representation of lessor Green's function for N at $\rv$, Eqs. (\ref{DysonGNsol1}) and (\ref{SigmaNdef}).
\label{FIGGN}}
\end{figure}

We are interested in the spin current in the normal region, given by 
\begin{align}
  j_{{\rm s},i}^\alpha(\rv,t) &= -\frac{1}{4m}(\nabla^{(\rv)}-\nabla^{(\rv')})_i \tr [ \sigma_\alpha G^<_{\rm N}(\rv,t,\rv',t)|_{\rv'=\rv},
  \label{spincurrentdef}
\end{align}
where $G^<_{\rm N}(\rv,t,\rv',t')\equiv i\average{c(\rv,t)c^\dagger(\rv',t')}$ denotes the lesser Green's function for the normal region.
It is calculated from the Dyson's equation for the path-ordered Green's function defined for a complex time along a complex contour $C$ 
\begin{align}
 G_{\rm N}(\rv,t,\rv',t')&=g_{\rm N}(\rv-\rv',t-t')\nnr
 & +\int_c d t_1\int_c d t_2   \intr_1 \intr_2 g_{\rm N}(\rv-\rv_1,t-t_1)\Sigma_{\rm N}(\rv_1,t_1,\rv_2,t_2)G_{\rm N}(\rv_2,t_2,\rv',t'), 
 \label{DysonGN}
\end{align}
where $g^<_{\rm N}$ denotes the Green's function without interface hopping and $\Sigma_{\rm N}(\rv_1,t_1,\rv_2,t_2)$ is the self-energy for N electron, given by the contour-ordered  Green's function in the ferromagnet as 
\begin{align}
 \Sigma_{\rm N}(\rv_1,t_1,\rv_2,t_2)&\equiv 
 \int_{{\rm I}_{\rm F}}d^3r_3  \int_{{\rm I}_{\rm F}}d^3r_4  
 t(\rv_1,\rv_3,t_1)G(\rv_3,t_1,\rv_4,t_2) t^*(\rv_2,\rv_4,t_2) .
\label{SigmaNdef}
\end{align}
Here $\rv_1$ and $\rv_2$ are coordinates at the interface ${\rm I}_{\rm N}$ in N region and $\rv_3$ and $\rv_4$ are those in  ${\rm I}_{\rm F}$ for F.
$G$ is the contour-ordered Green's function for F electron in the laboratory frame including the effect of spin gauge field.
We denote Green's functions of F electron by $G$ and $g$ without suffix and those of N electron with suffix N. 
The lesser component of the normal metal Green's function is obtained from Eq. (\ref{DysonGN}) as (suppressing the time and space coordinates)
\begin{align}
 G_{\rm N}^<=(1+G^\ret_{\rm N}\Sigma_{\rm N}^\ret) g^<_{\rm N}(1+\Sigma_{\rm N}^\adv G^\adv_{\rm N}) 
 + G^\ret_{\rm N}\Sigma_{\rm N}^<G^\adv_{\rm N}.
 \label{DysonGNsol}
\end{align}
For pumping effects, the last term on the right-hand side  is essential, as it contains the information of excitation in F region. We thus consider the second term only;
\begin{align}
 G_{\rm N}^<
 \simeq G^\ret_{\rm N}\Sigma_{\rm N}^<G^\adv_{\rm N},
 \label{DysonGNsol1}
\end{align}
and neglect spin-dependence of the normal region Green's functions, $G^\ret_{\rm N}$ and $G^\adv_{\rm N}$.
The contribution is diagramatically shown in Fig. \ref{FIGGN}.

%
%
%

\begin{figure}[tbh]
  \begin{center}
  \includegraphics[width=0.25\hsize]{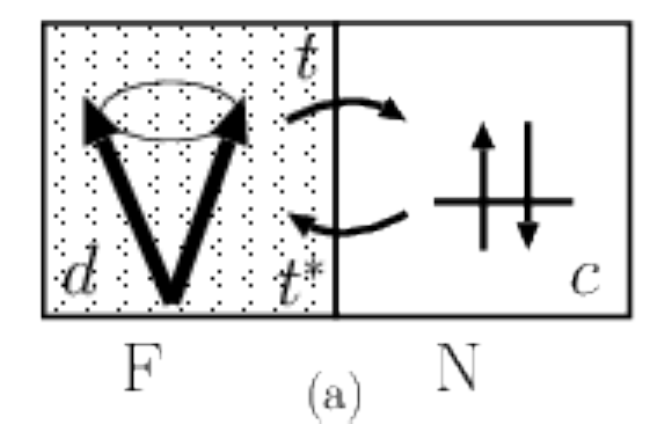}
  \includegraphics[width=0.25\hsize]{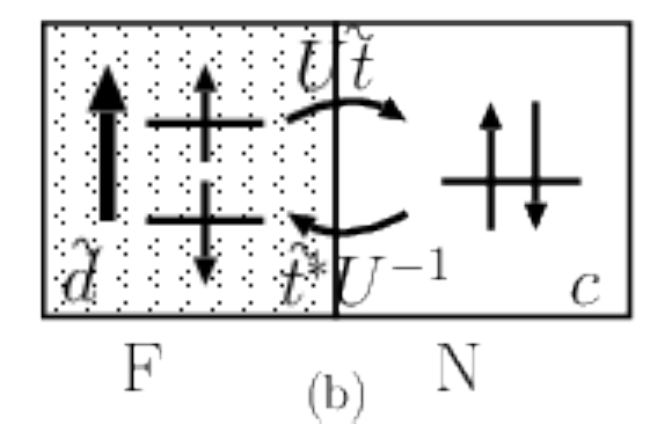}
  \includegraphics[width=0.25\hsize]{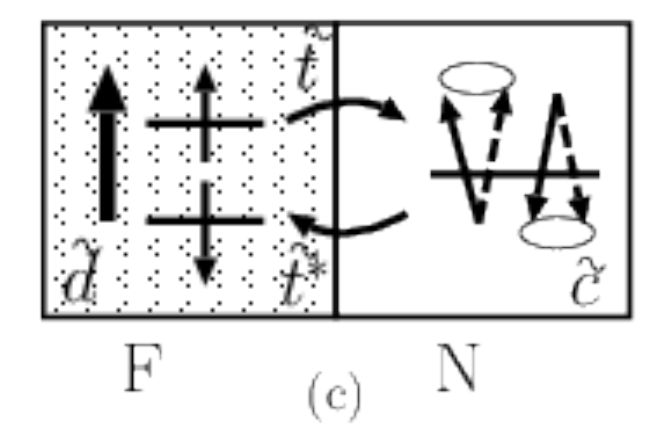}
  \end{center}
\caption{ Unitary transformation $U$ of F electron converts the original system with field operator $d$ (shown as (a)) to the rotated one with field operator $\dtil\equiv U^{-1}d$ (b). The hopping amplitude for representation in (b) is modified by $U$. 
If N electrons are also rotated as $\ctil\equiv U^{-1}c$, hopping becomes $\ttil\equiv U^{-1}t U$, while the N electron spin rotates with time, as shown as (c).
\label{FIGFN}}
\end{figure}
%

\subsection{Rotated frame}
To solve for the Green's function in the ferromagnet, rotated frame we used in Sec. \ref{SECQMF} is convenient. In the field representation, the unitary transformation is represented as (Fig. \ref{FIGFN}(c))
\begin{align}
 d&= U \dtil, \;\;\; c=U\ctil,
\end{align}
where $U$ is the same $2\times2$ matrix defined in Eq. (\ref{Udef}).
We rotate N electrons as well as F electrons, to simplify the following expressions.
The hopping interaction Hamiltonian reads 
\begin{align}
 H_{\rm I} & = \int_{{\rm I}_{\rm F}} d^3r \int_{{\rm I}_{\rm N}} d^3r'
 \lt( 
\ctil^\dagger(\rv')  \ttil(\rv',\rv)  \dtil(\rv) + 
  \dtil^\dagger(\rv) \ttil^*(\rv',\rv) \ctil(\rv') \rt),
 \label{FNhoppingrot}
\end{align}
where 
\begin{align}
\ttil(\rv',\rv) & \equiv U^\dagger (t) t(\rv',\rv,t) U(t) ,
\end{align}
is the hopping amplitude in the rotated frame.
The rotated amplitude (neglecting interface spin-orbit interaction) is diagonal in spin; 
\begin{align}
\ttil 
&=
\lt(\begin{array}{cc} 
\ttil_\uparrow  & 0 \\ 0 & \ttil_\downarrow
\end{array}\rt).
\end{align}

Including the interaction with spin gauge field,  the Hamiltonian for F and N electrons 
in the momentum representation is
\begin{align}
 H_{\rm F}+H_{\rm N} &= \sumkv \dtil_\kv^\dagger 
  \lt( \begin{array}{cc} 
         \epsilon_{k}-\spol -\Ascal{t}{z}  & \Ascal{t}{-} \\
                    \Ascal{t}{+}  & \epsilon_{k}+\spol +\Ascal{t}{z} 
        \end{array} \rt) \dtil_\kv 
        + \sumkv \epsilon_{\kv}^{\rm (N)} \ctil_\kv^\dagger \ctil_\kv 
        \label{HFF}
\end{align}
As for the hopping, we consider the case the interface is atomically sharp. 
The hopping Hamiltonian is then written in the momentum space as 
\begin{align}
 H_{\rm I} & = \sum_{\kv\kv'}\lt( 
  \ctil^\dagger(\kv) \ttil(\kv,\kv') \dtil(\kv') + 
   \dtil^\dagger(\kv') \ttil^*(\kv,\kv') \ctil(\kv) \rt),
 \label{FNhoppingk}
\end{align}
where $\kv=(k_x,k_y,k_z)$, $\kv'=(k_x',k_y, k_z)$, choosing the interface as the plane of $x=0$. Namely, the wave vectors parallel to the interface are conserved while $k_x$ and $k_x'$ are uncorrelated.

\subsection{Spin density induced by magnetization dynamics in F}
Pumped spin current in N is calculated by evaluating  $\Sigma_{\rm N}^<$ and using Eqs. (\ref{spincurrentdef})(\ref{DysonGNsol})(\ref{DysonGNsol1}). 
Before discussing the spin current, let us calculate spin density in ferromagnet induced by magnetization dynamics neglecting the effect of interface, $H_{\rm I}$. 
(Effects of $H_{\rm I}$ are discussed in Sec. \ref{SEC:damping}.)
The spin accumulation in N is discussed by extending the calculation here as shown in Sec. \ref{SEC:spinpolN}.

The lessor Green's function in F in the rotated frame  including the spin gauge field to the linear order is calculated from the Dyson's equation
\begin{align}
 G^< &= g^<+\gr (\Ascalv{t}\cdot\sigmav) g^< + g^<  (\Ascalv{t}\cdot\sigmav) \ga,
 \label{GlessF0}
\end{align}
where $g^\alpha$ ($\alpha=<,$r,a) represents Green's functions without spin gauge field.
The lessor Green's function satisfies  for static case $g^<=F(\ga-\gr)$, where $F\equiv \lt(\begin{array}{cc} f_\uparrow & 0 \\ 0 & f_\downarrow \end{array} \rt)$ is spin-dependent Fermi distribution function. 
We thus obtain the Green's function at the linear order as \cite{TataraSP16}
\begin{align}
 \delta G^< &= \gr [\Ascalv{t}\cdot\sigmav, F] \ga 
 + \ga F  (\Ascalv{t}\cdot\sigmav) \ga 
 - \gr  (\Ascalv{t}\cdot\sigmav) F \gr.
 \label{deltaGless}
\end{align}
The last two terms of the right-hand side are rapidly oscillating as function of position and are neglected.
The commutator is calculated as (sign $\pm$ denotes spin $\uparrow$ and $\downarrow$) 
\begin{align}
 [\Ascalv{t}\cdot\sigmav, F] = (f_+-f_-) \sum_\pm(\pm)\Ascal{t}{\pm}\sigma_\mp. 
 \label{AFcom}
 \end{align}
\begin{figure}[ht]
\begin{center}
  \includegraphics[width=0.3\hsize]{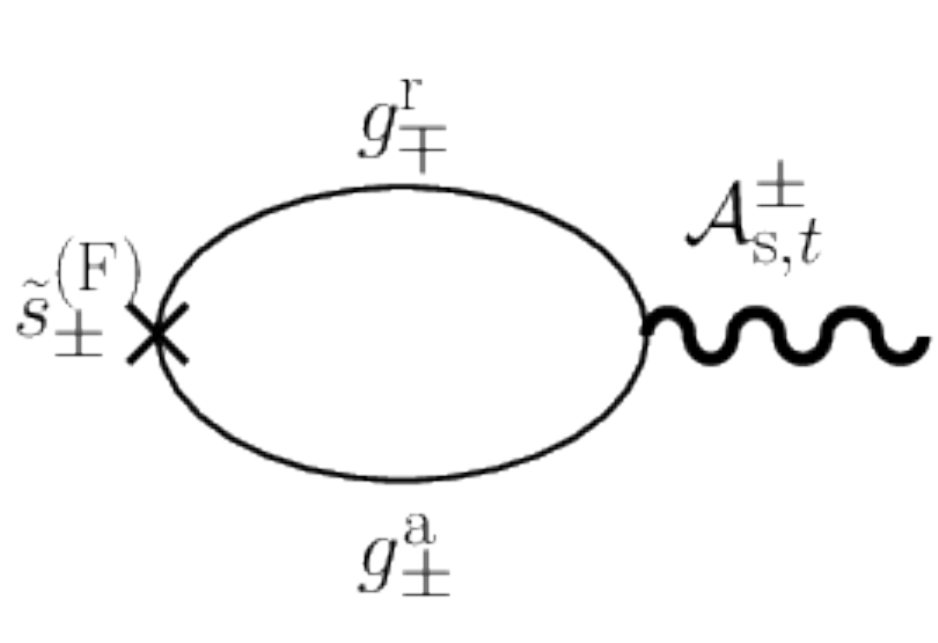}
\caption{Feynman diagram for electron spin density of ferromagnet induced by magnetization dynamics (represented by spin gauge field ${\cal A}_{\rm s}$) neglecting the effect of normal metal.
The amplitude is essentially given by the spin flip correlation function $\chi_\pm$
(Eq. (\ref{stilres1})).
\label{FIGsf0}}
\end{center}
\end{figure} 
In the rotated frame, the spin density in F pumped by the spin gauge field is therefore 
(diagrams shown in Fig. \ref{FIGsf0}) 
\begin{align}
\stil^{\rm (F)}_\alpha & (\kv,\kv') \equiv 
-i\sumom \tr[\sigma_\alpha  \delta G^< (\kv,\kv',\omega) ]
\nnr
&= -i\sumom \sum_{\kv''} (f_{\kv''+}-f_{\kv''-}) \sum_\pm(\pm)\Ascal{t}{\pm}
\tr[\sigma_\alpha \gr(\kv,\kv'',\omega) \sigma_\mp \ga(\kv'',\kv',\omega) ] \nnr
&=\lt\{ \begin{array}{cc}
         \mp i\sumom\sum_{\kv''}  (f_{\kv''+}-f_{\kv''-}) \Ascal{t}{\pm} \gr_\mp(\kv,\kv'',\omega) \ga_\pm(\kv'',\kv',\omega) & (\alpha=\pm) \\
         0 & (\alpha=z)
        \end{array}\rt. .
        \label{spindensity2}
\end{align}
Let us here neglect the effects of interface in dicussing spin polarization of F electrons; Then the Green's functions are translationally invariant, i.e., 
$g^{a}(\kv,\kv')=\delta_{\kv,\kv'} g^{a}(\kv)$ ($a=\ret, \adv$).
Using the explicit form of the free Green's function, 
$g^\adv_{\spinindex}(\kv,\omega)=\frac{1}{\omega- \epsilon_{\kv,\spinindex}-i0}$, and
\begin{align}
 \sumom  \gr_\mp(\kv,\kv'',\omega) \ga_\pm(\kv'',\kv',\omega) 
 &=\frac{i}{ \epsilon_{\kv,\pm}-\epsilon_{\kv,\mp} +i0},
\end{align}
the spin density in the rotated frame then reduces to 
\begin{align}
\stil^{\rm (F)}_\pm (\kv) 
&= -\Ascal{t}{\pm} \chi_{\pm},
\label{stilres1}
\end{align}
where 
\begin{align}
\chi_{\pm}
&\equiv 
- \sum_{\kv} 
\frac{f_{\kv,\pm}-f_{\kv,\mp}}
{ \epsilon_{\kv,\pm}-\epsilon_{\kv,\mp} +i0},
\end{align}
is the spin correlation function with spin flip, $+i0$ meaning an infinitesimal positive imaginary part.
Since we focus on adiabatic limit and spatially uniform magnetization, the correlation function is at zero momentum- and frequency-transfer.
We thus easily see that 
\begin{align}
\chi_{\pm}
= \frac{n_+-n_-}{2\spol}, \label{chipmresults2}
\end{align}
where $n_\pm=\sumkv f_{\kv\pm}$ is spin-resolved electron density.

The spin polarization of Eq. (\ref{stilres1}) in the rotated frame is proportional to 
$\Ascalv{t}^{\perp}$, and represents a renormalization of total spin in F. In fact, it corresponds  in the laboratory frame to $\sev^{\rm (F)}\propto \nv\times\dot{\nv}$, and exerts a torque proportional to $\dot{\nv}$ on $\nv$.

It may appear from Eq. (\ref{chipmresults2}) that a damping of spin, i.e.,  a torque proportional to $\nv\times \dot{\nv}$, arises when the imaginary part for the Green's function becomes finite, because $\frac{1}{\spol}$ is replaced by  $\frac{1}{\spol \mp i \eta_{\rm i}}$, where $\eta_{\rm i}$ is the imaginary part. 
This is not always the case. 
For example, nonmagnetic impurities introduce a finite imaginary part inversely proportional to the elastic lifetime ($\tau$),  $\frac{i}{2\tau}$. They should not, however,  cause damping of spin. 
The solution to this apparent controversy is that Eq. (\ref{spindensity2}) is not enough to discuss damping even including lifetime. 
In fact, there is an additional process called vertex correction contributing to the lesser Green's function, and it  gives rise to the same order of small correction as the lifetime does, and the sum of the two contributions vanishes. 
Similarly, we expect damping does not arise from spin-conserving component of spin gauge field, $\Ascal{t}{z}$.
This is indeed true as we explicitly demonstrate in Appendix \ref{SEC:nodamping}.
We shall show in Sec. \ref{SEC:damping} that damping arises from the spin-flip components of the self energy.

\subsection{Spin polarization and current in N \label{SEC:spinpolN}}

\begin{figure}[ht]
\begin{center}
  \includegraphics[width=0.4\hsize]{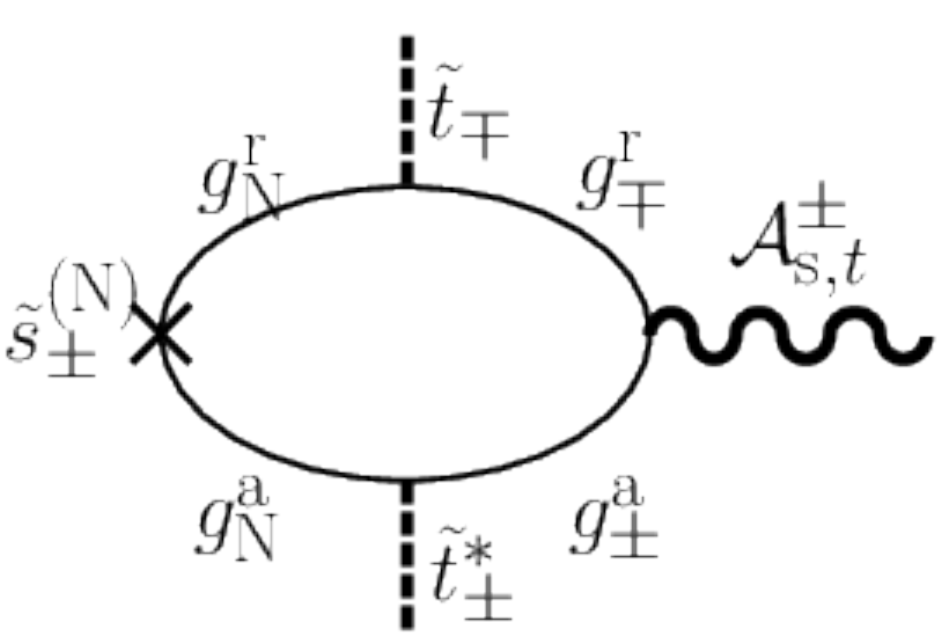}
\caption{Feynman diagram for electron spin density of normal metal driven by the spin gauge field of ferromagnetic metal, ${\cal A}_{\rm s}$.
The spin current is represented by the same diagram but with spin current vertex.
\label{FIGsn}}
\end{center}
\end{figure} 
The spin polarization  of N electron lesser Green's function including the self-energy to the linear order is calculated from Eqs. (\ref{DysonGNsol1}) (\ref{deltaGless})(\ref{AFcom}) as 
(diagram shown in Fig. \ref{FIGsn})
\begin{align}
-i\tr[ & \sigma_\pm G_{\rm N}^<(\rv,t,\rv',t)] 
=
-i \sum_{\kv\kv'\kv''} e^{i\kv\cdot\rv} e^{-i\kv'\cdot\rv'} 
g^\ret_{\rm N}(\kv,\omega) g^\adv_{\rm N}(\kv',\omega) \nnr
&\times
\sum_{\pm} (f_{\kv''\pm}-f_{\kv''\mp}) \Ascal{t}{\pm} \ttil_\mp(\kv,\kv'') \ttil^*_\pm(\kv'',\kv')  
g^\ret_\mp(\kv'',\omega) g^\adv_\pm(\kv'',\omega).
\end{align}
We assume that dependence of N Green's functions on $\omega$ is weak and use 
$\sum_{\kv} e^{i\kv\cdot\rv} g^\ret_{\rm N}(\kv,\omega)=-i\pi\dos_{\rm N}e^{i\kf x}e^{-|x|/\ell}\equiv g^\ret_{\rm N}(\rv)$, where $\ell$ is elastic mean free path, $\dos_{\rm N}$ and $\kf$ are the density of states at the Fermi energy and Fermi wave vector, respectively, whose $\omega$-dependences are neglected.
(For infinitely wide interface, the Green's function becomes one-dimensional.)
As a result of summation over wave vectors, the product of hopping amplitudes 
$\ttil_\mp(\kv,\kv'') \ttil^*_\pm(\kv'',\kv') $ is replaced by the  average over the Fermi surface, 
$\overline{\ttil_\mp  \ttil^*_\pm}\equiv \overline{T_{\pm\mp}}$, i.e.,
\begin{align}
 \ttil_\mp(\kv,\kv'') \ttil^*_\pm(\kv'',\kv')  \ra \overline{T_{\pm\mp}}.
\end{align}
The spin polarization  of N electron induced by magnetization dynamics (the spin gauge field) is therefore obtained 
in the rotated frame as  
\begin{align}
\tilde{\se}^{\rm (N)}_{\pm}(\rv,t)
&=
- |g^\ret_{\rm N}(\rv) |^2
\sum_{\pm}  \Ascal{t}{\pm} \chi_\pm \overline{T_{\pm\mp}} ,
\label{stilresunif}
\end{align}
or using $\chi_+^*=\chi_-$ 
\begin{align}
\tilde{\sev}^{\rm (N)}(\rv,t)
&=
- 2|g^\ret_{\rm N}(\rv) |^2
\lt[  \Ascalv{t}^{\perp} \Re[\chi_+ \overline{T_{+-}}]  +(\zvhat\times \Ascalv{t}^{\perp}) \Im[\chi_+ \overline{T_{+-}}]  \rt]  .
\end{align}
In the laboratory frame, we have (using ${\se}^{\rm (N)}_i={\cal R}_{ij}\tilde{\se}^{\rm (N)}_j$)
\begin{align}
{\sev}^{\rm (N)}(\rv,t)
&=
|g^\ret_{\rm N}(\rv) |^2
\lt[ \Re[\chi_+ \overline{T_{+-}}](\nv\times\dot{\nv})  + \Im[\chi_+ \overline{T_{+-}}]  \dot{\nv} \rt]  .
\end{align}
The spin current induced in N region is similarly given by (neglecting the contribution proportional to $\nv$) 
\begin{align}
{\jsv}(\rv,t)
&=
\frac{\kf}{m} |g^\ret_{\rm N}(\rv) |^2
\lt[ \Re[\chi_+ \overline{T_{+-}}](\nv\times\dot{\nv})  + \Im[\chi_+ \overline{T_{+-}}]  \dot{\nv}\rt]  \nnr
&=
e^{-|x|/\ell} \lt(\Re[\zeta^{\rm s}](\nv\times\dot{\nv})+\Im[\zeta^{\rm s}]\dot{\nv} \rt),
\label{spincurrentmetalFT}
\end{align}
where 
\begin{align}
  \zeta^{\rm s} &\equiv \pi^2\frac{\kf\dos_{\rm N}^2}{2m\spol}(n_+-n_-) \overline{T_{+-}}.
\end{align}
The coefficient $  \zeta^{\rm  s}$ is essentially the same as the one in Eq. (\ref{jsresultQM}) derived by quantum mechanical argument, as quantum mechanical dimensionless hopping amplitude corresponds to $\dos_{\rm N}\ttil$ of field representation.

For 3d ferromagnet, we may estimate the spin current by approximating roughly $\spol\sim 1/\dos_{\rm N}\sim\ef\sim 1$eV and $n_\sigma\sim \kf^3$.
The hopping amplitude $|\overline{T_{+-}}|$ in metallic case would be order of $\ef$.
The spin current density then is of the order of (including electric charge $e$ and recovering $\hbar$), 
$\js\sim e\frac{\hbar\kf}{m}\frac{h\hbar \omega}{\ef}\sim 5\times 10^{11}$ A/m$^2$ if precession  frequency is 10 GHz.

\section{Spin accumulation in ferromagnet \label{SEC:damping}}
The spin current pumping is equivalent to the increase of spin damping due to magnetization precession, as was discussed in Refs. \cite{Berger96,Tserkovnyak02}.
In this section, we confirm this fact by calculating the torque by evaluating the spin polarization of the conduction electron spin in F region. 

There are several ways to evaluate damping of magnetization. 
One way is to calculate the spin-flip probability of the electron as in Ref. \cite{Berger96}, which leads to damping of localized spin in the presence of strong $sd$ exchange interaction.
The second is to estimate the torque on the electron by use of equation motion \cite{TE08}.
The relation between the damping and spin current generation is clearly seen in this approach.
In fact, the total torque acting on conduction electron is ($\hbar$ times) the time-derivative of the electron spin density,  
\begin{align}
 \frac{d \sev}{dt} &=i \lt( \average{[H,d^\dagger]\sigmav d}+\average{d^\dagger \sigmav [H,d]} \rt) .
 \label{sdotgeneral}
\end{align}
At the interface, the right-hand side arises from the interface hopping.
Using the hopping Hamiltonian of Eq. (\ref{FNhoppingreal}), we have 
\begin{align}
\lt.\frac{d \sev}{dt}\rt|_{\rm interface} &=
 i \lt( \average{c^\dagger t\sigmav d}-\average{d^\dagger \sigmav t^\dagger  c} \rt) ,
\end{align} 
as the interface contribution.
As is natural, the the right -hand side agrees with the definition of the spin current passing through the interface. 
Evaluating the right-hand side, we obtain in general a term proportional to $\nv\times\dot{\nv}$, which gives the Gilbert damping, and term proportional to $\dot{\nv}$, which gives a renormalization of magnetization.
In contrast, away from the interface, the commutator $[H,d]$  arises from the kinetic term $H_0\equiv \intr \frac{|\nabla d|^2}{2m}$ describing electron propagation, resulting in
\begin{align}
 \frac{d \se_\alpha}{dt} &=i \lt( \average{[H_0,d^\dagger]\sigmav d}+\average{d^\dagger \sigmav [H,d]} \rt) \nnr
 &= \nabla\cdot\jsv^\alpha
 \end{align}
where $\jsv^\alpha(\rv)\equiv \frac{-i}{2m}(\nabla_{\rv}-\nabla_{\rv'}) 
 \average{d^\dagger(\rv') \sigma_\alpha d(\rv)} |_{\rv'=\rv}$ is the spin current. 
Away from the interface, the damping therefore occurs if the spin current has a source or a sink at the site of interest.

Here we use the third approach and estimate the torque on the localized spin by calculating the spin polarization of electrons as was done in Refs. \cite{KTS06,TKS_PR08}.
The electron spin polarization at position $\rv$ in the ferromagnet at time $t$ is 
$\sev^{\rm (F)}(\rv,t)\equiv \average{d^\dagger\sigmav d}$, which reads in the rotated frame 
$\se^{\rm (F)}_\alpha={\cal R}_{\alpha\beta}\stil^{\rm (F)}_\beta$, where
\begin{align}
 \stil^{\rm (F)}_\beta(\rv,t)=-i \tr[\sigma_\beta G^<(\rv,\rv,t,t)] ,
 \label{stildef} 
\end{align}
where $G_{\spinindex\spinindex'} ^<(\rv,\rv',t,t')\equiv i\average{\dtil_{\spinindex'}^\dagger \dtil_\spinindex}$ is the lesser Green's function in F region, which is a matrix in spin space ($\spinindex,\spinindex'=\pm$).
We are interested in the effect of the N region arising from the hopping.
We must note that the hopping interaction of Eq. (\ref{FNhoppingrot}) is not convenient for integrating out N electrons, since the $\ctil$ electrons' spins are time-dependent as a result of a unitary transformation, $U(t)$.
We thus use the following form (Fig. \ref{FIGFN}(b)), 
\begin{align}
 H_{\rm I} & = \int_{{\rm I}_{\rm F}} d^3r \int_{{\rm I}_{\rm N}} d^3r'
 \lt( 
c^\dagger(\rv') U \ttil(\rv',\rv)  \dtil(\rv) + 
  \dtil^\dagger(\rv) \ttil^*(\rv',\rv) U^{\dagger} c(\rv') \rt),
 \label{FNhoppingrothalf}
\end{align}
namely, the hopping amplitude between $\dtil$ and $c$ electrons includes unitary matrix $U$.

Let us argue in the rotated frame why the effect of damping arising from the interface. 
In the totally rotated frame of Fig. \ref{FIGFN}(c), the spin of F electron is static, while that of N electron varies with time. When F electron hops to N region and comes back, therefore, 
 electron spin gets rotated with the amount depending on the time it stayed in N region.  
This effect is in fact represented by a retardation effect of the matrices $U$ and $U^{-1}$ in Eq. (\ref{FNhoppingrothalf}). 
If off-diagonal nature of $U$ and $U^{-1}$ are neglected, the interface effects are all spin-conserving and do not induce damping for F electron (See Sec. \ref{SEC:nodamping}). 

\begin{figure}[tb]
  \begin{center}
  \includegraphics[width=0.6\hsize]{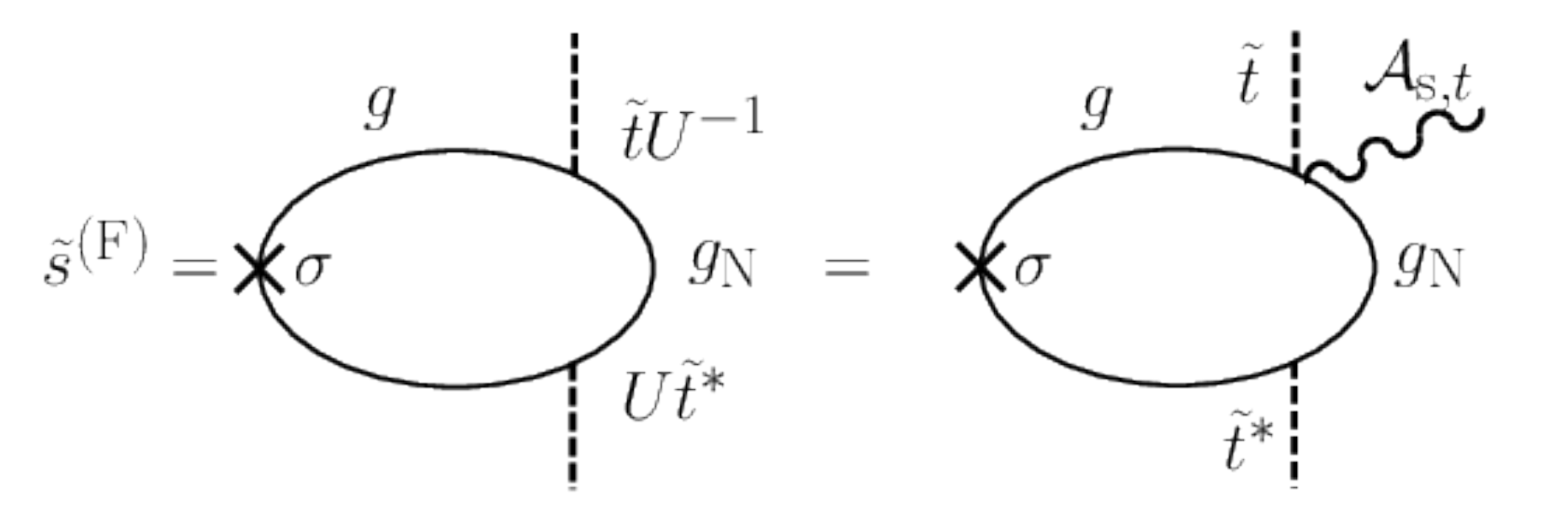}
  \end{center}
\caption{Diagramatic representation of the spin accumulation in ferromagnetic metal induced as a result of coupling to the normal metal (Eqs. (\ref{stildef})(\ref{SigmaF2})). Conduction electron Green's functions in ferromagnet and normal metal are denoted by $g$ and $g_{\rm N}$, respectively. 
Time-dependent matrix $U(t)$, defined by Eq. (\ref{Udef}), represents the effect of dynamic magnetization.
Expanding $U$ and $U^{-1}$ with respect to slow time-dependence of magnetization, we obtain gauge field representation, Eq. (\ref{SigmaF4}).
\label{FIGsF}} 
\end{figure}

We now proceed  calculation of induced spin density in the ferromagnetic metal.
Diagramatic representation of the contribution is in Fig. \ref{FIGsF}.
Writing spatial and temporal positions explicitly, the self-energy of F electron arising from the hopping to N region reads 
($\rv_1$ and $\rv_2$ are in F)
\begin{align}
\Sigma^{a}(\rv_1,\rv_2,t_1,t_2) & = \int_{{\rm I}_{\rm N}} d^3r_1'\int_{{\rm I}_{\rm N}} d^3r_2' \ttil(\rv_1,\rv_1') U^{-1}(t_1) g_{\rm N}^a(\rv_1',\rv_2',t_1-t_2)  U(t_2) \ttil^\dagger (\rv_2,\rv_2')
\label{SigmaF2}
\end{align}
where $a=\ret,\adv,<$.
We assume the Green's function in N region is spin-independent; i.e., we neglect higher order contribution of hopping. Moreover, we treat the hopping to occur only at the interface, i.e., at $x=0$. 
The self-energy is then represented as 
\begin{align}
\Sigma^a(\rv_1,\rv_2,t_1,t_2) & =  a^2\delta(x_1)\delta(x_2) 
\ttil  U^{-1}(t_1)   U(t_2) \ttil^\dagger 
\sum_{\kv}g^a_{\rm N}(\kv,t_1-t_2),
\label{SigmaF3}
\end{align}
where $a$ is the interface thickness, which we assume to be the order of the lattice constant.
Diagramatic representation of Eqs. (\ref{stildef})(\ref{SigmaF2}) are in Fig. \ref{FIGsF}.
Expanding the matrix using spin gauge field as 
$U^{-1}(t_1)   U(t_2)=1-i(t_1-t_2)\Ascal{t}{}+O((\Ascal{t}{})^2)$,
we  obtain the gauge field contribution of the self-energy as 
\begin{align}
\Sigma^a(\rv_1,\rv_2,t_1,t_2) 
& =  a^2 \delta(x_1)\delta(x_2) 
\sumom \frac{d e^{-i\omega(t_1-t_2)}}{d\omega} \ttil  A_{{\rm s},t} \ttil^\dagger 
\sum_{\kv}g_{\rm N}^a(\kv,\omega)
\nnr
&= - a^2 \delta(x_1)\delta(x_2) 
\sumom e^{-i\omega(t_1-t_2)} \ttil  A_{{\rm s},t} \ttil^\dagger 
\sum_{\kv}\frac{d}{d\omega} g_{\rm N}^a(\kv,\omega)
\label{SigmaF4}
\end{align}
The linear contribution of the lessor component of the off-diagonal self-energy is 
\begin{align}
 G^<(\rv,t,\rv',t) &= \gr \Sigma^\ret \ga +\gr \Sigma^<\ga +g^< \Sigma^\adv \ga  \nnr
 &= a^2 \sumom \sumkv \biggl[ \gr(\rv,\omega)  \frac{dg_{\rm N}^\ret(\kv,\omega)}{d\omega} \ttil  A_{{\rm s},t} \ttil^\dagger g^<(-\rv,\omega) 
 \nnr
 &
 + \gr(\rv,\omega)  \frac{d g_{\rm N}^<(\kv,\omega)}{d\omega}\ttil  A_{{\rm s},t} \ttil^\dagger g^\adv(-\rv,\omega) 
+
  g^<(\rv,\omega)  \frac{d g_{\rm N}^\adv(\kv,\omega)}{d\omega} \ttil  A_{{\rm s},t} \ttil^\dagger g^\adv(-\rv,\omega) 
 \biggr]
 \label{GFless1}
\end{align}
For finite distance from the interface, $r$, dominant contribution arises from the terms 
containing both $  g^\ret(\rv,\omega) $ and $g^\adv(-\rv,\omega) $, as they do not contain a  rapid oscillation  like $e^{i(\kfu+\kfd)r}$ and $e^{2i\kfs r}$.
Using an approximation $\sum_{\kv}g_{\rm N}^\ret(\kv,\omega)\sim -i\pi\dos_{\rm N}$ and partial integration with respect to $\omega$,
Eq. (\ref{GFless1}) reduces to 
\begin{align}
 G^<(\rv,t,\rv',t)
 &= 2\pi i \dos_{\rm N} a^2 \sumom f'_{\rm N}(\omega)
 \gr(\rv,\omega) \ttil  A_{{\rm s},t} \ttil^\dagger g^\adv(-\rv,\omega) 
 \label{GFless2}
\end{align}
For damping, off-diagonal contributions, $\Ascal{t}{\pm}$, are obviously essential.
The result of the spin density in F in the rotated frame, Eq. (\ref{stildef}),  is 
\begin{align}
\stil^{\rm (F)}_\alpha(\rv,t)
 &= 2\pi i \dos_{\rm N} a^2 \sumom f_{\rm N}'(\omega) A_{{\rm s},t}^\beta 
\tr[   \sigma_\alpha  \gr(\rv,\omega) \ttil \sigma_\beta  \ttil^\dagger g^\adv(-\rv,\omega) ]
\nnr
&=
 2\pi i \dos_{\rm N} a^2 \sumom f_{\rm N}'(\omega) A_{{\rm s},t}^\beta 
 \sum_{\kv\kv'}  e^{i(\kv-\kv')\cdot\rv}
 \tr[  \sigma_\alpha \gr(\kv,\omega) \ttil \sigma_\beta  \ttil^\dagger g^\adv(\kv',\omega) ]
\label{GFless3}
\end{align}
Evaluating the trace in spin space, we obtain 
\begin{align}
\tilde{\sev}^{\rm (F)}(\rv,t)
 &= - \dos_{\rm N}  
 \biggl[ \Av^\perp_{{\rm s},t} \gamma_1(\rv)
  + (\zvhat\times \Av^\perp_{{\rm s},t}) \gamma_2(\rv) \biggr]
  \label{GFless4}
\end{align}
where 
\begin{align}
 \gamma_1(\rv) &\equiv \sum_{\spinindex}\ttil_{-\spinindex} \ttil^\dagger_{\spinindex} \gr_{-\spinindex}(\rv) \ga_{\spinindex}(-\rv) \nnr
 \gamma_2(\rv) &\equiv \sum_{\spinindex}(-i\spinindex)\ttil_{-\spinindex} \ttil^\dagger_{\spinindex} \gr_{-\spinindex}(\rv) \ga_{\spinindex}(-\rv).
\end{align}
We consider an interface with infinite area and consider spin accumulation averaged over the plane parallel to the interface. 
The wave vector contributing is then those with finite $k_x$ but with $k_y=k_z=0$ and Green's function become one-dimensional like
\begin{align}
\sumkv e^{i\kv\cdot\rv}\gr_{\spinindex}(\kv)
&= \frac{im}{\kfs}e^{i\kfs|x|}e^{-|x|/(2\ell_\spinindex)},
\end{align}
where $\ell_{\spinindex}\equiv v_{{\rm F}\spinindex}\tau_\spinindex $ 
($v_{{\rm F}\spinindex}\equiv k_{{\rm F}\spinindex}/m$) is electron mean free path for spin $\spinindex$.
The induced spin density in the ferromagnet is finally obtained from Eq. (\ref{GFless4}) as 
\begin{align}
{\sev}^{\rm (F)}(\rv,t)
 &= \frac{ m^2\dos_{\rm N}a^2}{2\kfu\kfd}  \sum_{\spinindex}
 \biggl[ (\nv\times\dot{\nv}) \overline{T_{\spinindex,-\spinindex}} e^{-i\spinindex(\kfu-\kfd)x}
   + \dot{\nv}(-i\spinindex)  \overline{T_{\spinindex,-\spinindex}} e^{-i\spinindex(\kfu-\kfd)x} \biggr]
\nnr
&=
 \frac{ m^2\dos_{\rm N}a^2 }{2\kfu\kfd}   \sum_{\spinindex}
 \biggl[ (\nv\times\dot{\nv}) 
 \lt[\Re[ \overline{T_{\uparrow,\downarrow}}]\cos((\kfu-\kfd)x)
      +\Im[ \overline{T_{\uparrow,\downarrow}}]\sin((\kfu-\kfd)x) \rt] 
 \nnr
 &  + \dot{\nv}
 \lt[\Im[ \overline{T_{\uparrow,\downarrow}}]\cos((\kfu-\kfd)x)
    -\Re[ \overline{T_{\uparrow,\downarrow}}]\sin((\kfu-\kfd)x) \rt]   
\label{spinresult1}
\end{align}
and the torque on localized spin, $-\spol \nv\times \sev^{\rm (F)}$,  is
\begin{align}
{\torquev}(\rv,t)
&=
- \frac{ m^2\dos_{\rm N}a^2\spol}{2\kfu\kfd}    \sum_{\spinindex}
 \biggl[
 -\dot{\nv} 
 \lt[\Re[ \overline{T_{\uparrow,\downarrow}}]\cos((\kfu-\kfd)x)
      +\Im[ \overline{T_{\uparrow,\downarrow}}]\sin((\kfu-\kfd)x) \rt] 
 \nnr
 &  +(\nv\times \dot{\nv})
 \lt[\Im[ \overline{T_{\uparrow,\downarrow}}]\cos((\kfu-\kfd)x)
     -\Re[ \overline{T_{\uparrow,\downarrow}}]\sin((\kfu-\kfd)x) \rt]  . 
\label{torqueresult1}
\end{align}

\subsection{Enhanced damping and spin renormalization of ferromagnetic metal \label{SECmetaldamping}}

The total induced spin accumulation density in ferromagnet is
\begin{align}
 \overline{{\sev}^{\rm (F)}} & \equiv \frac{1}{d}\int_{-d}^0 dx {\sev}^{\rm (F)}(x) \nnr
 &= \frac{1}{\spol}\lt( (\nv\times \dot{\nv}) \lt[-\Im[\delta](1-\cos\tilde{d}) +\Re[\delta]\sin\tilde{d} \rt] 
 + \dot{\nv} \lt[ \Re[ \delta ](1-\cos \tilde{d}) +\Im[\delta]\sin \tilde{d} \rt]\rt),
 \label{totalaccumulation}
\end{align}
where $\tilde{d}\equiv (\kfu-\kfd)d$, 
$d$ is the thickness of ferromagnet and 
\begin{align}
 \delta \equiv \frac{ m^2\dos_{\rm N}a^2\spol}{\kfu\kfd(\kfu-\kfd)d}  \overline{T_{\uparrow,\downarrow}}.
\end{align} 
As a result of this induced electron spin density, $\overline{{\sev}^{\rm (F)}} $, 
the equation of motion for the averaged magnetization is modified to be \cite{Berger96}
\begin{align}
\dot{\nv}=-\alpha \nv\times\dot{\nv}-\gamma\Bv\times\nv - \spol \nv\times  \overline{{\sev}^{\rm (F)}} ,                                                                      
\label{LLGeq}
\end{align}
where $\Bv$ is the external magnetic field.

Let us first discuss thin ferromagnet case, $d \gg |\kfu-\kfd|^{-1}$, where oscillating part with respect  to $\tilde{d}$ is neglected.
The spin density then reads $
 \overline{{\sev}^{\rm (F)}}\simeq  \frac{1}{\spol}\lt(
  -\Im[\delta] (\nv\times \dot{\nv}) +\Re[\delta] \dot{\nv} \rt)
$
and 
the equation of motion becomes  
\begin{align}
 (1+\Im \delta ) \dot{\nv}= -\tilde{\alpha} \nv\times\dot{\nv} - \gamma\Bv\times\nv,
\end{align}
where 
\begin{align}
\tilde{\alpha} \equiv \alpha+ \Re \delta, \label{alphaenhance}
\end{align}
is the Gilbert damping including the enhancement due to  the spin pumping effect. 
The precession angular frequency $\omega_B$ is modified by the imaginary part of $T_{\uparrow,\downarrow}$, i.e., by the spin current proportional to $\dot{\nv}$, as 
\begin{align}
 \omega_B=\frac{\gamma B}{1+ \Im\delta}.\label{resonancefreq}
\end{align}
This is equivalent to the modification of the gyromagnetic ratio ($\gamma$) or the $g$-factor.

For most 3d ferromagnets, we may approximate $\frac{ m^2\dos_{\rm N}a\spol\ef^2}{2\kfu\kfd(\kfu-\kfd)} \simeq O(1)$ (as $\kfu-\kfd\propto \spol$), resulting in $\delta\sim \frac{a}{d} T_{\uparrow,\downarrow}$.
As discussed in Sec. \ref{SECQMSO}, when interface spin-orbit interaction is taken into account, 
we have $T_{\uparrow,\downarrow}=\ttil^0_{\uparrow} \ttil^0_{\downarrow} 
 +i\widetilde{\gamma}_{xz}  (\ttil^0_\uparrow+\ttil^0_{\downarrow})+O((\widetilde{\gamma})^2)$, where $\ttil^0_\spinindex$ and $\widetilde{\gamma}_{xz}$ are assumed to be real.
Moreover, $\ttil^0_\spinindex$ can be chosen as positive and thus $T_{\uparrow,\downarrow}>0$.
($\ttil^0_\spinindex$ here is field-representaion, and has unit of energy.)  
Equations (\ref{alphaenhance}) and (\ref{resonancefreq}) indicates that the strength of the hopping amplitude $\ttil^0_\spinindex$ and interface spin-orbit interaction $\widetilde{\gamma}_{xz}$ are experimentally accessible by measuring Gilbert damping and shift of resonance frequency as has been known \cite{Tserkovnyak02}.
A significant consequence of Eq. (\ref{alphaenhance}) is that the enhancement of the Gilbert damping, 
\begin{align}
\delta \alpha\sim \frac{a}{d}  \frac{1}{\ef^2} \ttil^0_{\uparrow} \ttil^0_{\downarrow} ,
\end{align}
can exceed in thin ferromagnets the intrinsic damping parameter $\alpha$, as the two contributions are governed by different material parameters.
In contrast to the positive enhancement of damping, the shift of the resonant frequency or $g$-factor can be positive or negative, as it is linear in the interface spin-orbit parameter $\widetilde{\gamma}_{xz}$.

Experimentally, enhancement of the Gilbert damping and frequency shift has been measured in many systems \cite{Mizukami01}.
In the case of Py/Pt junction, enhancement of damping is observed to be proportional to $1/d$ in the range of 2nm$<d<10$nm, and the enhancement was large, $\delta \alpha/\alpha\simeq 4$ at $d=2$ nm \cite{Mizukami01}.
These results appears to be consistent with our analysis.
Same $1/d$ dependence was observed in the shift of $g$-factor.
The shift was positive and magnitude is about 2\% for Py/Pt and Py/Pd with $d=2$nm, while it was negative for Ta/Pt \cite{Mizukami01}.
The existence of both signs suggests that the shift is due to the linear effect of spin-orbit interaction, and the interface spin-orbit interaction we discuss is one of possible mechanisms. 

For thin ferromagnet, $\tilde{d}\lesssim1$,   the spin accumulation of Eq. (\ref{totalaccumulation}) reads 
\begin{align}
 \overline{{\sev}^{\rm (F)}} 
 &= \frac{1}{\spol}\lt(  (\nv\times \dot{\nv}) \Re[\delta_{\rm thin}] + \dot{\nv} \Im[\delta_{\rm thin}]\rt),
 \label{spinaccumuthin}
\end{align}
where 
\begin{align}
 \delta_{\rm thin} \equiv \delta \tilde{d}= \frac{ m^2\dos_{\rm N}a^2\spol}{2\kfu\kfd}  \overline{T_{\uparrow,\downarrow}}.
\end{align} 
Equation (\ref{spinaccumuthin}) indicates that the roles of imaginary and real part of $  \overline{T_{\uparrow,\downarrow}}$ are interchanged for thick and thin ferromagnet, resulting in
\begin{align}
\tilde{\alpha} = \alpha+ \Im \delta_{\rm thin} \nnr
 \omega_B  =\frac{\gamma B}{1-\Re\delta_{\rm thin}}, \label{thin}
\end{align}
for thin ferromagnet.
Thus, for weak interface spin-orbit interaction,  positive shift of resonance frequency is expected (as $\Re\delta_{\rm thin}>0$).
Significant feature is that the damping can be smallened or even be negative if  strong interface spin-orbit interaction exists with negative sign of $\Im \delta_{\rm thin}$. 
Our result indicates that 'spin mixing conductance' description of Ref. \cite{Tserkovnyak02} breaks down in thin metallic ferromagnet (and insulator case as we shall see in Sec. \ref{SEC:alphaInsulator}).

In this section, we have discussed spin accumulation and enhanced Gilbert damping in ferromagnet attached to a normal metal.
In the field-theoretic description, the damping enhancement arises from the imaginary part of the self-energy due to the interface. Thus a randomness like the interface scattering changing the electron momentum is essential for the damping effect, which sounds physically reasonable. 
The same is true for the reaction, namely, spin current pumping effect into N region, and thus spin current pumping requires randomness, too. 
(In the quantum mechanical treatment of Sec. \ref{SECQM}, change of electron wave vector at the interface is essential.)
The spin current pumping effect therefore appears different from general pumping effects, where randomness does not play essential roles apparently \cite{Moskalets12}.

Spin accumulation and enhanced Gilbert damping was discussed by Berger\cite{Berger96} based on a quantum mechanical argument. There $1/d$ dependence was pointed out and the damping effect was calculated by evaluating the decay rate of magnons. 
Comparison of enhanced Gilbert damping with experiments was carried out in Ref. \cite{Tserkovnyak02} but in a phenomenological manner.

\section{Case  with magnetization structure \label{SEC:structure}}
Field theoretic approach has an advantage that generalization of the results is straightforward. 
Here we discuss briefly the case of ferromagnet with spatially-varying magnetization.
The excitations in metallic ferromagnet consist of spin waves (magnons) and Stoner excitation.
While spin waves usually have gap as a result of magnetic anisotropy, Stoner excitation is gapless for finite wave vector, 
$(\kfu-\kfd)<|q|<(\kfu+\kfd) $, and it may be expected to have significant contribution for magnetization structures having wavelength larger than $\kfu-\kfd$. Let us look into this possibility.

Our result of spin accumulation in ferromagnet, represented in the rotated frame, Eq. (\ref{stilresunif}), indicates that when the magnetization has a spatial profile, the accumulation is determined by the spin gauge field and spin correlation function depending on the wave vector $\qv$ as 
\begin{align}
\sum_{\qv}  \Ascal{t}{\pm}(q) \chi_\pm(q,0),
\label{stilstruc}
\end{align}
where 
\begin{align}
\chi_{\pm}(\qv,\Omega)
&\equiv 
- \sum_{\kv} 
\frac{f_{\kv+\qv,\pm}-f_{\kv,\mp}}
{ \epsilon_{\kv+\qv,\pm}-\epsilon_{\kv,\mp} +\Omega+i0},
\end{align}
is the correlation function with finite momentum transfer $\qv$ and finite angular frequency $\Omega$.
For the case of free electron with quadratic dispersion, the correlation function is \cite{TF94_JPSJ}
\begin{align}
\chi_{\pm}(\qv,\Omega)&= A_q+i\Omega B_q \theta_{\rm st}(q)+O(\Omega^2) ,
\label{chiomega}
\end{align}
where 
\begin{align}
 A_q&=\frac{ma^3}{8\pi^2}\left[(\kfu+\kfd)
 \left(1+\frac{(\kfu-\kfd)^2}{q^2}\right)  
 \right. \nonumber\\
& +  \left. \frac{1}{2q^3}((\kfu+\kfd)^2-q^2)(q^2-(\kfu-\kfd)^2) \ln
  \left| \frac{q+(\kfu+\kfd)}{q-(\kfu+\kfd)} \right|  \right]  \nonumber\\
B_q &=  \frac{m^2a^3}{4\pi |q|},  \label{B} 
\end{align}
and 
\begin{align}
\theta_{\rm st}(q) &\equiv \left\{ \begin{array}{ccc} 1  & \mbox{  }  &
  (\kfu-\kfd)<|q|<(\kfu+\kfd)   \\
                                            0  &   &
  {\rm otherwise} \end{array} \right..\label{st}
\end{align}
describes the wave vectors where Stoner excitation exists.
As we see from Eq. (\ref{chiomega}), the Stoner excitation contribution vanishes to the lowest order in $\Omega$, and thus the spin pumping effect in the adiabatic limit ($\Omega\ra0$) is not affected. 
Moreover, the real part of the correlation function,   $A_q$,  is a decreasing function of $q$ and thus the spin pumping efficiency would decrease when ferromagnet has a structure.
However, for rigorous argument, we need to include the spatial component of the spin gauge field arising form the spatial derivative of the magnetization profile.

As for the effect of the Stoner excitation on spin damping (Gilbert damping), it was demonstrated for the case of a domain wall that the effect is negligibly small for a wide wall with thickness $\lambda\gg(\kfu-\kfd)^{-1}$  (Refs. \cite{TF94,TF94_JPSJ}). 
Simanek and Heinrich presented a result of the Gilbert damping as the linear term in the frequency of the imaginary part of the spin correlation function integrated over the wave vector \cite{SimanekHeinrich03}. 
The result  is, however, obtained for a model where ferromagnet is atomically thin layer (a sheet), and would not be applicable for most experimental situations. 
Discussion of Gilbert damping including finite wave vector and the impurity scattering was given in Ref. \cite{Umetsu12}. 
Inhomogenuity effects of damping of a domain wall was studied recently in detail \cite{Yuan16}.

\section{Insulator ferromagnet \label{SEC:insulator}}

In this section, we discuss the case of ferromagnetic insulator.
It turns out that the generation mechanisms for spin current  in the insulating and metallic cases are distinct. 

\subsection{Magnon and adiabatic gauge field}

The Lagrangian for the insulating ferromagnet is 
\begin{align}
 L_{\rm IF} &= \intr \lt[S\dot{\phi}(\cos\theta-1) -\frac{J}{2}(\nabla\Sv)^2 \rt]-H_{K},
 \label{LIF}
\end{align}
where $J$ is the exchange interaction between the localized spin, $\Sv$, and $H_{K}$ denotes the magnetic anisotropy energy.

We first study low energy magnon dynamics induced by slow magnetization dynamics. 
For  separating the classical variable and fluctuation (magnon), rotating coordinate description used in the metallic case is convenient. 
For magnons described by the Holstein-Primakov boson,  the unitary transformation is a $3\times3$ matrix defined as follows \cite{TataraDW15}.
\begin{align}
 \Sv&= U \widetilde{\Sv},
 \label{Unitary3x3Tr}
\end{align}
where 
\begin{align}
  U &= \lt( \begin{array}{ccc}
         \cos\theta\cos\phi & -\sin\phi & \sin\theta\cos\phi  \\
         \cos\theta\sin\phi &  \cos\phi & \sin\theta\sin\phi  \\
         -\sin\theta & 0 & \cos \theta 
       \end{array}  \rt) = \lt( \begin{array}{ccc}
         \ev_\theta & \ev_\phi & \nv 
       \end{array}  \rt).
 \label{Unitary3x3}
\end{align}
The diagonalized spin $\widetilde{\Sv}$ is represented in terms of annihilation and creation operators for the Holstein-Primakov boson, $\boson$ and $\boson^\dagger$, as 
\cite{Kittel63}
\begin{align}
  \widetilde{\Sv} &= \lt( \begin{array}{c}
         \sqrt{\frac{S}{2}} (\boson^\dagger +\boson) \\
         i\sqrt{\frac{S}{2}}(\boson^\dagger-\boson) \\
        S - \boson^\dagger \boson 
       \end{array}  \rt).\label{HPboson}
\end{align}
We neglect the terms that are third- and higher-order in boson operators.
Derivatives of the localized spin then read  
\begin{align}
  \partial_\mu\Sv=U(\partial_\mu+iA_{U,\mu}) \widetilde{\Sv},
  \label{partialS}
\end{align}
where 
\begin{align}
 A_{U,\mu}   & \equiv -i U^{-1} \nabla_\mu U,
\end{align}
is the spin gauge field represented as a $3\times3$ matrix.
The spin Berry's phase of the Lagrangian (\ref{LIF}) is 
written in terms of magnon as (derivation is in Sec. \ref{SEC:rotatedmagnon})
\begin{align}
 L_{\rm m} 
 &=  2S\gamma^2\intr
  i [\boson^\dagger (\partial_t+i\Ascal{t}{z})\boson-\boson^\dagger(\stackrel{\leftarrow}{\partial}_t -i\Ascal{t}{z})) \boson] ,
 \end{align}
namely,  magnons interacts with the adiabatic component of the same spin gauge field for electrons, $\Ascal{t}{z}$,  defined in Eq. (\ref{Aexpression}).
As magnon is a single-component field, the gauge field is also single-component, i.e., a U(1) gauge field.
This is a significant difference between insulating and metallic ferromagnet; In the metallic case, conduction electron couples to an SU(2) gauge field with spin-flip components, which turned out to be essential for spin current generation. In contrast, in the insulating case, the magnon has diagonal gauge field, i.e., a spin chemical potential, which simply induces diagonal spin polarization.
Pumping of magnon was discussed in a different approach by evaluating magnon source term in Ref. \cite{Nakata11}.

The exchange interaction at the interface is represented by a Hamiltonian
\begin{align}
 H_{\rm I} &= J_{\rm I} \intr_{\rm I} \Sv(\rv)\cdot c^\dagger \sigmav c,
 \label{HIdef}
\end{align}
where $J_{\rm I}$ is the strength of the interface $sd$ exchange interaction and the integral is over the interface. We consider a sharp  interface at $x=0$.
Using Eq. (\ref{Unitary3x3Tr}), the interaction is represented in terms of magnon operators up to the second order as
\begin{align}
 H_{\rm I} &=  \Jint \intr_{\rm I} \lt[(S-\boson^\dagger\boson)c^\dagger(\nv\cdot\sigmav)c
 +\sqrt{\frac{S}{2}} \lt[ \boson^\dagger c^\dagger \Phiv\cdot\sigmav c + \boson c^\dagger \Phiv^*\cdot\sigmav c \rt]\rt],
 \label{HIboson}
\end{align}
where 
\begin{align}
 \Phiv&\equiv \evth+i\evph
 = \lt( \begin{array}{c}\cos\theta\cos\phi-i\sin\phi \\ \cos\theta\sin\phi+i\cos\phi \\ -\sin\theta \end{array}\rt).
 \label{Phidef}
\end{align}
Equation (\ref{HIboson}) indicates that there are two mechanisms for spin current generation; namely, the one due to the magnetization at the interface (the term proportional to $\nv$) and the one due to the magnon spin scattering at the interface (described by the term linear in magnon operators). 

Let us briefly demonstrate based on the expression of Eq. (\ref{HIboson}) that spin-flip processes due to magnon creation or annihilation lead to generation of spin current in the normal metal. 
At the second order, the interaction induces a factor on the electron wave function 
$(\Phiv^*(t)\cdot\sigmav) (\Phiv(t')\cdot\sigmav)$ for magnon creation and 
$(\Phiv(t)\cdot\sigmav) (\Phiv^*(t')\cdot\sigmav)$ for annihilation 
(we allow an infinitesimal difference in time $t$ and $t'$).
The factor for the creation has charge and spin contributions,  
$(\Phiv^*(t)\cdot\sigmav) (\Phiv(t')\cdot\sigmav)=\Phiv^*(t)\cdot\Phiv(t')+i\sigmav\cdot(\Phiv^*(t)\times\Phiv(t')) $.
For magnon annihilation, we have 
$(\Phiv^*(t)\times\Phiv(t'))^*$, and thus the sum of the magnon creation and annihilation processes give arise to a factor 
\begin{align}
 \sum_{\qv}[(n_\qv+1) (\Phiv^*(t)\times\Phiv(t')) + n_\qv (\Phiv^*(t)\times\Phiv(t'))^*]
 &= 
 \sum_{\qv}[(2n_\qv+1) \Re[\Phiv^*(t)\times\Phiv(t') ] +i \Im [\Phiv^*(t)\times\Phiv(t') ].
 \label{spinfactor}
\end{align}
For adiabatic change the amplitude is expanded as  
\begin{align}
(\Phiv^*(t)\times\Phiv(t')) 
&= 2i(1+i(t-t')\cos\theta\dot{\phi})\nv
-(t-t')(\nv\times\dot{\nv}-i\dot{\nv}) +O((\partial_t)^2) ,
\end{align}
where we see that an retardation effect from the adiabatic change of magnetization (represented by the second term on the right-hand side) gives rise to a magnon state change  proportional to  $\nv\times\dot{\nv}$ and $\dot{\nv}$.
The retardation contribution for the spin part (Eq. (\ref{spinfactor})) is 
\begin{align}
 (t-t')\sum_{\qv}[-(2n_\qv+1) (\nv\times\dot{\nv})+i\dot{\nv}].
 \label{magnonsimple}
\end{align}
We therefore expect that a spin current proportional to $\nv\times\dot{\nv}$ emerges proportional to the magnon creation and annihilation number, $\sum_{\qv}(2n_\qv+1)$. 
(As we shall see below, the factor $t-t'$ reduces to a derivative with respect to angular frequency of the Green's function.)
A rigorous estimation using Green's function method is presented in Sec. \ref{SEC:magnonspincurrent}. 

In Eq. (\ref{magnonsimple}), the last term proportional to $\dot{\nv}$ is an imaginary part arising from the difference of magnon creation and annihilation probabilities of vacuum, $n_\qv+1$ and $n_\qv$. 
The term is, however, unphysical one corresponding to a real energy shift due to magnon interaction, and is removed by redefinition of the Fermi energy.

\subsection{Spin current pumped by the interface exchange interaction \label{SEC:interfacespincurrent}}

Here we study the spin current pumped by the classical magnetization at the interface, namely, the one driven by 
the term proportional to $S\nv$ in Eq. (\ref{HIboson}).
We treat the exchange interaction perturbatively to the second order as the exchange interaction between conduction electron and insulator ferromagnet is localized at the interface and is expected to be weak.  
The weak coupling scheme employed here is in the opposite limit as the strong coupling (adiabatic) approach used in the metallic ferromagnet (Sec. \ref{SEC:GF}). 

In the perturbative regime, the issue of adiabaticity needs to be argued carefully.
In the strong $sd$ coupling limit, the adiabaticity is trivially satisfied, as the time needed for the electron spin to follow the localized spin is the fastest timescale.
In the weak coupling limit, this timescale is long. Nevertheless, the adiabatic condition is satisfied if the electron spin relaxation is strong so that the electron spin relaxes quickly to the local equilibrium state determined by the localized spin.
Thus the adiabatic condition is expected to be $\spol_{\rm I}\tau_{\rm sf} /\hbar\ll 1$, where $\spol_{\rm I}$ and $\tau_{\rm sf}$ are  the interface spin splitting energy, and conduction electron spin relaxation time, respectively.
In the following calculation, we consider the case of $\ef \tau_{\rm sf}/\hbar \gg 1$, i.e., $\hbar(\tau_{\rm sf})^{-1} \ll \ef$, as the spin flip lifetime is by definition longer than the elastic electron lifetime $\tau$, which satisfies $\ef\tau/\hbar \gg 1$ in metal.
Our results therefore cover both adiabatic and nonadiabatic limits.

\begin{figure}[ht]
  \begin{center}
  \includegraphics[width=0.5\hsize]{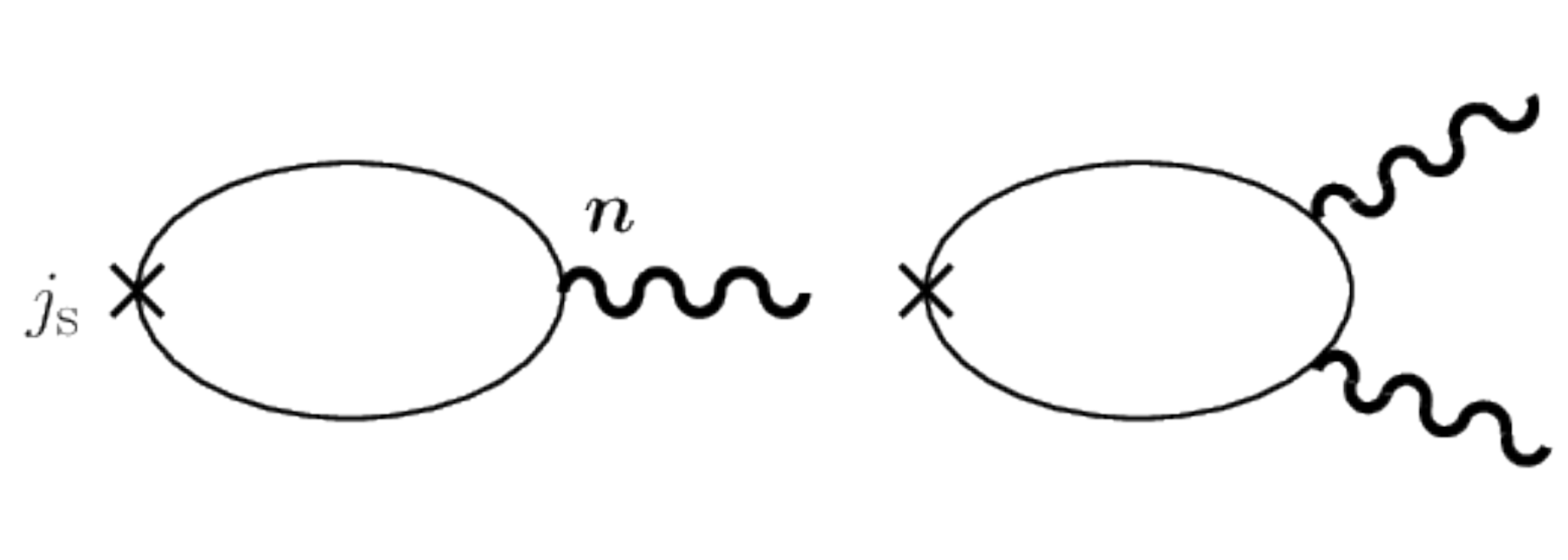}
  \end{center}
 {\caption The Feynmann diagrams for spin current pumped by interface $sd$ exchange interaction.  \label{FIGInsulatorSC}}
\end{figure}

The calculation is carried out by evaluating Feynmann diagrams of Fig. \ref{FIGInsulatorSC}, similar to the study of Refs. \cite{Takeuchi08,Hosono_LT09}. 
A difference is that while Refs. \cite{Takeuchi08,Hosono_LT09} assumed a smooth magnetization structure and used a gradient expansion, the exchange interaction we consider is localized.

The lesser Green's function for normal metal including the interface exchange interaction to the linear order is 
\begin{align}
 G_{\rm N}^{(1)<}(\rv,t,\rv,t) &= \spol_{\rm I} \sumom \sumOm \sum_{\kv\kv'} e^{-i\Omega t} e^{i(\kv'-\kv)\cdot\rv} 
 \lt[ (f(\omega+\Omega)-f(\omega))g^\ret_{\kv',\omega+\Omega} g^\adv_{\kv\omega} 
   -f(\omega)g^\ret_{\kv',\omega+\Omega} g^\ret_{\kv\omega} 
  +f(\omega+\Omega)g^\adv_{\kv',\omega+\Omega} g^\adv_{\kv\omega}  \rt](\nv_{\Omega}\cdot\sigmav) ,
\end{align}
where $\spol_{\rm I}\equiv \Jint S$ is the local spin polarization at the interface.
Expanding the expression with respect to $\Omega$ and keeping the dominant contribution at long distance, i.e., the terms containing both $g^\adv$ and $g^\ret$.
Using $\sumkv g^\adv_{\kv\omega}e^{i\kv\cdot\rv}\simeq \frac{im}{\kf}e^{ikr}e^{-\frac{|x|}{\ell}}(\equiv g^\adv(\rv))$, the result of spin current is 
\begin{align}
 j_{\rm s} ^{(1)}(\rv,t) & = - \spol_{\rm I} \frac{m}{\kf} \dot{\nv}e^{-|x|/\ell}.
\end{align}

The second-order contribution is similarly calculated to obtain 
\begin{align}
 G_{\rm N}^{(2)<}(\rv,t,\rv,t) &\simeq  
 (\spol_{\rm I})^2 \sumom \int\frac{d\Omega_1}{2\pi} \int\frac{d\Omega_2}{2\pi}  
\nnr & \times \sum_{\kv\kv'\kv''}
 e^{-i(\Omega_1+\Omega_2) t} e^{i(\kv'-\kv)\cdot\rv} 
 f'(\omega)g^\ret_{\kv',\omega} g^\adv_{\kv\omega}
 (\Omega_1 g^\adv_{\kv''\omega}+\Omega_2 g^\ret_{\kv''\omega})(\nv_{\Omega_1}\cdot\sigmav)  (\nv_{\Omega_2}\cdot\sigmav) 
 \nnr
 &= - 2\pi i \dos  (\spol_{\rm I})^2 |g^\ret(\rv)|^2 (\nv\times\dot{\nv}) \cdot\sigmav.
\end{align}
The corresponding spin current at the interface ($x=0$) is thus 
\begin{align}
 j_{\rm s} ^{(2)}(x=0,t) & = \dos  (\spol_{\rm I})^2  \frac{m}{\kf}  (\nv\times\dot{\nv}),
\end{align}
and the total spin current reads 
\begin{align}
 j_{\rm s} (x=0,t) & = -\spol_{\rm I} \frac{m}{\kf} \dot{\nv} -2 \dos  (\spol_{\rm I})^2  \frac{m}{\kf}  (\nv\times\dot{\nv}).
\label{totaljsinsulator}
\end{align}
In the perturbation regime, the spin current proportional to $\dot{\nv}$ is dominant (larger by a factor of $(\dos  \spol_{\rm I})^{-1}$) compared to the 
one  proportional to $\nv\times\dot{\nv}$.

Expression of spin current induced by the interface exchange interaction was presented in Ref. \cite{Kajiwara10} in the limit of strong spin relaxation, $\spol_{\rm I}\tau_{\rm sf} \ll 1$, where $\tau_{\rm sf}$ is the spin relaxation time of electron. 
By solving the Landau-Lifshitz-Gilbert equation for the electron spin, they obtained an result of Eq. (\ref{totaljsinsulator}) with $\dos \spol_{\rm I}$ replaced by $\spol_{\rm I} \tau_{\rm sf}$.

\subsection{Calculation of magnon-induced spin current \label{SEC:magnonspincurrent}}
Here magnon-induced spin current due to the magnon interaction in Eq. (\ref{HIboson}) is calculated.
As magnon is a small fluctuation of magnetization, the contribution here is a small correction to the contribution of Eq. (\ref{totaljsinsulator}).
Nevertheless, the magnon contribution has a typical linear dependence on the temperature, and is expected to be experimentally identified easily. 

\begin{figure}[ht]
  \begin{center}
  \includegraphics[width=0.25\hsize]{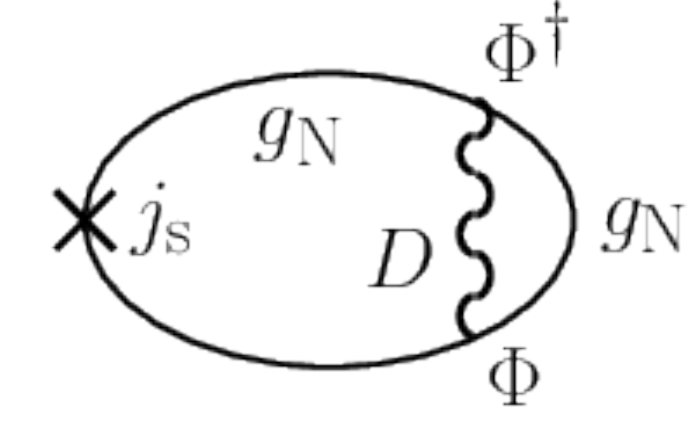}
  \end{center}
{\caption %
The Feynmann diagrams for spin current pumped by magnons at the interface.  
 Green's functions for magnons and electrons in the normal metal are denoted by $D$ and $g_{\rm N}$, respectively. $\Phi$ represents the effects of magnetization dynamics (Eq. (\ref{Phidef})). 
 \label{FIGJsMagnon}}
\end{figure}

Spin current induced in normal metal is evaluated by calculating the self-energy arising from the interface magnon scattering of Eq. (\ref{HIboson}).
The contribution to the path-ordered Green's function of N electron from the magnon scattering to the second order is 
\begin{align}
 G_{\rm N}(\rv,t,\rv'.t') &= 
 \int_C dt_1 \int_C dt_2 \sum_{\rv_1\rv_2} g_{\rm N}(\rv,t,\rv_1,t_1)
 \Sigma_{\rm I}(\rv_1,t_1,\rv_2,t_2) g_{\rm N}(\rv_2,t_2,\rv',t') ,
 \label{GlessIdef}
\end{align}
where 
\begin{align}
 \Sigma_{\rm I}(\rv_1,t_1,\rv_2,t_2) 
 &\equiv 
 i {\frac{S\Jint^2}{2}} {\cal D}_{\alpha\beta}(\rv_1,t_1,\rv_2,t_2)  \sigma_\alpha g_{\rm N}(\rv_1,t_1,\rv_2,t_2)  \sigma_\beta,
 \label{SigmaIdef}
\end{align}
represents the self energy.
Here 
\begin{align}
{\cal D}_{\alpha\beta}(\rv_1,t_1,\rv_2,t_2)&\equiv
-i\average{T_C {\cal B}_\alpha(\rv_1,t_1) {\cal B}_{\beta}(\rv_2,t_2)},
\label{calDdef}
\end{align}
is the Green's function for magnon dressed by the magnetization structure ($\Phiv$ is defined in Eq. (\ref{Phidef})),
\begin{align}
 {\cal B}_{\alpha}(\rv,t)
 &\equiv \Phi_\alpha(t) \boson^\dagger(\rv,t)+\Phi^\dagger_\alpha(t) \boson(\rv,t).
 \label{calBdef}
\end{align}
Diagramatic representation is in Fig. \ref{FIGJsMagnon}.
In the present approximation including the interface scattering to the second order, the electron Green's function in Eq. (\ref{SigmaIdef}) is treated as  spin-independent, resulting in a self energy (defined on complex time contour)
\begin{align}
 \Sigma_{\rm I}(\rv_1,t_1,\rv_2,t_2) 
 &=
 i {\frac{S\Jint^2}{2}}
 (\delta_{\alpha\beta}+i\epsilon_{\alpha \beta\gamma} \sigma_\gamma)
 {\cal D}_{\alpha\beta}(\rv_1,t_1,\rv_2,t_2)  g_{\rm N}(\rv_1,t_1,\rv_2,t_2) .
 \label{SigmaI1}
\end{align}
We focus on the spin-polarized contribution,
\begin{align}
 \Sigma_{{\rm I},\gamma}(\rv_1,t_1,\rv_2,t_2) 
 &\equiv 
 - {\frac{S\Jint^2}{2}}\widetilde{\cal D}_{\gamma}(\rv_1,t_1,\rv_2,t_2)  g_{\rm N}(\rv_1,t_1,\rv_2,t_2) ,
 \label{SigmaIspin}
\end{align}
where $\widetilde{\cal D}_{\gamma}\equiv \epsilon_{\alpha \beta\gamma} {\cal D}_{\alpha\beta}$.
The  spin-dependent contribution of lessor Green's function, Eq. (\ref{GlessIdef}), reads
(time and spatial coordinates partially suppressed)
\begin{align}
 G_{\rm N}^<(\rv,t,\rv',t') &= \sigma_\gamma 
 \int_{-\infty}^\infty  dt_1 \int_{-\infty}^\infty dt_2 
 \biggl[ g_{\rm N}^\ret(t-t_1)
 \Sigma_{{\rm I},\gamma}^\ret(t_1,t_2) g_{\rm N}^<(t_2-t') + g_{\rm N}^\ret \Sigma_{{\rm I},\gamma}^< g_{\rm N}^\adv
+ g_{\rm N}^< \Sigma_{{\rm I},\gamma}^\adv g_{\rm N}^\adv  \biggr]
\equiv 
\sigma_\gamma G_{{\rm N},\gamma}^<(\rv,t,\rv'.t').
 \label{GlessI2}
\end{align}
For the self energy type of the Green's functions, depending on two time as $g(t_1-t_2){\cal D}(t_1-t_2)$ (Eq. (\ref{SigmaIspin})), real-time components are written as (suppressing time and suffix of N) (See Sec. \ref{SEC:SEdecomp})
\begin{align}
[g(t_1-t_2){\cal D}(t_1-t_2)]^\ret 
&= g^\ret {\cal D}^<+g^> {\cal D}^\ret =  g^< {\cal D}^\ret+g^\ret {\cal D}^> 
\nonumber\\
[g(t_1-t_2) {\cal D} (t_1-t_2)]^{\adv}
&= g^\adv {\cal D}^>+g^< {\cal D}^\adv
= g^\adv {\cal D}^<+g^> {\cal D}^\adv 
\nonumber\\
[g(t_1-t_2){\cal D}(t_1-t_2)]^<
&= g^< {\cal D}^<.
\end{align}
The Green's function $ \widetilde{\cal D}$ is that of a composite field 
${\cal B}_{\alpha}$ defined in Eq. (\ref{calBdef}), and is decomposed to elementary magnon Green's function, $D$, as 
\begin{align}
\widetilde{\cal D}_{\gamma}(\rv_1,t_1,\rv_2,t_2)
&= 
[\Phiv^\dagger (t_1) \times \Phiv(t_2)]_\gamma D(\rv_1,t_1,\rv_2,t_2) 
-
[\Phiv^\dagger (t_2) \times \Phiv(t_1)]_\gamma D(\rv_2,t_2,\rv_1,t_1) ,
\label{calDdef2}
\end{align}
where 
\begin{align}
D(\rv_1,t_1,\rv_2,t_2) 
& \equiv 
-i\average{T_C \boson(\rv_1,t_1) \boson^\dagger(\rv_2,t_2)}.
\end{align}
The spin-dependent factor in Eq. (\ref{calDdef2}) is calculated for adiabatic dynamics as 
\begin{align}
\Phiv^\dagger (t_1) \times \Phiv(t_2)
&=
2i\nv(t_1)+(t_2-t_1) [\Psiv + i\dot{\nv}]+O((\partial_t)^2 ), 
\end{align}
where 
\begin{align}
 \Psiv \equiv 2\cos\theta \dot{\phi}\nv + \nv\times\dot{\nv}.
 \label{Psivdef}
\end{align}
The real-time Green's functions are therefore ($D(1,2)\equiv D(\rv_1,t_1,\rv_2,t_2)$)
\begin{align}
 \widetilde{\cal D}_{\gamma} ^< &  (\rv_1,t_1,\rv_2,t_2)
= 
 2i\nv(t_1)[D^<(\rv_1,t_1,\rv_2,t_2) -D^>(\rv_2,t_2,\rv_1,t_1) ]
 \nnr &
 +(t_2-t_1) \biggl[
 \Psiv [D^<(\rv_1,t_1,\rv_2,t_2) +D^>(\rv_2,t_2,\rv_1,t_1)]
 +i\dot{\nv}[D^<(\rv_1,t_1,\rv_2,t_2) -D^>(\rv_2,t_2,\rv_1,t_1)]
 \biggr]
 \nnr
 \widetilde{\cal D}_{\gamma}^\ret & (1,2)
= \theta(t_1-t_2)(\widetilde{\cal D}_{\gamma}^<(1,2) -\widetilde{\cal D}_{\gamma}^>(1,2)) \nnr
\widetilde{\cal D}_{\gamma}^\adv & (1,2)
= -\theta(t_2-t_1)\epsilon_{\alpha\beta\gamma}({\cal D}_{\alpha\beta}^<(1,2) -{\cal D}_{\alpha\beta}^>(1,2)) , \label{tildecalDs}
\end{align}
and $ \widetilde{\cal D}_{\gamma} ^<$ is obtained by exchanging $<$ and $>$ in  $\widetilde{\cal D}_{\gamma} ^<$.
Elementary Green's functions are calculated as  
\begin{align}
 D^<(\rv_1,t_1,\rv_2,t_2)  &= -i\sum_{\qv}e^{i\qv\cdot(\rv_1-\rv_2)} n_{\qv}e^{-i\omega_\qv(t_1-t_2)} \nnr
 D^>(\rv_1,t_1,\rv_2,t_2)  &= -i\sum_{\qv}e^{i\qv\cdot(\rv_1-\rv_2)} (n_{\qv}+1)e^{-i\omega_\qv(t_1-t_2)} ,
\end{align}
where $\omega_\qv$ is magnon energy and $n_\qv\equiv \frac{1}{e^{\beta\omega_\qv}-1}$.
In our model, the interface is atomically flat and has an infinite area, and thus  $\rv_i$($i=1,2$)  are at $x=0$. 
Fourier components defined as ($a=\ret,\adv,<,>$) 
\begin{align}
 \widetilde{\cal D}_{\gamma}^a(x_1=0,t_1,x_2=0,t_2) &\equiv 
\sum_{\qv}\sumOm e^{-i\Omega(t_1-t_2)} 
 \widetilde{\cal D}_{\gamma}^a({\qv,\Omega}),
 \end{align}
 are calculated  from Eq. (\ref{tildecalDs}) as 
\begin{align}
 \widetilde{\cal D}_{\gamma}^<({\qv,\Omega})
 &= 
  -i\biggl[2\nv(D^<_--D_+^>) +\frac{d}{d\Omega}\lt[ \Psiv(D^<_-+D_+^>) +i\dot{\nv}(D^<_--D_+^>) \rt] \biggr]\nnr
\widetilde{\cal D}_{\gamma}^\ret({\qv,\Omega})
 &=
 -i\biggl[2\nv(D^\ret_-+D_+^\ret) +\frac{d}{d\Omega}\lt[ \Psiv(D^\ret_--D_+^\ret) +i\dot{\nv}(D^\ret_-+D_+^\ret) \rt] \biggr]\nnr
 \widetilde{\cal D}_{\gamma}^\adv({\qv,\Omega})
&= 
 -i\biggl[2\nv(D^\adv_-+D_+^\adv) +\frac{d}{d\Omega}\lt[ \Psiv(D^\adv_--D_+^\adv) +i\dot{\nv}(D^\adv_-+D_+^\adv) \rt] \biggr],
 \label{calDs}
\end{align}
where 
\begin{align}
  D_\pm^\adv &\equiv \frac{1}{\Omega\pm \omega_{\qv}-i0} , 
 & D_\pm^\ret \equiv \frac{1}{\Omega\pm \omega_{\qv}+i0} \nnr
  D_-^< &\equiv n_\qv (D_-^\adv-D_-^\ret), & D_+^> \equiv (1+n_\qv) (D_+^\adv-D_+^\ret).
\end{align}

 The spin part of the Green's function, Eq. (\ref{GlessI2}), is  
\begin{align}
 G_{{\rm N},\gamma} ^< & (\rv,t,\rv',t) 
=  
- \frac{SJ_{\rm I}^2}{2} \sumom \sumOm \sum_{\kv\kv'}\sum_{\kv''\qv}
 \biggl[ g^\ret_{{\rm N},\kv\omega}
 \lt(\widetilde{\cal D}_{\gamma}^\ret(\qv,\Omega) g^>_{{\rm N},\kv'',\omega-\Omega}
 +\widetilde{\cal D}_{\gamma}^<(\qv,\Omega) g^\ret_{{\rm N},\kv'',\omega-\Omega}\rt)
 g^<_{{\rm N},\kv'\omega} \nnr
&
+ g^\ret_{{\rm N},\kv\omega}  \widetilde{\cal D}_{\gamma}^\ret(\qv,\Omega) g^>_{{\rm N},\kv'',\omega-\Omega} g^\adv_{{\rm N},\kv'\omega} 
+ g^<_{{\rm N},\kv\omega} \lt(\widetilde{\cal D}_{\gamma}^\adv(\qv,\Omega) g^>_{{\rm N},\kv'',\omega-\Omega}
 +\widetilde{\cal D}_{\gamma}^<(\qv,\Omega) g^\adv_{{\rm N},\kv'',\omega-\Omega}\rt) g^\adv_{{\rm N},\kv'\omega}   \biggr]
.
 \label{GlessI3}
\end{align}
The contribution survives at long distance is the one containing $g^\ret_{{\rm N},\omega}(\rv)$ and $g^\adv_{{\rm N},\omega}(-\rv)$, i.e., 
\begin{align}
 G_{{\rm N},\gamma} ^< & (\rv,t,\rv',t) \simeq 
 \sumom \sum_{\kv\kv'} 
 g^\ret_{{\rm N},\kv\omega}  g^\adv_{{\rm N},\kv'\omega} e^{i\kv\cdot\rv} e^{-i\kv'\cdot\rv'} 
 \widetilde{\Sigma}_{{\rm I},\gamma},
 \end{align}
 where 
\begin{align}
\widetilde{\Sigma}_{{\rm I},\gamma}&\equiv   
- {\frac{SJ_{\rm I}^2}{2}}  \sumOm \sum_{\kv''\qv} 
 \biggl[
\lt(  f_{\kv'}\widetilde{\cal D}_{\gamma}^\ret(\qv,\Omega) -f_{\kv}\widetilde{\cal D}_{\gamma}^\adv(\qv,\Omega)  \rt)(f_{\kv''}-1)(  g^\adv_{{\rm N},\kv'',\omega-\Omega} -g^\ret_{{\rm N},\kv'',\omega-\Omega})
\nnr& 
 +\widetilde{\cal D}_{\gamma}^<(\qv,\Omega) 
 (f_{\kv'}g^\ret_{{\rm N},\kv'',\omega-\Omega}-f_{\kv}g^\adv_{{\rm N},\kv'',\omega-\Omega}
 +f_{\kv''}(g^\adv_{{\rm N},\kv'',\omega-\Omega}-g^\ret_{{\rm N},\kv'',\omega-\Omega})  )   \biggr]
.
 \label{GlessI4}
\end{align}
We focus on the pumped contribution, containing derivative with respect to $\Omega$ in Eq. (\ref{calDs}).
The result is, using partial integration with respect to $\Omega$ 
($\widetilde{\bm{\Sigma}}_{{\rm I}}$ is a vector representation of 
$\widetilde{\Sigma}_{{\rm I},\gamma}$), 
\begin{align}
\widetilde{\bm{\Sigma}}_{{\rm I}}
&\simeq 
-i {\frac{SJ_{\rm I}^2}{2}}  \sumOm \sum_{\kv''\qv} 
 \biggl[ \nnr
& \lt(  f_{\kv'}[\Psiv(D_-^\ret -D_+^\ret) +i\dot{\nv}(D_-^\ret +D_+^\ret)]
  -f_{\kv}[\Psiv(D_-^\adv -D_+^\adv) +i\dot{\nv}(D_-^\adv +D_+^\adv)] \rt)
  (f_{\kv''}-1)\frac{d}{d\Omega}(  g^\adv_{{\rm N},\kv'',\omega-\Omega} -g^\ret_{{\rm N},\kv'',\omega-\Omega})
\nnr& 
 +[\Psiv(D_-^< +D_+^>) +i\dot{\nv}(D_-^< -D_+^>)]
\frac{d}{d\Omega}[ (f_{\kv''}-f_{\kv})g^\adv_{{\rm N},\kv'',\omega-\Omega}
-(f_{\kv''}-f_{\kv'})g^\ret_{{\rm N},\kv'',\omega-\Omega}]
\biggr]
.
 \label{GlessI5}
\end{align}
Using $\frac{d}{d\Omega}g^\adv_{\kv'',\omega-\Omega}=(g^\adv_{\kv'',\omega})^2+O(\Omega)$ and  an  approximation, we obtain 
$\sum_{\kv''}(g^\adv_{\kv'',\omega})^2\simeq -\pi i\frac{\dos}{2\ef}$, 
\begin{align}
\widetilde{\bm{\Sigma}}_{{\rm I}}
&\simeq 
\frac{\pi\dos}{\ef} {\frac{SJ_{\rm I}^2}{2}}  \sumOm \sum_{\qv\kv''} 
 \biggl[\nnr
 & \Psiv \biggl( (f_{\kv''}-1)[ f_{\kv'}(D_-^\ret -D_+^\ret) -f_{\kv}(D_-^\adv -D_+^\adv) ]
 +\frac{1}{2}(2f_{\kv''}-f_{\kv}-f_{\kv'})(D_-^< +D_+^>)  \biggr) \nnr 
 &
 +i\dot{\nv} \biggl( 
 (f_{\kv''}-1)[ f_{\kv'}(D_-^\ret +D_+^\ret) -f_{\kv}(D_-^\adv +D_+^\adv) ]
 +\frac{1}{2}(2f_{\kv''}-f_{\kv}-f_{\kv'})(D_-^< -D_+^>)  \biggr) \biggr]
.
 \label{GlessI6}
\end{align}
As argued for Eq. (\ref{magnonsimple}), only the imaginary part of self energy contributes to the induced spin current, as the real part, the shift of the chemical potential, is compensated by redistribution of electrons.
The result is thus 
\begin{align}
\widetilde{\bm{\Sigma}}_{{\rm I}}
&\simeq 
i \Psiv \frac{\pi\dos}{\ef}{\frac{SJ_{\rm I}^2}{2}}  \sum_{\qv\kv''} 
 (1+2n_\qv) (2f_{\kv''}-f_{\kv}-f_{\kv'}).
 \label{GlessIres}
\end{align}
We further note that the component of $\Psiv$ proportional to $\nv$ (Eq. (\ref{Psivdef})) does not contribute to the current generation, as a result of gauge invariance. (In other words, the contribution cancels with the one arising from the effective gauge field for magnon.)

The final result of the spin current pumped by the magnon scattering is therefore 
\begin{align}
 \jv_{\rm s}^{{\rm m}}(\rv,t) &= 
 \frac{\pi\dos}{\ef} {\frac{SJ_{\rm I}^2}{2}}  |g^\ret(\rv)|^2 \sum_{\qv} 
 (1+2n_\qv) (\nv\times\dot{\nv}).
 \label{magnonjs}
\end{align}
At high temperature compared to magnon energy, $\beta \omega_q \ll1$,  
$ 1+2n_\qv\simeq \frac{2\kb T}{\omega_\qv}$, and the magnon-induced spin current depends linearly on temperature.
The result (\ref{magnonjs}) agrees with previous study carried out in the context of thermally-induced spin current \cite{Adachi11}.

\subsection{Correction to Gilbert damping in  the insulating case \label{SEC:alphaInsulator}}
In this subsection, we calculate the correction to the Gilbert damping and $g$-factor of insulating ferromagnet as a result of spin pumping effect. 
We study the torque on the ferromagnetic magnetization arising from the effect of conduction electron of normal metal, given by
\begin{align}
 \tauv_{\rm I} &= \Bv_{\rm I}\times \nv = \spol_{\rm I}(\nv\times \sev_{\rm I}) ,
\end{align}
where 
\begin{align}
 \Bv_{\rm I} &\equiv -\frac{\delta H_{\rm I}}{\delta \nv}=-\spol_{\rm I} \sev_{\rm I},
\end{align}
is the effective magnetic field arising from the interface electron spin polarization, 
$\sev_{\rm I}(t)\equiv -i \tr[\sigmav G_{\rm N}^<(0,t)]$.
The contribution to the electron spin density linear in the interface exchange interaction, Eq. (\ref{HIdef}),  is
\begin{align}
\se_{\rm I}^{(1),\alpha}(t) &= -i \int dt_1 \spol_{\rm I} n_\beta(t_1) \tr[\sigma_\alpha g_{\rm N}(t,t_1) \sigma_\beta g_{\rm N}(t_1,t)]^< ,
\end{align}
where the Green's functions connect positions at the interface, i.e., from $x=0$ to $x=0$, and are  spin unpolarized.
(The Feynman diagrams for the spin density are the same as the one for the spin current, Fig. \ref{FIGInsulatorSC} with the vertex $\js$ replaced by the Pauli matrix.)  
Pumped contribution proportional to the time variation of magnetization is obtained as
\begin{align}
\sev_{\rm I}^{(1)}(t) &= - \spol_{\rm I} \dot{\nv} \sumom\sum_{\kv\kv'} f'(\omega)
 (\ga_{{\rm N},\kv'}-\gr_{{\rm N},\kv'}) (\ga_{{\rm N},\kv}-\gr_{{\rm N},\kv}) \nnr
 &= - \spol_{\rm I} (\pi\dos)^2 \dot{\nv} .
\end{align}
The second order contribution similarly reads
\begin{align}
\se_{\rm I}^{(2),\alpha}(t) &= -\frac{i}{2}  \int dt_1\int dt_2 (\spol_{\rm I})^2 n_\beta(t_1) n_\gamma(t_2) \tr[\sigma_\alpha g_{\rm N}(t,t_1) \sigma_\beta g_{\rm N}(t_1,t_2) \sigma_\gamma g_{\rm N}(t_2,t)]^<  \nnr
& \simeq -2( \spol_{\rm I})^2 (\pi\dos)^3 (\nv\times \dot{\nv}) .
\end{align}
The interface torque is therefore 
\begin{align}
\torquev_{\rm I} &= -( \spol_{\rm I}\pi\dos)^2 (\nv\times\dot{\nv}) +2( \spol_{\rm I}\pi\dos)^3  \dot{\nv} .
\end{align}
Including this torque in the LLG equation, 
$\dot{\nv}=-\alpha \nv\times\dot{\nv} - \gamma\Bv\times\nv+\torquev$, 
we have 
\begin{align}
(1-\delta_{\rm I}) \dot{\nv} =-\alpha_{\rm I}(\nv\times\dot{\nv}) - \gamma\Bv\times\nv,
\end{align}
where 
\begin{align}
\delta_{\rm I} &= 2\mu_d(\pi\spol_{\rm I}\dos)^3 \nnr       
\alpha_{\rm I} &= \alpha+\mu_d(\pi\spol_{\rm I}\dos)^2,       \label{alphaIresult}
\end{align}
where $\mu_d\sim d_{\rm mp}/d$ is the ratio of the length of magnetic proximity ($d_{\rm mp}$) and thickness of the ferromagnet, $d$.
The Gilbert damping constant therefore increases as far as the interface spin-orbit interaction is neglected.
The resonance frequency is 
$\omega_B=\frac{\gamma B}{1-\delta_{\rm I}}$, and the shift can have both signs depending on the sign of interface exchange interaction, $\spol_{\rm I}$.

There may be a possibility that magnon excitation induce torque that corresponds to effective damping. In fact, such torque arises of $\average{b}$ or $\average{b^\dagger}$ are finite, i.e., if magnon Bose condensation glows, as seen from Eq. (\ref{HPboson}). 
Such condensation can in principle develop from the interface interaction of  magnon creation or annihilation induced by electron spin flip, Eq. (\ref{HIboson}). 
However, conventional spin relaxation processes arising from the second order of random spin scattering do not contribute to such magnon condensation and  additional damping.

Comparing the result of pumped spin current, Eq. (\ref{totaljsinsulator}), and that of damping coefficient, Eq. (\ref{alphaIresult}), we notice that the 'spin mixing conductance' argument \cite{Tserkovnyak02}, where the coefficients for the spin current component proportional to $\nv\times\dot{\nv}$ and the enhancement of the Gilbert damping constant are governed by the same quantity (the imaginary part of 'spin mixing conductance') does not hold for the insulator case.
In fact, our result indicates that the spin current component proportional to $\nv\times\dot{\nv}$ arises from the second order correction to the interaction (the second diagram of Fig. \ref{FIGInsulatorSC}), while the damping correction arises from the first order process  (the first diagram of Fig. \ref{FIGInsulatorSC}).
Although the magnitudes of the two effects happen to be both second order of interface spin splitting, $\spol_{\rm I}$,  physical origins appear to be distinct.
From our analysis, we see that the 'spin mixing conductance' description is not general and applies only to the case of thick metallic ferromagnet (see Sec. \ref{SECmetaldamping} for metallic case).

\section{Discussion\label{SECdiscussion}}

\begin{table}[ht]
\begin{center}
\begin{tabular}{c|cccccc}
 Ferromagnet(F)  & $A_{\rm i}$ & $A_{\rm r}$ & $\delta \alpha$ & $\delta \omega_B$ & Assumption & Equations   \\ \hline
Metal & $\Im \overline{T_{+-}}$ & $\Re \overline{T_{+-}}$ &  
  \begin{tabular}{c} 
    $\Re \overline{T_{+-}}$ \\    $\Im \overline{T_{+-}}$ \end{tabular}
    &
  \begin{tabular}{c} 
    $\Im \overline{T_{+-}}$ \\    $\Re \overline{T_{+-}}$ \end{tabular}
    &
  \begin{tabular}{c} 
    Thick F \\   Thin F \end{tabular}
    &
  \begin{tabular}{c} 
     (\ref{jsresultQM})(\ref{spincurrentmetalFT}) (\ref{alphaenhance})(\ref{resonancefreq})\\    (\ref{thin})\end{tabular}
\\
Insulator & $\spol_{\rm I}\dos $ & $(\spol_{\rm I}\dos)^2 $ & $(\spol_{\rm I}\dos)^2 $ & $(\spol_{\rm I}\dos)^3 $ & Weak spin relaxation $^*$ &   (\ref{totaljsinsulator}) (\ref{alphaIresult})  \\
 & - & $(\spol_{\rm I}\dos)^2\sumqv(1+2n_\qv)$ & - & - &   Magnon &  (\ref{magnonjs}) \\
\hline 
 \end{tabular}
\end{center}
\caption{ Summary of essential parameters determining spin current $\js$, corrections to the Gilbert damping $\delta \alpha$ and resonance frequency shift $\delta \omega_B$ for metallic and insulating ferromagnets.
Coefficients $A_{\rm i}$ and $A_{\rm r}$  are for the spin current, defined by Eq. (\ref{Jsphenom}).
Label $-$ indicates that it is not discussed in the present paper. 
$^*$ : For strong spin relaxation case, the density of states $\dos$ is replaced by inverse of electron spin-flip time, $\tau_{\rm sf}$. \cite{Kajiwara10} 
\label{TABLEresult}}
\end{table}
Our results are summarized in table \ref{TABLEresult}.
Let us discuss experimental results in the light of our results.
In the early ferromagnetic resonance (FMR) experiments, consistent studies of $g$-factor and the Gilbert damping were carried out on metallic ferromagnets \cite{Mizukami01}.
The results appear to be consistent with theories (Refs. \cite{Berger96,Tserkovnyak02} and the present paper).
Both the damping constant and the $g$ factor have $1/d$-dependence on the thickness of ferromagnet in the range of 2nm$<d<$10nm \cite{Mizukami01}.
The maximum additional damping reaches $\delta \alpha\sim 0.1$ at $d=$2nm, which exceeds the original value of $\alpha\sim 0.01$. 
The $g$-factor modulation is about 1\% at $d=2$nm, and its sign depends on the material; 
the $g$-factor increases for Pd/Py/Pd and Pt/Py/Pt while decreases for Ta/Py/Ta.
These results appear consistent with ours, because $\delta \omega_B$ is governed by $\Im \overline{T_{+-}}$, whose sign depends on the sign of interface spin-orbit interaction.
In contrast, damping enhancement proportional to  $\Re \overline{T_{+-}}$ is positive for thick metals.
However, other possibilities like the effect of a large interface orbital moment playing a role in the $g$ factor, cannot be ruled out at present.

Recently, inverse spin Hall measurement has become common for detecting the spin current. In this method, however, only the dc component proportional to $\nv\times\dot{\nv}$ is accessible so far and there remains an ambiguity for  qualitative estimates because another phenomenological parameter, the conversion efficiency from spin to charge, enters. 
Qualitatively, the values of $A_{\rm r}$ obtained by the inverse spin Hall measurements \cite{Czeschka11} and FMR measurements are  consistent with each other.

The cases of insulating ferromagnets have been studied recently. 
In the early experiments, orders of magnitude smaller value of $A_{\rm r}$ compared to metallic cases were reported \cite{Kajiwara10}, while those small values are now understood as due to poor interface quality.
In fact, FMR measurements on epitaxially grown samples like YIG/Au/Fe turned out to show $A_{\rm r}$ of $ 1\sim 5\times 10^{18}$m$^{-2}$ (Refs.  \cite{Heinrich11,Burrowes12}), which is the same order as in the metallic cases. 
Inverse spin Hall measurements on YIG/Pt reports similar values \cite{Qiu13}, and the value is consistent with the first principles calculation \cite{Jia11}.
Systematic studies of YIG/NM with NM=Pt, Ta, W, Au, Ag, Cu, Ti, V, Cr, Mn etc. were carried out with the result of 
$ A_{\rm r}\sim 10^{17}-10^{18}$m$^{-2}$ (Refs. \cite{WangPRL14,Du14,DuPRB14,Wang17}).
If we use naive phenomenological relation, Eq. (\ref{alphag}), $A_{\rm r}=10^{18}$m$^{-2}$ corresponds to $\delta \alpha =3\times 10^{-4}$ if $a=2\AA$, $S=1$ and $d=20\AA$. 
Assuming interface $sd$ exchange interaction, the value indicates $M_{\rm I}\dos\sim 0.01$, which appears reasonable at least by the order of magnitude from the result of x-ray magnetic circular dichroism (XMCD) suggesting spin polarization of interface  Pt of $0.05\mub$ \cite{Lu13}.

On the other hand, FMR frequency shift of insulators cannot be explained by our theory.
In fact, the shift for YIG/Pt is $\delta \omega_B/\omega_B\sim 1.6\times 10^{-2}$, which is larger than $\delta \alpha\sim 2\times 10^{-3}$, while our perturbation theory assuming weak interface $sd$ interaction  predicts  $\delta \omega_B/\omega_B  < \delta\alpha$.
We expect that the discrepancy arises from the interface spin-orbit interaction that would be present at insulator-metal interface, which modifies the magnetic proximity effect and damping torque significantly.
It would be necessary to introduce anomalous $sd$ coupling at the interface like the one  discussed in Ref. \cite{Xia97}.
Experimentally, influence of interface spin-orbit interaction \cite{Caminale16} and proximity effect 
needs to be carefully characterized  by using the microscopic technique,
such as MCD, to compare with theories.

\section{Summary \label{SECsummary}}
We have presented a microscopic study of spin pumping effects, generation of spin current in ferromagnet-normal metal junction by magnetization dynamics, for both metallic and insulating ferromagnets.  
As for the case of metallic ferromagnet, a simple quantum mechanical picture was developed using a unitary transformation to diagonal the time-dependent $sd$ exchange interaction.
The problem of dynamic magnetization is thereby mapped to the one with static magnetization and off-diagonal spin gauge field, which mixes the electron spin.
In the slowly-varying limit, spin gauge field becomes static, and the conventional spin pumping formula is derived simply by evaluating the spin accumulation formed in the normal metal as a result of interface hopping. 
The effect of interface spin-orbit interaction was discussed.
Rigorous field-theoretical derivation was also presented, and the enhancement of spin damping (Gilbert damping) in the ferromagnet as a result of spin pumping effect was discussed.
The case of insulating ferromagnet was  studied based on a model where spin current is driven locally by the interface exchange interaction as a result of magnetic proximity effect.
The dominant contribution turns out to be the one proportional to $\dot{\nv}$, while magnon contribution leads to $\nv\times\dot{\nv}$, whose amplitude depends linearly on the temperature.
Our analysis clearly demonstrate the difference in the spin current generation mechanism  for metallic and insulating ferromagnet.

The influence of atomic-scale interface structure on the spin pumping effect  
are open and urgent issues, in particular for the case of ferrimagnetic 
insulators which have two sub-lattice magnetic moments.


\acknowledgements
GT thanks 
H. Kohno, C. Uchiyama, K. Hashimoto and A. Shitade for valuable discussions.
This work was supported by 
a Grant-in-Aid for Exploratory Research (No.16K13853) 
and 
a Grant-in-Aid for Scientific Research (B) (No. 17H02929) from the Japan Society for the Promotion of Science 
and  
a Grant-in-Aid for Scientific Research on Innovative Areas (No.26103006) from The Ministry of Education, Culture, Sports, Science and Technology (MEXT), Japan.

\appendix
\section{Effect of spin-conserving spin gauge field on spin density \label{SEC:nodamping}}
Here we calculate contribution of spin-conserving spin gauge field, $\Ascal{t}{z}$,  on the interface effects of spin density in F.
It turns out that  spin-conserving spin gauge field combined with interface effects does not induce damping.
This result is consistent with a naive expectation that only the nonadiabatic components of spin current should contribute to damping.

\begin{figure}[tbh]
  \begin{center}
  \includegraphics[width=0.2\hsize]{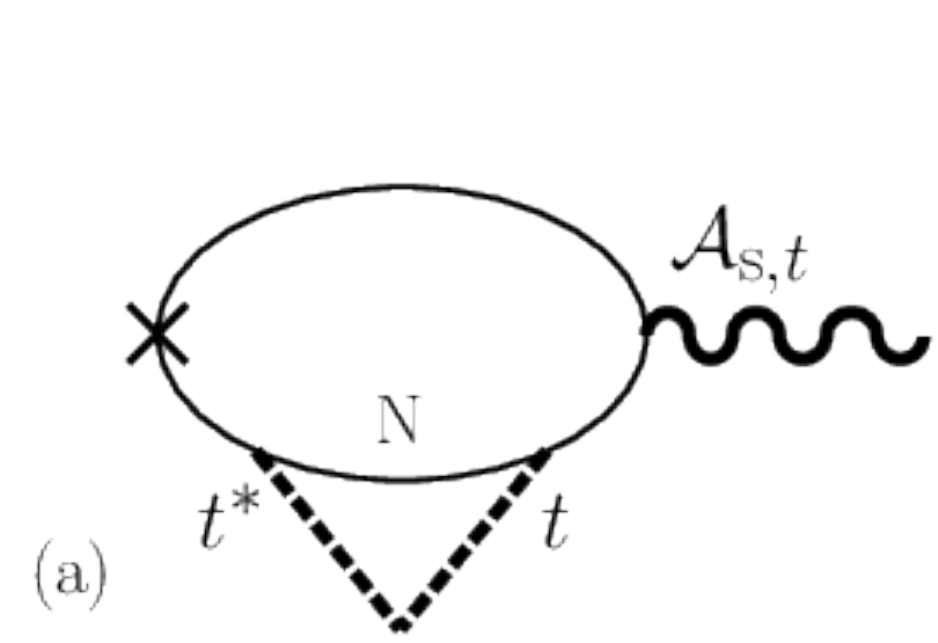}
  \includegraphics[width=0.2\hsize]{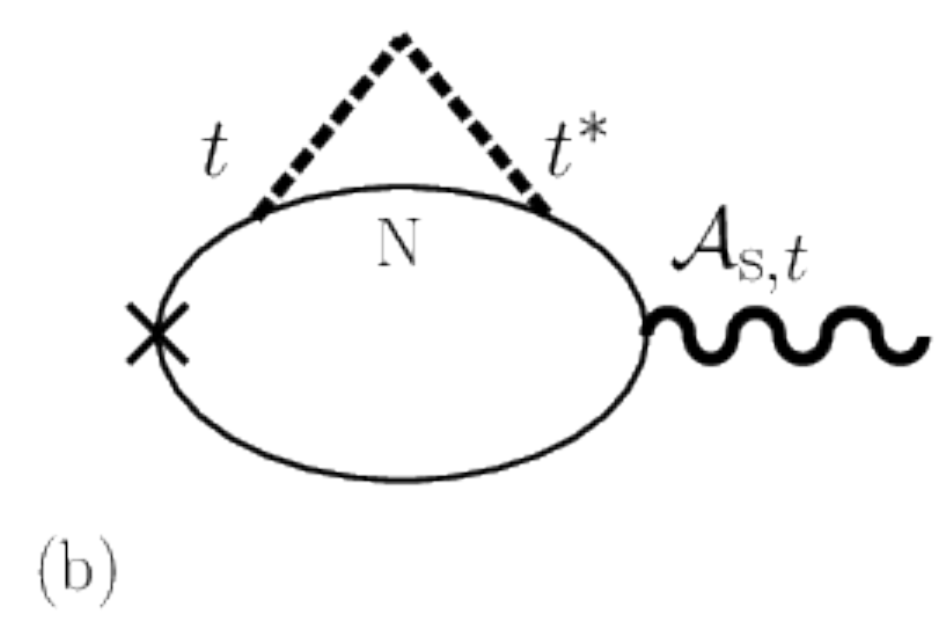}
  \includegraphics[width=0.2\hsize]{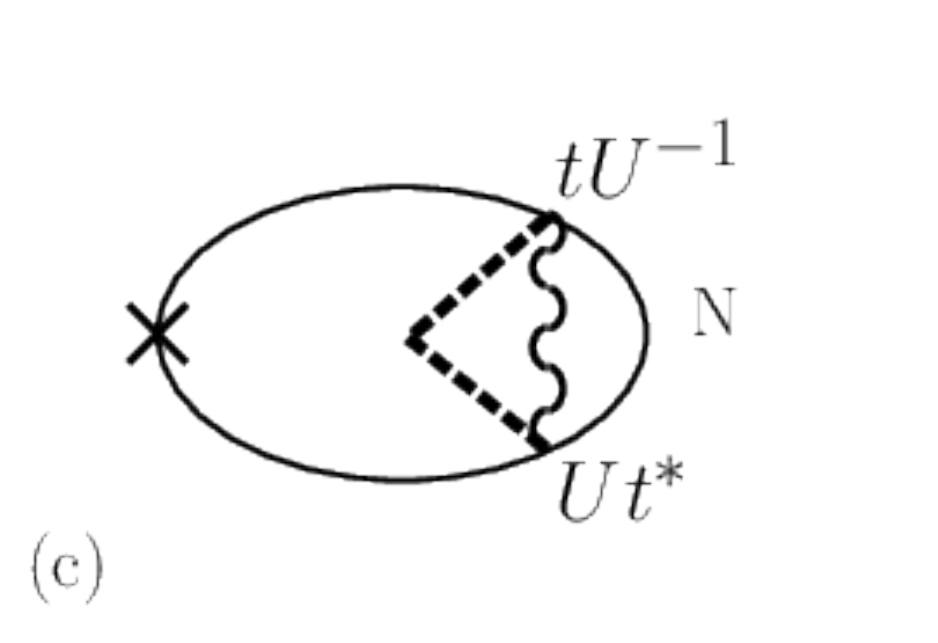}
  \end{center}
\caption{ Diagramatic representation of the contribution to the lessor Green's function for F electron arising from the interface hopping (represented by $t$ and $t^*$) and spin gauge field ($\Ascal{t}{}$).
The diagram (c) includes the spin gauge field implicitly in unitary matrices $U$ and $U^{-1}$. 
\label{FIGGFF}}
\end{figure}
%
The contribution to the lesser Green's function in F  from the interface hopping (lowest, the second-order in the hopping) at the  linear order in the spin gauge field reads (diagramatically shown in Fig. \ref{FIGGFF}) 
\begin{align}
 \delta G^< &=  \delta G^{<}_{\rm (a)} +\delta G^{<}_{\rm (b)}+\delta G^{<}_{\rm (c)} \nnr
\delta G^{<}_{\rm (a)}
 &=
    \gr  (\Ascalv{t}\cdot\sigmav) \gr \Sigma_0^\ret   g^< 
 + \gr  (\Ascalv{t}\cdot\sigmav) \gr \Sigma_0^<  g^\adv 
 + \gr  (\Ascalv{t}\cdot\sigmav) \gless \Sigma_0^\adv  g^\adv 
 + \gless  (\Ascalv{t}\cdot\sigmav) \ga \Sigma_0^\adv  g^\adv  \nnr
\delta G^{<}_{\rm (b)}
 &=
    \gr  \Sigma_0^\ret \gr (\Ascalv{t}\cdot\sigmav)   g^< 
 + \gr  \Sigma_0^<   \gless (\Ascalv{t}\cdot\sigmav) g^\adv 
 + \gr   \Sigma_0^\adv \ga (\Ascalv{t}\cdot\sigmav) g^\adv 
 + \gless \Sigma_0^\adv  \ga  (\Ascalv{t}\cdot\sigmav) g^\adv  \nnr
 \delta G^{<}_{\rm (c)}
 &=
    \gr  \Sigma^\ret    g^< 
 + \gr  \Sigma^<   g^\adv 
 + \gless \Sigma^\adv  \ga .
 \end{align}
Here 
\begin{align}
  \Sigma^{a} & \equiv \ttil U^{-1} g^{a}_{\rm N} U \ttil^\dagger ,\;\;\;  (a=\adv,\ret,<) \nnr
  \Sigma^{a}_0 & \equiv \ttil  g^{a}_{\rm N} ,
\end{align}
are self energy due to the interface hopping, where $ \Sigma^{a} $ is the full self energy including the time-dependent unitary matrix $U$, which includes spin gauge field. 
$\Sigma^{a}_0$ is the contribution of $\Sigma^{a}$ with the spin gauge field neglected. 
We here focus on the contribution of the adiabatic ($z$) component, $\Ascal{t}{z}$.
Using $g^<=F(\ga-\gr)$ for F ($F$ is a $2\times2$ matrix of the spin-polarized  Fermi distribution function) and $g^<_{\rm N}=f_{\rm N}(\ga_{\rm N}-\gr_{\rm N})$
and noting that all the angular frequencies of the Green's function are equal, we obtain 
\begin{align}
 \delta G^{<}_{\rm (a)}+
 \delta G^{<}_{\rm (b)}
 & \simeq 
 \Ascal{t}{z}\sigma_z\bigl[ 
   -2F [(\gr)^3\Sigma_0^\ret - (\ga)^3\Sigma_0^\adv]
   -(F-f_{\rm N}) [(\gr)^2\ga+\gr(\ga)^2](\Sigma_0^\adv-\Sigma_0^\ret) \biggr].
\end{align}
The contribution  $\delta G^{<}_{\rm (c)}$ is calculated noting that 
\begin{align}
\ttil U^{-1} g^{a}_{\rm N} U \ttil^\dagger & = g^{a}_{\rm N} \ttil \ttil^\dagger  - \frac{dg^{a}_{\rm N}}{d\omega}  \ttil   (\Ascalv{t}\cdot\sigmav) \ttil^\dagger +O( (\Ascalv{t}) ^2).
\end{align}
The linear contribution with respect to the $z$ component of  the  gauge field turns out to be  
\begin{align}
 \delta G^{<}_{\rm (c)}
 & \simeq 
 \Ascal{t}{z}\sigma_z\biggl[ 
   F \lt[(\gr)^2\delpo{\omega}\Sigma_0^\ret - (\ga)^2\delpo{\omega}\Sigma_0^\adv\rt]
   +(F-f_{\rm N}) \gr\ga\delpo{\omega}(\Sigma_0^\adv-\Sigma_0^\ret) \biggr].
\end{align}
We therefore obtain the  effect of spin-conserving  gauge field as 
\begin{align}
 \delta G^{<}
 & =
 \Ascal{t}{z}\sigma_z\delpo{\omega}\biggl[ 
   F \lt[(\gr)^2\Sigma_0^\ret - (\ga)^2\Sigma_0^\adv\rt]
   +(F-f_{\rm N}) \gr\ga(\Sigma_0^\adv-\Sigma_0^\ret) \biggr],
\end{align}
which vanishes after integration over $\omega$.
Therefore, contribution from spin-conserving  gauge field and interface hopping vanishes in the spin density, leaving the damping unaffected.

\section{Magnon representation of spin Berry's phase term \label{SEC:rotatedmagnon}}
Here we derive the expression for the spin Berry's phase term of the Lagrangian (\ref{LIF}) in terms of magnon operator.
The time-integral of the term is written by introducing an artificial variable $u$ as \cite{Auerbach94}
\begin{align}
 \int dt L_{\rm B} &= S \int dt  \dot{\phi}(\cos\theta-1) 
 = S^{-2} \int dt \int_0^1du\Sv\cdot(\partial_t\Sv\times\partial_u \Sv),
\end{align}
where $\Sv(t,u)$ is extended to a function of $t$ and $u$, but only $\Sv(t,u=1)$ is physical.
Noting that the unitary transformation matrix element of Eq. (\ref{Unitary3x3}) is written as
\begin{align}
U_{ij}=(\ev_j )_i,  
\end{align}
where $\rv_1\equiv\evth$, $\ev_2\equiv \evph$ and $\ev_3\equiv \nv$, we obtain 
\begin{align}
 \Sv\cdot(\partial_t\Sv\times\partial_u \Sv)
 &= \widetilde{\Sv}\cdot[(\partial_t+i A_{U,t})\widetilde{\Sv}\times(\partial_u+i A_{U,u})\widetilde{\Sv})].
\end{align}
Evaluating to the second order in the magnon operators, we have 
\begin{align}
 \partial_t\widetilde{\Sv}\times \partial_u\widetilde{\Sv}
 &= 2i\gamma \zvhat [(\partial_u\boson^\dagger)(\partial_t\boson)-(\partial_t\boson^\dagger)(\partial_u\boson) ].
\end{align}
Using the explicit form of $A_{U,\mu}$, the gauge field contribution is
\begin{align}
 \partial_u\widetilde{\Sv}\cdot[\widetilde{\Sv}\times i A_{U,t}\widetilde{\Sv})]
 &=S^2\gamma [ (\partial_u \boson^\dagger)(-\sin\theta \dot{\phi}+i\dot{\theta}) 
 + (\partial_u \boson)(-\sin\theta \dot{\phi}-i\dot{\theta}) ]
 -2S\gamma^2
 \cos\theta (\partial_t{\phi}) \partial_u(\boson^\dagger \boson) .
\end{align}
The terms linear in the boson operators vanish by the equation of motion, and the second-order contribution is
\begin{align}
 \Sv\cdot(\partial_t\Sv\times\partial_u \Sv)
 &= 2S\gamma^2 \biggl[i \partial_u[\boson^\dagger (\partial_t\boson)-(\partial_t\boson^\dagger) \boson] \nnr
 & 
 -\partial_u[\cos\theta (\partial_t \phi)\boson^\dagger \boson ] 
 +\partial_t[\cos\theta (\partial_u \phi)\boson^\dagger \boson ] 
 +\sin\theta ((\partial_t \theta)(\partial_u \phi)-(\partial_u \theta)(\partial_t \phi))\boson^\dagger \boson
 \biggr].
 \label{solidangleboson}
\end{align}
Integrating over $t$ and $u$, the term total derivative with respect to $t$ of Eq. (\ref{solidangleboson}) vanishes, resulting in
\begin{align}
 \int dt\int_0^{1}du \Sv\cdot(\partial_t\Sv\times\partial_u \Sv)
 &=  2S\gamma^2\int dt \biggl[ 
  i [\boson^\dagger (\partial_t\boson)-(\partial_t\boson^\dagger) \boson] 
 -\cos\theta (\partial_t \phi)\boson^\dagger \boson \nnr
 & 
 +\sin\theta ((\partial_t \theta)(\partial_u \phi)-(\partial_u \theta)(\partial_t \phi))\boson^\dagger \boson
 \biggr].
 \label{solidanglebosonres}
\end{align}
The last term of Eq. (\ref{solidanglebosonres}) represents the renormalization of spin Berry's phase term, i.e., the effect $S\ra S-\boson^\dagger\boson$, which we neglect below.
The Lagrangian for magnon thus reads
\begin{align}
 L_{\rm m} 
 &=  2S\gamma^2\intr
  i [\boson^\dagger (\partial_t+i\Ascal{t}{z})\boson-\boson^\dagger(\stackrel{\leftarrow}{\partial}_t -i\Ascal{t}{z})) \boson] ,
 \end{align}
namely,  magnons interacts with the adiabatic component of spin gauge field, $\Ascal{t}{z}$.

\section{ Decomposition of contour-ordered self energy
\label{SEC:SEdecomp}}
Here we summarize decomposition formula of self energy.
Obviously, we have 
\begin{align}
[g{\cal D}]^<
&= g^< {\cal D}^<.
\end{align}
Retarded component is defined as 
\begin{align}
[g{\cal D}]^\ret 
&\equiv 
[g{\cal D}]^{\rm t} -
[g{\cal D}]^<,
\end{align}
where the time-ordered one is 
\begin{align}
[g(t_1-t_2){\cal D}(t_1-t_2)]^{\rm t}
&\equiv 
\theta(t_1-t_2)  g^> {\cal D}^> +\theta(t_2-t_1)  g^< {\cal D}^< \nonumber\\
&=g^\ret {\cal D}^\ret +g^\ret {\cal D}^< + g^< {\cal D}^\ret  +g^< {\cal D}^<.
\end{align}
We thus obtain 
\begin{align}
[g{\cal D}]^\ret 
& = g^\ret {\cal D}^\ret +g^\ret {\cal D}^< + g^< {\cal D}^\ret.
\end{align}
Noting that $g^\ret {\cal D}^\adv=0$, we can write it as 
\begin{align}
[g{\cal D}]^\ret &=g^\ret {\cal D}^<+g^> {\cal D}^\ret =  g^< {\cal D}^\ret+g^\ret {\cal D}^> 
.
\end{align}
The advanced component is similarly written as 
\begin{align}
[g {\cal D} ]^{\adv}
&=  - g^\adv {\cal D}^\adv +g^\adv {\cal D}^< + g^< {\cal D}^\adv
\nonumber\\ 
&= g^\adv {\cal D}^>+g^< {\cal D}^\adv
= g^\adv {\cal D}^<+g^> {\cal D}^\adv .
\end{align}

\input{spinpump.bbl}
\end{document}

%% file: spinpump.bbl
%

%% file: spinpump_q.bbl
\begin{thebibliography}{44}%
\makeatletter
\providecommand \@ifxundefined [1]{%
 \@ifx{#1\undefined}
}%
\providecommand \@ifnum [1]{%
 \ifnum #1\expandafter \@firstoftwo
 \else \expandafter \@secondoftwo
 \fi
}%
\providecommand \@ifx [1]{%
 \ifx #1\expandafter \@firstoftwo
 \else \expandafter \@secondoftwo
 \fi
}%
\providecommand \natexlab [1]{#1}%
\providecommand \enquote  [1]{``#1''}%
\providecommand \bibnamefont  [1]{#1}%
\providecommand \bibfnamefont [1]{#1}%
\providecommand \citenamefont [1]{#1}%
\providecommand \href@noop [0]{\@secondoftwo}%
\providecommand \href [0]{\begingroup \@sanitize@url \@href}%
\providecommand \@href[1]{\@@startlink{#1}\@@href}%
\providecommand \@@href[1]{\endgroup#1\@@endlink}%
\providecommand \@sanitize@url [0]{\catcode `\\12\catcode `\$12\catcode
  `\&12\catcode `\#12\catcode `\^12\catcode `\_12\catcode `\%12\relax}%
\providecommand \@@startlink[1]{}%
\providecommand \@@endlink[0]{}%
\providecommand \url  [0]{\begingroup\@sanitize@url \@url }%
\providecommand \@url [1]{\endgroup\@href {#1}{\urlprefix }}%
\providecommand \urlprefix  [0]{URL }%
\providecommand \Eprint [0]{\href }%
\providecommand \doibase [0]{http://dx.doi.org/}%
\providecommand \selectlanguage [0]{\@gobble}%
\providecommand \bibinfo  [0]{\@secondoftwo}%
\providecommand \bibfield  [0]{\@secondoftwo}%
\providecommand \translation [1]{[#1]}%
\providecommand \BibitemOpen [0]{}%
\providecommand \bibitemStop [0]{}%
\providecommand \bibitemNoStop [0]{.\EOS\space}%
\providecommand \EOS [0]{\spacefactor3000\relax}%
\providecommand \BibitemShut  [1]{\csname bibitem#1\endcsname}%
\let\auto@bib@innerbib\@empty
\bibitem [{\citenamefont {Saitoh}\ \emph {et~al.}(2006)\citenamefont {Saitoh},
  \citenamefont {Ueda}, \citenamefont {Miyajima},\ and\ \citenamefont
  {Tatara}}]{Saitoh06}%
  \BibitemOpen
  \bibfield  {author} {\bibinfo {author} {\bibfnamefont {E.}~\bibnamefont
  {Saitoh}}, \bibinfo {author} {\bibfnamefont {M.}~\bibnamefont {Ueda}},
  \bibinfo {author} {\bibfnamefont {H.}~\bibnamefont {Miyajima}}, \ and\
  \bibinfo {author} {\bibfnamefont {G.}~\bibnamefont {Tatara}},\ }\href@noop {}
  {\bibfield  {journal} {\bibinfo  {journal} {Appl. Phys. Lett.}\ }\textbf
  {\bibinfo {volume} {88}},\ \bibinfo {pages} {182509} (\bibinfo {year}
  {2006})}\BibitemShut {NoStop}%
\bibitem [{\citenamefont {Tserkovnyak}\ \emph {et~al.}(2002)\citenamefont
  {Tserkovnyak}, \citenamefont {Brataas},\ and\ \citenamefont
  {Bauer}}]{Tserkovnyak02}%
  \BibitemOpen
  \bibfield  {author} {\bibinfo {author} {\bibfnamefont {Y.}~\bibnamefont
  {Tserkovnyak}}, \bibinfo {author} {\bibfnamefont {A.}~\bibnamefont
  {Brataas}}, \ and\ \bibinfo {author} {\bibfnamefont {G.~E.~W.}\ \bibnamefont
  {Bauer}},\ }\href {http://link.aps.org/abstract/PRL/v88/e117601} {\bibfield
  {journal} {\bibinfo  {journal} {Phys. Rev. Lett.}\ }\textbf {\bibinfo
  {volume} {88}},\ \bibinfo {eid} {117601} (\bibinfo {year}
  {2002})}\BibitemShut {NoStop}%
\bibitem [{\citenamefont {Moskalets}(2012)}]{Moskalets12}%
  \BibitemOpen
  \bibfield  {author} {\bibinfo {author} {\bibfnamefont {M.~V.}\ \bibnamefont
  {Moskalets}},\ }\href@noop {} {\emph {\bibinfo {title} {Scattering matrix
  approach to non-stationary quantum transport}}}\ (\bibinfo  {publisher}
  {Imperial College Press},\ \bibinfo {year} {2012})\BibitemShut {NoStop}%
\bibitem [{\citenamefont {Büttiker}\ \emph {et~al.}(1994)\citenamefont
  {Büttiker}, \citenamefont {Thomas},\ and\ \citenamefont
  {Prêtre}}]{Buttiker94}%
  \BibitemOpen
  \bibfield  {author} {\bibinfo {author} {\bibfnamefont {M.}~\bibnamefont
  {B\"{u}ttiker}}, \bibinfo {author} {\bibfnamefont {H.}~\bibnamefont {Thomas}}, \
  and\ \bibinfo {author} {\bibfnamefont {A.}~\bibnamefont {Pretre}},\ }\href
  {\doibase 10.1007/BF01307664} {\bibfield  {journal} {\bibinfo  {journal}
  {Zeitschrift f\"{u}r Physik B Condensed Matter}\ }\textbf {\bibinfo {volume}
  {94}},\ \bibinfo {pages} {133} (\bibinfo {year} {1994})}\BibitemShut
  {NoStop}%
\bibitem [{\citenamefont {Brouwer}(1998)}]{Brouwer98}%
  \BibitemOpen
  \bibfield  {author} {\bibinfo {author} {\bibfnamefont {P.~W.}\ \bibnamefont
  {Brouwer}},\ }\href {\doibase 10.1103/PhysRevB.58.R10135} {\bibfield
  {journal} {\bibinfo  {journal} {Phys. Rev. B}\ }\textbf {\bibinfo {volume}
  {58}},\ \bibinfo {pages} {R10135} (\bibinfo {year} {1998})}\BibitemShut
  {NoStop}%
\bibitem [{\citenamefont {Tatara}\ and\ \citenamefont {Kohno}(2003)}]{TK03}%
  \BibitemOpen
  \bibfield  {author} {\bibinfo {author} {\bibfnamefont {G.}~\bibnamefont
  {Tatara}}\ and\ \bibinfo {author} {\bibfnamefont {H.}~\bibnamefont {Kohno}},\
  }\href {http://link.aps.org/abstract/PRB/v67/e113316} {\bibfield  {journal}
  {\bibinfo  {journal} {Phys. Rev. B}\ }\textbf {\bibinfo {volume} {67}},\
  \bibinfo {eid} {113316} (\bibinfo {year} {2003})}\BibitemShut {NoStop}%
\bibitem [{\citenamefont {Tatara}\ \emph {et~al.}(2008)\citenamefont {Tatara},
  \citenamefont {Kohno},\ and\ \citenamefont {Shibata}}]{TKS_PR08}%
  \BibitemOpen
  \bibfield  {author} {\bibinfo {author} {\bibfnamefont {G.}~\bibnamefont
  {Tatara}}, \bibinfo {author} {\bibfnamefont {H.}~\bibnamefont {Kohno}}, \
  and\ \bibinfo {author} {\bibfnamefont {J.}~\bibnamefont {Shibata}},\ }\href
  {\doibase doi:10.1016/j.physrep.2008.07.003} {\bibfield  {journal} {\bibinfo
  {journal} {Physics Reports}\ }\textbf {\bibinfo {volume} {468}},\ \bibinfo
  {pages} {213} (\bibinfo {year} {2008})}\BibitemShut {NoStop}%
\bibitem [{\citenamefont {Dugaev}\ \emph {et~al.}(2005)\citenamefont {Dugaev},
  \citenamefont {Bruno}, \citenamefont {Canals},\ and\ \citenamefont
  {Lacroix}}]{Dugaev05}%
  \BibitemOpen
  \bibfield  {author} {\bibinfo {author} {\bibfnamefont {V.~K.}\ \bibnamefont
  {Dugaev}}, \bibinfo {author} {\bibfnamefont {P.}~\bibnamefont {Bruno}},
  \bibinfo {author} {\bibfnamefont {B.}~\bibnamefont {Canals}}, \ and\ \bibinfo
  {author} {\bibfnamefont {C.}~\bibnamefont {Lacroix}},\ }\href {\doibase
  10.1103/PhysRevB.72.024456} {\bibfield  {journal} {\bibinfo  {journal} {Phys.
  Rev. B}\ }\textbf {\bibinfo {volume} {72}},\ \bibinfo {eid} {024456}
  (\bibinfo {year} {2005})}\BibitemShut {NoStop}%
\bibitem [{\citenamefont {Silsbee}\ \emph {et~al.}(1979)\citenamefont
  {Silsbee}, \citenamefont {Janossy},\ and\ \citenamefont {Monod}}]{Silsbee79}%
  \BibitemOpen
  \bibfield  {author} {\bibinfo {author} {\bibfnamefont {R.~H.}\ \bibnamefont
  {Silsbee}}, \bibinfo {author} {\bibfnamefont {A.}~\bibnamefont {Janossy}}, \
  and\ \bibinfo {author} {\bibfnamefont {P.}~\bibnamefont {Monod}},\ }\href
  {\doibase 10.1103/PhysRevB.19.4382} {\bibfield  {journal} {\bibinfo
  {journal} {Phys. Rev. B}\ }\textbf {\bibinfo {volume} {19}},\ \bibinfo
  {pages} {4382} (\bibinfo {year} {1979})}\BibitemShut {NoStop}%
\bibitem [{\citenamefont {Berger}(1996)}]{Berger96}%
  \BibitemOpen
  \bibfield  {author} {\bibinfo {author} {\bibfnamefont {L.}~\bibnamefont
  {Berger}},\ }\href {\doibase 10.1103/PhysRevB.54.9353} {\bibfield  {journal}
  {\bibinfo  {journal} {Phys. Rev. B}\ }\textbf {\bibinfo {volume} {54}},\
  \bibinfo {pages} {9353} (\bibinfo {year} {1996})}\BibitemShut {NoStop}%
\bibitem [{\citenamefont {Mizukami}\ \emph {et~al.}(2001)\citenamefont
  {Mizukami}, \citenamefont {Ando},\ and\ \citenamefont
  {Miyazaki}}]{Mizukami01}%
  \BibitemOpen
  \bibfield  {author} {\bibinfo {author} {\bibfnamefont {S.}~\bibnamefont
  {Mizukami}}, \bibinfo {author} {\bibfnamefont {Y.}~\bibnamefont {Ando}}, \
  and\ \bibinfo {author} {\bibfnamefont {T.}~\bibnamefont {Miyazaki}},\ }\href
  {\doibase 10.1143/JJAP.40.580} {\bibfield  {journal} {\bibinfo  {journal}
  {Japanese J. Appl. Phys.}\ }\textbf {\bibinfo {volume} {40}},\
  \bibinfo {pages} {580} (\bibinfo {year} {2001})}\BibitemShut {NoStop}%
\bibitem [{\citenamefont {\ifmmode~\check{S}\else \v{S}\fi{}im\'anek}\ and\
  \citenamefont {Heinrich}(2003)}]{SimanekHeinrich03}%
  \BibitemOpen
  \bibfield  {author} {\bibinfo {author} {\bibfnamefont {E.}~\bibnamefont
  {\ifmmode~\check{S}\else \v{S}\fi{}im\'anek}}\ and\ \bibinfo {author}
  {\bibfnamefont {B.}~\bibnamefont {Heinrich}},\ }\href {\doibase
  10.1103/PhysRevB.67.144418} {\bibfield  {journal} {\bibinfo  {journal} {Phys.
  Rev. B}\ }\textbf {\bibinfo {volume} {67}},\ \bibinfo {pages} {144418}
  (\bibinfo {year} {2003})}\BibitemShut {NoStop}%
\bibitem [{\citenamefont {\ifmmode~\check{S}\else
  \v{S}\fi{}im\'anek}(2003)}]{Simanek03}%
  \BibitemOpen
  \bibfield  {author} {\bibinfo {author} {\bibfnamefont {E.}~\bibnamefont
  {\ifmmode~\check{S}\else \v{S}\fi{}im\'anek}},\ }\href {\doibase
  10.1103/PhysRevB.68.224403} {\bibfield  {journal} {\bibinfo  {journal} {Phys.
  Rev. B}\ }\textbf {\bibinfo {volume} {68}},\ \bibinfo {pages} {224403}
  (\bibinfo {year} {2003})}\BibitemShut {NoStop}%
\bibitem [{\citenamefont {Chen}\ and\ \citenamefont {Zhang}(2015)}]{Chen15}%
  \BibitemOpen
  \bibfield  {author} {\bibinfo {author} {\bibfnamefont {K.}~\bibnamefont
  {Chen}}\ and\ \bibinfo {author} {\bibfnamefont {S.}~\bibnamefont {Zhang}},\
  }\href {\doibase 10.1103/PhysRevLett.114.126602} {\bibfield  {journal}
  {\bibinfo  {journal} {Phys. Rev. Lett.}\ }\textbf {\bibinfo {volume} {114}},\
  \bibinfo {pages} {126602} (\bibinfo {year} {2015})}\BibitemShut {NoStop}%
\bibitem [{\citenamefont {Tatara}(2016)}]{TataraSP16}%
  \BibitemOpen
  \bibfield  {author} {\bibinfo {author} {\bibfnamefont {G.}~\bibnamefont
  {Tatara}},\ }\href {\doibase 10.1103/PhysRevB.94.224412} {\bibfield
  {journal} {\bibinfo  {journal} {Phys. Rev. B}\ }\textbf {\bibinfo {volume}
  {94}},\ \bibinfo {pages} {224412} (\bibinfo {year} {2016})}\BibitemShut
  {NoStop}%
\bibitem [{\citenamefont {Fisher}\ and\ \citenamefont {Lee}(1981)}]{Fisher81}%
  \BibitemOpen
  \bibfield  {author} {\bibinfo {author} {\bibfnamefont {D.~S.}\ \bibnamefont
  {Fisher}}\ and\ \bibinfo {author} {\bibfnamefont {P.~A.}\ \bibnamefont
  {Lee}},\ }\href {\doibase 10.1103/PhysRevB.23.6851} {\bibfield  {journal}
  {\bibinfo  {journal} {Phys. Rev. B}\ }\textbf {\bibinfo {volume} {23}},\
  \bibinfo {pages} {6851} (\bibinfo {year} {1981})}\BibitemShut {NoStop}%
\bibitem [{\citenamefont {Takeuchi}\ and\ \citenamefont
  {Tatara}(2008)}]{Takeuchi08}%
  \BibitemOpen
  \bibfield  {author} {\bibinfo {author} {\bibfnamefont {A.}~\bibnamefont
  {Takeuchi}}\ and\ \bibinfo {author} {\bibfnamefont {G.}~\bibnamefont
  {Tatara}},\ }\href {\doibase 10.1143/JPSJ.77.074701} {\bibfield  {journal}
  {\bibinfo  {journal} {J. Phys. Soc.  Japan}\ }\textbf
  {\bibinfo {volume} {77}},\ \bibinfo {pages} {074701} (\bibinfo {year}
  {2008})} \BibitemShut {NoStop}%
\bibitem [{\citenamefont {Hosono}\ \emph {et~al.}(2009)\citenamefont {Hosono},
  \citenamefont {Takeuchi},\ and\ \citenamefont {Tatara}}]{Hosono_LT09}%
  \BibitemOpen
  \bibfield  {author} {\bibinfo {author} {\bibfnamefont {K.}~\bibnamefont
  {Hosono}}, \bibinfo {author} {\bibfnamefont {A.}~\bibnamefont {Takeuchi}}, \
  and\ \bibinfo {author} {\bibfnamefont {G.}~\bibnamefont {Tatara}},\ }\href
  {http://stacks.iop.org/1742-6596/150/022029} {\bibfield  {journal} {\bibinfo
  {journal} {Journal of Physics: Conference Series}\ }\textbf {\bibinfo
  {volume} {150}},\ \bibinfo {pages} {022029} (\bibinfo {year}
  {2009})}\BibitemShut {NoStop}%
\bibitem [{\citenamefont {Adachi}\ \emph {et~al.}(2011)\citenamefont {Adachi},
  \citenamefont {Ohe}, \citenamefont {Takahashi},\ and\ \citenamefont
  {Maekawa}}]{Adachi11}%
  \BibitemOpen
  \bibfield  {author} {\bibinfo {author} {\bibfnamefont {H.}~\bibnamefont
  {Adachi}}, \bibinfo {author} {\bibfnamefont {J.-i.}\ \bibnamefont {Ohe}},
  \bibinfo {author} {\bibfnamefont {S.}~\bibnamefont {Takahashi}}, \ and\
  \bibinfo {author} {\bibfnamefont {S.}~\bibnamefont {Maekawa}},\ }\href
  {\doibase 10.1103/PhysRevB.83.094410} {\bibfield  {journal} {\bibinfo
  {journal} {Phys. Rev. B}\ }\textbf {\bibinfo {volume} {83}},\ \bibinfo
  {pages} {094410} (\bibinfo {year} {2011})}\BibitemShut {NoStop}%
\bibitem [{\citenamefont {Sakurai}(1994)}]{Sakurai94}%
  \BibitemOpen
  \bibfield  {author} {\bibinfo {author} {\bibfnamefont {J.~J.}\ \bibnamefont
  {Sakurai}},\ }\href@noop {} {\emph {\bibinfo {title} {Modern Quantum
  Mechanics}}}\ (\bibinfo  {publisher} {Addison Wesley},\ \bibinfo {year}
  {1994})\BibitemShut {NoStop}%
\bibitem [{\citenamefont {Hashimoto}\ \emph {et~al.}(2017)\citenamefont
  {Hashimoto}, \citenamefont {Tatara},\ and\ \citenamefont
  {Uchiyama}}]{Hashimoto17}%
  \BibitemOpen
  \bibfield  {author} {\bibinfo {author} {\bibfnamefont {K.}~\bibnamefont
  {Hashimoto}}, \bibinfo {author} {\bibfnamefont {G.}~\bibnamefont {Tatara}}, \
  and\ \bibinfo {author} {\bibfnamefont {C.}~\bibnamefont {Uchiyama}},\
  }\href@noop {} {\bibfield  {journal} {\bibinfo  {journal} {arXiv:1706.00583}\ }
  (\bibinfo {year} {2017})}\BibitemShut {NoStop}%
\bibitem [{\citenamefont {Tatara}\ and\ \citenamefont {Entel}(2008)}]{TE08}%
  \BibitemOpen
  \bibfield  {author} {\bibinfo {author} {\bibfnamefont {G.}~\bibnamefont
  {Tatara}}\ and\ \bibinfo {author} {\bibfnamefont {P.}~\bibnamefont {Entel}},\
  }\href {\doibase 10.1103/PhysRevB.78.064429} {\bibfield  {journal} {\bibinfo
  {journal} {Phys. Rev. B}\ }\textbf {\bibinfo {volume} {78}},\ \bibinfo {eid}
  {064429} (\bibinfo {year} {2008})}\BibitemShut {NoStop}%
\bibitem [{\citenamefont {Kohno}\ \emph {et~al.}(2006)\citenamefont {Kohno},
  \citenamefont {Tatara},\ and\ \citenamefont {Shibata}}]{KTS06}%
  \BibitemOpen
  \bibfield  {author} {\bibinfo {author} {\bibfnamefont {H.}~\bibnamefont
  {Kohno}}, \bibinfo {author} {\bibfnamefont {G.}~\bibnamefont {Tatara}}, \
  and\ \bibinfo {author} {\bibfnamefont {J.}~\bibnamefont {Shibata}},\ }\href
  {\doibase 10.1143/JPSJ.75.113706} {\bibfield  {journal} {\bibinfo  {journal}
  {J. Phys.  Soc. Japan}\ }\textbf {\bibinfo {volume}
  {75}},\ \bibinfo {pages} {113706} (\bibinfo {year} {2006})}\BibitemShut
  {NoStop}%
\bibitem [{\citenamefont {Tatara}\ and\ \citenamefont
  {Fukuyama}(1994{\natexlab{a}})}]{TF94_JPSJ}%
  \BibitemOpen
  \bibfield  {author} {\bibinfo {author} {\bibfnamefont {G.}~\bibnamefont
  {Tatara}}\ and\ \bibinfo {author} {\bibfnamefont {H.}~\bibnamefont
  {Fukuyama}},\ }\href {\doibase 10.1143/JPSJ.63.2538} {\bibfield  {journal}
  {\bibinfo  {journal} {J.  Phys.l Soc. Japan}\ }\textbf
  {\bibinfo {volume} {63}},\ \bibinfo {pages} {2538} (\bibinfo {year}
  {1994}{\natexlab{a}})}\BibitemShut {NoStop}%
\bibitem [{\citenamefont {Tatara}\ and\ \citenamefont
  {Fukuyama}(1994{\natexlab{b}})}]{TF94}%
  \BibitemOpen
  \bibfield  {author} {\bibinfo {author} {\bibfnamefont {G.}~\bibnamefont
  {Tatara}}\ and\ \bibinfo {author} {\bibfnamefont {H.}~\bibnamefont
  {Fukuyama}},\ }\href {\doibase 10.1103/PhysRevLett.72.772} {\bibfield
  {journal} {\bibinfo  {journal} {Phys. Rev. Lett.}\ }\textbf {\bibinfo
  {volume} {72}},\ \bibinfo {pages} {772} (\bibinfo {year}
  {1994}{\natexlab{b}})}\BibitemShut {NoStop}%
\bibitem [{\citenamefont {Umetsu}\ \emph {et~al.}(2012)\citenamefont {Umetsu},
  \citenamefont {Miura},\ and\ \citenamefont {Sakuma}}]{Umetsu12}%
  \BibitemOpen
  \bibfield  {author} {\bibinfo {author} {\bibfnamefont {N.}~\bibnamefont
  {Umetsu}}, \bibinfo {author} {\bibfnamefont {D.}~\bibnamefont {Miura}}, \
  and\ \bibinfo {author} {\bibfnamefont {A.}~\bibnamefont {Sakuma}},\ }\href
  {\doibase 10.1143/JPSJ.81.114716} {\bibfield  {journal} {\bibinfo  {journal}
  {J. Phys. Soc. Japan}\ }\textbf {\bibinfo {volume}
  {81}},\ \bibinfo {pages} {114716} (\bibinfo {year} {2012})}
  \BibitemShut {NoStop}%
\bibitem [{\citenamefont {Yuan}\ \emph {et~al.}(2016)\citenamefont {Yuan},
  \citenamefont {Yuan}, \citenamefont {Xia},\ and\ \citenamefont
  {Wang}}]{Yuan16}%
  \BibitemOpen
  \bibfield  {author} {\bibinfo {author} {\bibfnamefont {H.~Y.}\ \bibnamefont
  {Yuan}}, \bibinfo {author} {\bibfnamefont {Z.}~\bibnamefont {Yuan}}, \bibinfo
  {author} {\bibfnamefont {K.}~\bibnamefont {Xia}}, \ and\ \bibinfo {author}
  {\bibfnamefont {X.~R.}\ \bibnamefont {Wang}},\ }\href {\doibase
  10.1103/PhysRevB.94.064415} {\bibfield  {journal} {\bibinfo  {journal} {Phys.
  Rev. B}\ }\textbf {\bibinfo {volume} {94}},\ \bibinfo {pages} {064415}
  (\bibinfo {year} {2016})}\BibitemShut {NoStop}%
\bibitem [{\citenamefont {Tatara}(2015)}]{TataraDW15}%
  \BibitemOpen
  \bibfield  {author} {\bibinfo {author} {\bibfnamefont {G.}~\bibnamefont
  {Tatara}},\ }\href {\doibase 10.1103/PhysRevB.92.064405} {\bibfield
  {journal} {\bibinfo  {journal} {Phys. Rev. B}\ }\textbf {\bibinfo {volume}
  {92}},\ \bibinfo {pages} {064405} (\bibinfo {year} {2015})}\BibitemShut
  {NoStop}%
\bibitem [{\citenamefont {Kittel}(1963)}]{Kittel63}%
  \BibitemOpen
  \bibfield  {author} {\bibinfo {author} {\bibfnamefont {C.}~\bibnamefont
  {Kittel}},\ }\href@noop {} {\emph {\bibinfo {title} {Quantum theory of
  solids}}}\ (\bibinfo  {publisher} {Wiley},\ \bibinfo {address} {New York},\
  \bibinfo {year} {1963})\BibitemShut {NoStop}%
\bibitem [{\citenamefont {Nakata}\ and\ \citenamefont
  {Tatara}(2011)}]{Nakata11}%
  \BibitemOpen
  \bibfield  {author} {\bibinfo {author} {\bibfnamefont {K.}~\bibnamefont
  {Nakata}}\ and\ \bibinfo {author} {\bibfnamefont {G.}~\bibnamefont
  {Tatara}},\ }\href {\doibase 10.1143/JPSJ.80.054602} {\bibfield  {journal}
  {\bibinfo  {journal} {J. Phys. Soc. Japan}\ }\textbf
  {\bibinfo {volume} {80}},\ \bibinfo {pages} {054602} (\bibinfo {year}
  {2011})}\BibitemShut {NoStop}%
\bibitem [{\citenamefont {Kajiwara}\ \emph {et~al.}(2010)\citenamefont
  {Kajiwara}, \citenamefont {Harii}, \citenamefont {Takahashi}, \citenamefont
  {Ohe}, \citenamefont {Uchida}, \citenamefont {Mizuguchi}, \citenamefont
  {Umezawa}, \citenamefont {Kawai}, \citenamefont {Ando}, \citenamefont
  {Takanashi}, \citenamefont {Maekawa},\ and\ \citenamefont
  {Saitoh}}]{Kajiwara10}%
  \BibitemOpen
  \bibfield  {author} {\bibinfo {author} {\bibfnamefont {Y.}~\bibnamefont
  {Kajiwara}}, \bibinfo {author} {\bibfnamefont {K.}~\bibnamefont {Harii}},
  \bibinfo {author} {\bibfnamefont {S.}~\bibnamefont {Takahashi}}, \bibinfo
  {author} {\bibfnamefont {J.}~\bibnamefont {Ohe}}, \bibinfo {author}
  {\bibfnamefont {K.}~\bibnamefont {Uchida}}, \bibinfo {author} {\bibfnamefont
  {M.}~\bibnamefont {Mizuguchi}}, \bibinfo {author} {\bibfnamefont
  {H.}~\bibnamefont {Umezawa}}, \bibinfo {author} {\bibfnamefont
  {H.}~\bibnamefont {Kawai}}, \bibinfo {author} {\bibfnamefont
  {K.}~\bibnamefont {Ando}}, \bibinfo {author} {\bibfnamefont {K.}~\bibnamefont
  {Takanashi}}, \bibinfo {author} {\bibfnamefont {S.}~\bibnamefont {Maekawa}},
  \ and\ \bibinfo {author} {\bibfnamefont {E.}~\bibnamefont {Saitoh}},\ }\href
  {\doibase 10.1038/nature08876} {\bibfield  {journal} {\bibinfo  {journal}
  {Nature}\ }\textbf {\bibinfo {volume} {464}},\ \bibinfo {pages} {262}
  (\bibinfo {year} {2010})}\BibitemShut {NoStop}%
\bibitem [{\citenamefont {Czeschka}\ \emph {et~al.}(2011)\citenamefont
  {Czeschka}, \citenamefont {Dreher}, \citenamefont {Brandt}, \citenamefont
  {Weiler}, \citenamefont {Althammer}, \citenamefont {Imort}, \citenamefont
  {Reiss}, \citenamefont {Thomas}, \citenamefont {Schoch}, \citenamefont
  {Limmer}, \citenamefont {Huebl}, \citenamefont {Gross},\ and\ \citenamefont
  {Goennenwein}}]{Czeschka11}%
  \BibitemOpen
  \bibfield  {author} {\bibinfo {author} {\bibfnamefont {F.~D.}\ \bibnamefont
  {Czeschka}}, \bibinfo {author} {\bibfnamefont {L.}~\bibnamefont {Dreher}},
  \bibinfo {author} {\bibfnamefont {M.~S.}\ \bibnamefont {Brandt}}, \bibinfo
  {author} {\bibfnamefont {M.}~\bibnamefont {Weiler}}, \bibinfo {author}
  {\bibfnamefont {M.}~\bibnamefont {Althammer}}, \bibinfo {author}
  {\bibfnamefont {I.-M.}\ \bibnamefont {Imort}}, \bibinfo {author}
  {\bibfnamefont {G.}~\bibnamefont {Reiss}}, \bibinfo {author} {\bibfnamefont
  {A.}~\bibnamefont {Thomas}}, \bibinfo {author} {\bibfnamefont
  {W.}~\bibnamefont {Schoch}}, \bibinfo {author} {\bibfnamefont
  {W.}~\bibnamefont {Limmer}}, \bibinfo {author} {\bibfnamefont
  {H.}~\bibnamefont {Huebl}}, \bibinfo {author} {\bibfnamefont
  {R.}~\bibnamefont {Gross}}, \ and\ \bibinfo {author} {\bibfnamefont
  {S.~T.~B.}\ \bibnamefont {Goennenwein}},\ }\href {\doibase
  10.1103/PhysRevLett.107.046601} {\bibfield  {journal} {\bibinfo  {journal}
  {Phys. Rev. Lett.}\ }\textbf {\bibinfo {volume} {107}},\ \bibinfo {pages}
  {046601} (\bibinfo {year} {2011})}\BibitemShut {NoStop}%
\bibitem [{\citenamefont {Heinrich}\ \emph {et~al.}(2011)\citenamefont
  {Heinrich}, \citenamefont {Burrowes}, \citenamefont {Montoya}, \citenamefont
  {Kardasz}, \citenamefont {Girt}, \citenamefont {Song}, \citenamefont {Sun},\
  and\ \citenamefont {Wu}}]{Heinrich11}%
  \BibitemOpen
  \bibfield  {author} {\bibinfo {author} {\bibfnamefont {B.}~\bibnamefont
  {Heinrich}}, \bibinfo {author} {\bibfnamefont {C.}~\bibnamefont {Burrowes}},
  \bibinfo {author} {\bibfnamefont {E.}~\bibnamefont {Montoya}}, \bibinfo
  {author} {\bibfnamefont {B.}~\bibnamefont {Kardasz}}, \bibinfo {author}
  {\bibfnamefont {E.}~\bibnamefont {Girt}}, \bibinfo {author} {\bibfnamefont
  {Y.-Y.}\ \bibnamefont {Song}}, \bibinfo {author} {\bibfnamefont
  {Y.}~\bibnamefont {Sun}}, \ and\ \bibinfo {author} {\bibfnamefont
  {M.}~\bibnamefont {Wu}},\ }\href {\doibase 10.1103/PhysRevLett.107.066604}
  {\bibfield  {journal} {\bibinfo  {journal} {Phys. Rev. Lett.}\ }\textbf
  {\bibinfo {volume} {107}},\ \bibinfo {pages} {066604} (\bibinfo {year}
  {2011})}\BibitemShut {NoStop}%
\bibitem [{\citenamefont {Burrowes}\ \emph {et~al.}(2012)\citenamefont
  {Burrowes}, \citenamefont {Heinrich}, \citenamefont {Kardasz}, \citenamefont
  {Montoya}, \citenamefont {Girt}, \citenamefont {Sun}, \citenamefont {Song},\
  and\ \citenamefont {Wu}}]{Burrowes12}%
  \BibitemOpen
  \bibfield  {author} {\bibinfo {author} {\bibfnamefont {C.}~\bibnamefont
  {Burrowes}}, \bibinfo {author} {\bibfnamefont {B.}~\bibnamefont {Heinrich}},
  \bibinfo {author} {\bibfnamefont {B.}~\bibnamefont {Kardasz}}, \bibinfo
  {author} {\bibfnamefont {E.~A.}\ \bibnamefont {Montoya}}, \bibinfo {author}
  {\bibfnamefont {E.}~\bibnamefont {Girt}}, \bibinfo {author} {\bibfnamefont
  {Y.}~\bibnamefont {Sun}}, \bibinfo {author} {\bibfnamefont {Y.-Y.}\
  \bibnamefont {Song}}, \ and\ \bibinfo {author} {\bibfnamefont
  {M.}~\bibnamefont {Wu}},\ }\href {\doibase 10.1063/1.3690918} {\bibfield
  {journal} {\bibinfo  {journal} {Applied Physics Letters}\ }\textbf {\bibinfo
  {volume} {100}},\ \bibinfo {pages} {092403} (\bibinfo {year} {2012})}
  \BibitemShut {NoStop}%
\bibitem [{\citenamefont {Qiu}\ \emph {et~al.}(2013)\citenamefont {Qiu},
  \citenamefont {Ando}, \citenamefont {Uchida}, \citenamefont {Kajiwara},
  \citenamefont {Takahashi}, \citenamefont {Nakayama}, \citenamefont {An},
  \citenamefont {Fujikawa},\ and\ \citenamefont {Saitoh}}]{Qiu13}%
  \BibitemOpen
  \bibfield  {author} {\bibinfo {author} {\bibfnamefont {Z.}~\bibnamefont
  {Qiu}}, \bibinfo {author} {\bibfnamefont {K.}~\bibnamefont {Ando}}, \bibinfo
  {author} {\bibfnamefont {K.}~\bibnamefont {Uchida}}, \bibinfo {author}
  {\bibfnamefont {Y.}~\bibnamefont {Kajiwara}}, \bibinfo {author}
  {\bibfnamefont {R.}~\bibnamefont {Takahashi}}, \bibinfo {author}
  {\bibfnamefont {H.}~\bibnamefont {Nakayama}}, \bibinfo {author}
  {\bibfnamefont {T.}~\bibnamefont {An}}, \bibinfo {author} {\bibfnamefont
  {Y.}~\bibnamefont {Fujikawa}}, \ and\ \bibinfo {author} {\bibfnamefont
  {E.}~\bibnamefont {Saitoh}},\ }\href {\doibase 10.1063/1.4819460} {\bibfield
  {journal} {\bibinfo  {journal} {Applied Physics Letters}\ }\textbf {\bibinfo
  {volume} {103}},\ \bibinfo {pages} {092404} (\bibinfo {year} {2013})}
  \BibitemShut {NoStop}%
\bibitem [{\citenamefont {Jia}\ \emph {et~al.}(2011)\citenamefont {Jia},
  \citenamefont {Liu}, \citenamefont {Xia},\ and\ \citenamefont
  {Bauer}}]{Jia11}%
  \BibitemOpen
  \bibfield  {author} {\bibinfo {author} {\bibfnamefont {X.}~\bibnamefont
  {Jia}}, \bibinfo {author} {\bibfnamefont {K.}~\bibnamefont {Liu}}, \bibinfo
  {author} {\bibfnamefont {K.}~\bibnamefont {Xia}}, \ and\ \bibinfo {author}
  {\bibfnamefont {G.~E.~W.}\ \bibnamefont {Bauer}},\ }\href
  {http://stacks.iop.org/0295-5075/96/i=1/a=17005} {\bibfield  {journal}
  {\bibinfo  {journal} {EPL (Europhysics Letters)}\ }\textbf {\bibinfo {volume}
  {96}},\ \bibinfo {pages} {17005} (\bibinfo {year} {2011})}\BibitemShut
  {NoStop}%
\bibitem [{\citenamefont {Wang}\ \emph {et~al.}(2014)\citenamefont {Wang},
  \citenamefont {Du}, \citenamefont {Pu}, \citenamefont {Adur}, \citenamefont
  {Hammel},\ and\ \citenamefont {Yang}}]{WangPRL14}%
  \BibitemOpen
  \bibfield  {author} {\bibinfo {author} {\bibfnamefont {H.~L.}\ \bibnamefont
  {Wang}}, \bibinfo {author} {\bibfnamefont {C.~H.}\ \bibnamefont {Du}},
  \bibinfo {author} {\bibfnamefont {Y.}~\bibnamefont {Pu}}, \bibinfo {author}
  {\bibfnamefont {R.}~\bibnamefont {Adur}}, \bibinfo {author} {\bibfnamefont
  {P.~C.}\ \bibnamefont {Hammel}}, \ and\ \bibinfo {author} {\bibfnamefont
  {F.~Y.}\ \bibnamefont {Yang}},\ }\href {\doibase
  10.1103/PhysRevLett.112.197201} {\bibfield  {journal} {\bibinfo  {journal}
  {Phys. Rev. Lett.}\ }\textbf {\bibinfo {volume} {112}},\ \bibinfo {pages}
  {197201} (\bibinfo {year} {2014})}\BibitemShut {NoStop}%
\bibitem [{\citenamefont {Du}\ \emph {et~al.}(2014{\natexlab{a}})\citenamefont
  {Du}, \citenamefont {Wang}, \citenamefont {Yang},\ and\ \citenamefont
  {Hammel}}]{Du14}%
  \BibitemOpen
  \bibfield  {author} {\bibinfo {author} {\bibfnamefont {C.}~\bibnamefont
  {Du}}, \bibinfo {author} {\bibfnamefont {H.}~\bibnamefont {Wang}}, \bibinfo
  {author} {\bibfnamefont {F.}~\bibnamefont {Yang}}, \ and\ \bibinfo {author}
  {\bibfnamefont {P.~C.}\ \bibnamefont {Hammel}},\ }\href {\doibase
  10.1103/PhysRevApplied.1.044004} {\bibfield  {journal} {\bibinfo  {journal}
  {Phys. Rev. Applied}\ }\textbf {\bibinfo {volume} {1}},\ \bibinfo {pages}
  {044004} (\bibinfo {year} {2014}{\natexlab{a}})}\BibitemShut {NoStop}%
\bibitem [{\citenamefont {Du}\ \emph {et~al.}(2014{\natexlab{b}})\citenamefont
  {Du}, \citenamefont {Wang}, \citenamefont {Yang},\ and\ \citenamefont
  {Hammel}}]{DuPRB14}%
  \BibitemOpen
  \bibfield  {author} {\bibinfo {author} {\bibfnamefont {C.}~\bibnamefont
  {Du}}, \bibinfo {author} {\bibfnamefont {H.}~\bibnamefont {Wang}}, \bibinfo
  {author} {\bibfnamefont {F.}~\bibnamefont {Yang}}, \ and\ \bibinfo {author}
  {\bibfnamefont {P.~C.}\ \bibnamefont {Hammel}},\ }\href {\doibase
  10.1103/PhysRevB.90.140407} {\bibfield  {journal} {\bibinfo  {journal} {Phys.
  Rev. B}\ }\textbf {\bibinfo {volume} {90}},\ \bibinfo {pages} {140407}
  (\bibinfo {year} {2014}{\natexlab{b}})}\BibitemShut {NoStop}%
\bibitem [{\citenamefont {Wang}\ \emph {et~al.}(2017)\citenamefont {Wang},
  \citenamefont {Du}, \citenamefont {Hammel},\ and\ \citenamefont
  {Yang}}]{Wang17}%
  \BibitemOpen
  \bibfield  {author} {\bibinfo {author} {\bibfnamefont {H.}~\bibnamefont
  {Wang}}, \bibinfo {author} {\bibfnamefont {C.}~\bibnamefont {Du}}, \bibinfo
  {author} {\bibfnamefont {P.~C.}\ \bibnamefont {Hammel}}, \ and\ \bibinfo
  {author} {\bibfnamefont {F.}~\bibnamefont {Yang}},\ }\href {\doibase
  10.1063/1.4975704} {\bibfield  {journal} {\bibinfo  {journal} {Applied
  Physics Letters}\ }\textbf {\bibinfo {volume} {110}},\ \bibinfo {pages}
  {062402} (\bibinfo {year} {2017})}
  \BibitemShut {NoStop}%
\bibitem [{\citenamefont {Lu}\ \emph {et~al.}(2013)\citenamefont {Lu},
  \citenamefont {Choi}, \citenamefont {Ortega}, \citenamefont {Cheng},
  \citenamefont {Cai}, \citenamefont {Huang}, \citenamefont {Sun},\ and\
  \citenamefont {Chien}}]{Lu13}%
  \BibitemOpen
  \bibfield  {author} {\bibinfo {author} {\bibfnamefont {Y.~M.}\ \bibnamefont
  {Lu}}, \bibinfo {author} {\bibfnamefont {Y.}~\bibnamefont {Choi}}, \bibinfo
  {author} {\bibfnamefont {C.~M.}\ \bibnamefont {Ortega}}, \bibinfo {author}
  {\bibfnamefont {X.~M.}\ \bibnamefont {Cheng}}, \bibinfo {author}
  {\bibfnamefont {J.~W.}\ \bibnamefont {Cai}}, \bibinfo {author} {\bibfnamefont
  {S.~Y.}\ \bibnamefont {Huang}}, \bibinfo {author} {\bibfnamefont
  {L.}~\bibnamefont {Sun}}, \ and\ \bibinfo {author} {\bibfnamefont {C.~L.}\
  \bibnamefont {Chien}},\ }\href {\doibase 10.1103/PhysRevLett.110.147207}
  {\bibfield  {journal} {\bibinfo  {journal} {Phys. Rev. Lett.}\ }\textbf
  {\bibinfo {volume} {110}},\ \bibinfo {pages} {147207} (\bibinfo {year}
  {2013})}\BibitemShut {NoStop}%
\bibitem [{\citenamefont {Xia}\ \emph {et~al.}(1997)\citenamefont {Xia},
  \citenamefont {Zhang}, \citenamefont {Lu},\ and\ \citenamefont
  {Zhai}}]{Xia97}%
  \BibitemOpen
  \bibfield  {author} {\bibinfo {author} {\bibfnamefont {K.}~\bibnamefont
  {Xia}}, \bibinfo {author} {\bibfnamefont {W.}~\bibnamefont {Zhang}}, \bibinfo
  {author} {\bibfnamefont {M.}~\bibnamefont {Lu}}, \ and\ \bibinfo {author}
  {\bibfnamefont {H.}~\bibnamefont {Zhai}},\ }\href {\doibase
  10.1103/PhysRevB.55.12561} {\bibfield  {journal} {\bibinfo  {journal} {Phys.
  Rev. B}\ }\textbf {\bibinfo {volume} {55}},\ \bibinfo {pages} {12561}
  (\bibinfo {year} {1997})}\BibitemShut {NoStop}%
\bibitem [{\citenamefont {Caminale}\ \emph {et~al.}(2016)\citenamefont
  {Caminale}, \citenamefont {Ghosh}, \citenamefont {Auffret}, \citenamefont
  {Ebels}, \citenamefont {Ollefs}, \citenamefont {Wilhelm}, \citenamefont
  {Rogalev},\ and\ \citenamefont {Bailey}}]{Caminale16}%
  \BibitemOpen
  \bibfield  {author} {\bibinfo {author} {\bibfnamefont {M.}~\bibnamefont
  {Caminale}}, \bibinfo {author} {\bibfnamefont {A.}~\bibnamefont {Ghosh}},
  \bibinfo {author} {\bibfnamefont {S.}~\bibnamefont {Auffret}}, \bibinfo
  {author} {\bibfnamefont {U.}~\bibnamefont {Ebels}}, \bibinfo {author}
  {\bibfnamefont {K.}~\bibnamefont {Ollefs}}, \bibinfo {author} {\bibfnamefont
  {F.}~\bibnamefont {Wilhelm}}, \bibinfo {author} {\bibfnamefont
  {A.}~\bibnamefont {Rogalev}}, \ and\ \bibinfo {author} {\bibfnamefont
  {W.~E.}\ \bibnamefont {Bailey}},\ }\href {\doibase
  10.1103/PhysRevB.94.014414} {\bibfield  {journal} {\bibinfo  {journal} {Phys.
  Rev. B}\ }\textbf {\bibinfo {volume} {94}},\ \bibinfo {pages} {014414}
  (\bibinfo {year} {2016})}\BibitemShut {NoStop}%
\bibitem [{\citenamefont {Auerbach}(1994)}]{Auerbach94}%
  \BibitemOpen
  \bibfield  {author} {\bibinfo {author} {\bibfnamefont {A.}~\bibnamefont
  {Auerbach}},\ }\href@noop {} {\emph {\bibinfo {title} {Intracting Electrons
  and Quantum Magnetism}}}\ (\bibinfo  {publisher} {Springer Verlag},\ \bibinfo
  {year} {1994})\BibitemShut {NoStop}%
\end{thebibliography}
